\documentclass[reprint,aps,prc,amsmath,amssymb,nofootinbib,superscriptaddress]{revtex4-2}

\usepackage{mathtools}
\usepackage{braket} 
\usepackage[dvipsnames]{xcolor}  

\usepackage[T1]{fontenc}
\usepackage[utf8]{inputenc}
\usepackage{graphicx,color,rotating,pifont}
\usepackage{amsmath,amssymb}
\usepackage{mathtools}
\usepackage{siunitx}
\usepackage{bm}
\usepackage{ae}
\usepackage{dcolumn}
\usepackage{txfonts}
\usepackage{tensor}
\usepackage{braket}
\usepackage{booktabs}
\usepackage{xspace}
\usepackage{soul}
\setstcolor{red}
\setul{}{.2ex} 

\usepackage[normalem]{ulem}

\usepackage{tikz}
\usetikzlibrary{positioning,calc,shapes,decorations.markings,decorations.pathmorphing}
\tikzset{asg/.cd,
  omega-vertex/.style={circle,solid,draw=black,fill=white,minimum size=5pt, inner sep=0pt},
  dbd-vertex/.style={coordinate},
  pline/.style={thick, postaction={decorate}, decoration={markings, mark=at position .5 with {\arrow[xshift=2pt]{stealth}}}},
  hline/.style={thick, postaction={decorate}, decoration={markings, mark=at position .5 with {\arrowreversed[xshift=-2pt]{stealth}}}},
  shift arrow/.style={/pgf/decoration/transform={xshift=#1}},
  shift arrow/.default=-2pt,
  dbd-2b/.style={decorate, decoration=snake},
  omega-2b/.style={densely dashed},
  neutron/.style={draw=blue},
  proton/.style={draw=red},
}

\usepackage[colorlinks,linkcolor=blue,anchorcolor=blue,citecolor=blue]{hyperref} 
\hypersetup{
   colorlinks=true, 
   linkcolor=blue,  
   citecolor=blue,  
   filecolor=blue,  
   urlcolor=blue    
}

\DeclareMathSymbol{\NS}{\mathord}{AMSb}{"4E}

\DeclareSIUnit{\fm}{\femto\meter}

\newcommand{\dd}{\ensuremath{\mathrm{d}}}

\renewcommand{\vec}[1]{\ensuremath{\bm{#1}}}

\newcommand{\beq}{\begin{equation}}
\newcommand{\eeq}{\end{equation}}
\newcommand{\beqn}{\begin{eqnarray}}
\newcommand{\eeqn}{\end{eqnarray}}
\newcommand{\bsub}{\begin{subequations}}
\newcommand{\esub}{\end{subequations}}
\newcommand{\bpm}{\begin{pmatrix}}
\newcommand{\epm}{\end{pmatrix}}

\begin{document}

\title{{\it Ab initio} studies of double Gamow-Teller transition and its correlation with neutrinoless double beta decay}

\author{J. M. Yao}   
\email{yaojm8@mail.sysu.edu.cn}
\affiliation{School of Physics and Astronomy, Sun Yat-sen University, Zhuhai 519082, P.R. China}
\affiliation{Facility for Rare Isotope Beams, Michigan State University, East Lansing, Michigan 48824-1321, USA}

\author{I. Ginnett}%
\affiliation{TRIUMF 4004 Wesbrook Mall, Vancouver BC V6T 2A3, Canada}%
 \affiliation{Department of Physics \& Astronomy, Michigan State University, East Lansing, Michigan 48824-1321, USA}

\author{A. Belley}%
\email{abelley@triumf.ca}
\affiliation{TRIUMF 4004 Wesbrook Mall, Vancouver BC V6T 2A3, Canada}%
\affiliation{Department of Physics \& Astronomy, University of British Columbia, Vancouver, British Columbia V6T 1Z1, Canada}

\author{T. Miyagi}%
\affiliation{Technische Universit\"at Darmstadt, Department of Physics, 64289 Darmstadt, Germany}
\affiliation{ExtreMe Matter Institute EMMI, GSI Helmholtzzentrum f\"ur Schwerionenforschung GmbH, 64291 Darmstadt, Germany}

 \author{R. Wirth}   
  \affiliation{Facility for Rare Isotope Beams, Michigan State University, East Lansing, Michigan 48824-1321, USA}

 \author{S. Bogner} 
  \affiliation{Facility for Rare Isotope Beams, Michigan State University, East Lansing, Michigan 48824-1321, USA}
 \affiliation{Department of Physics \& Astronomy, Michigan State University, East Lansing, Michigan 48824-1321, USA}

\author{J. Engel} 
\address{Department of Physics and Astronomy, University of North Carolina, Chapel Hill, North Carolina 27516-3255, USA}

 \author{H. Hergert} 
\email{hergert@frib.msu.edu}
  \affiliation{Facility for Rare Isotope Beams, Michigan State University, East Lansing, Michigan 48824-1321, USA}
 \affiliation{Department of Physics \& Astronomy, Michigan State University, East Lansing, Michigan 48824-1321, USA}

\author{J. D. Holt}%
\email{jholt@triumf.ca}
\affiliation{TRIUMF 4004 Wesbrook Mall, Vancouver BC V6T 2A3, Canada}%
\affiliation{Department of Physics, McGill University, 3600 Rue University, Montr\'eal, QC H3A 2T8, Canada}%
 
\author{S. R. Stroberg}%
\affiliation{Physics Division, Argonne National Laboratory, Lemont, IL, 60439, USA}
 
\date{\today}

\begin{abstract} 
 We use chiral interactions and several {\em ab initio} methods to compute the  nuclear matrix elements (NMEs) for ground-state to ground-state double Gamow-Teller  transitions in a range of isotopes, and explore the correlation of these NMEs with those for neutrinoless double beta decay produced by the exchange of a light Majorana neutrino. When all the NMEs of both isospin-conserving and isospin-changing transitions from the {\em ab initio} calculations are considered, the correlation is strong.
 For the experimentally relevant isospin-changing transitions by themselves, however, the correlation is weaker and may not be helpful for reducing the uncertainty in the NMEs for neutrinoless double beta decay.

\end{abstract}

\maketitle

\section{Introduction}

 Neutrinoless double beta ($0\nu\beta\beta$) decay is a hypothetical lepton-number-violating process~\cite{Furry:1939} in which two neutrons in a parent nucleus decay into two protons in its daughter nucleus via the emission of only two electrons.  The hunt for  $0\nu\beta\beta$ decay is of particular importance as its observation would demonstrate the Majorana nature of neutrinos and provide a key ingredient for generating the matter-antimatter asymmetry in the Universe. If $0\nu\beta\beta$ decay is driven by  the  standard mechanism of exchanging light Majorana neutrinos, its half-life can also be used to determine the effective neutrino mass $\langle m_{\beta\beta}\rangle=\sum_iU^2_{ei}m_i$, where $m_i$ are the masses of light neutrinos  and $U_{ei}$ are elements of the neutrino-mixing matrix~\cite{PDG:2018}. The $0\nu\beta\beta$ decay rate is governed by a nuclear matrix element (NME) that must be computed.  The precise determination of the NMEs for candidate nuclei, which are vital for interpreting and planning the current-~\cite{EXO-200:2019,CUORE:2019,GERDA:2020,KamLAND-Zen:2022} and next-generation ton-scale~\cite{Xie:2020,LEGEND:2021,CUPID:2022} experiments, is thus being pursued energetically by theoretical physicists~\cite{Cirigliano:2022}.
  
 A wide range of conventional nuclear models have been applied to compute the NMEs in nuclei of interest to experiment~\cite{Menendez:2009,Rodriguez:2010,Barea:2013,Mustonen:2013,Holt:2013,Kwiatkowski:2014,Song:2014,Yao:2015,Hyvarinen:2015,Horoi:2016,Song:2017,Jiao:2017,Yoshinaga:2018,Fang:2018,Rath:2019,Terasaki:2019,Coraggio:2020,Deppisch:2020ztt,Wang:2021,Coraggio:2022}, under the assumption that light-neutrino exchange is the dominant decay mechanism. The discrepancy between these predictions is as large as a factor of about three, causing uncertainty at the level of an order of magnitude in the half-life for a given value of the neutrino mass. Resolving this discrepancy has been one of the most significant objectives in the nuclear community~\cite{Menendez:2014,Menendez:2016}; see for instance the recent reviews in Refs.~\cite{LongRangePlan2015,Engel:2017,Yao:2021Review,Agostini:2022RMP}.   Unfortunately, the systematic uncertainty turns out to be difficult to reduce because each model has its phenomenological assumptions and uncontrolled approximations.  In recent years, remarkable progress has been achieved in first-principles calculations of nuclear structure and reactions~\cite{Hergert:2020}. It has enabled the first wave of multi-method results for $0\nu\beta\beta$-decay NMEs in light nuclei~\cite{Pastore:2018,Cirigliano:2019PRC,Basili2020,Yao:2021PRC}, the lightest experimental candidate $^{48}$Ca~\cite{Yao:2020PRL,Novario:2021PRL}, and in one case even in the heavier candidates $^{76}$Ge and $^{82}$Se~\cite{Belley2021PRL}. These calculations start with realistic two-nucleon-plus-three-nucleon (NN+3N) interactions from either phenomenological parametrization or chiral effective field theory (EFT).  In heavier candidate nuclei, such calculations that include fully controllable uncertainties are still challenging, however~\cite{Yao:2021Review}.  Under these circumstances, it is worthwhile to explore correlations between the $0\nu\beta\beta$-decay NMEs and other observables.  Such correlations may provide model-independent constraints on the NMEs.  
 
 Recently, Shimizu {\em et al.}~\cite{Shimizu:2018PRL} found a linear correlation between the NMEs of $0\nu\beta\beta$ decay and those that govern the ground-state to ground-state double Gamow-Teller (DGT)  transition, a double spin-isospin flip excitation mode accessible in  high-energy heavy-ion double-charge-exchange (HIDCE) processes~\cite{Vogel:1988,Auerbach:1989,Zheng:1989,Zheng:1990}.  The correlation, which appeared in both medium-mass and heavy nuclei in calculations based on the large-scale nuclear shell-model,  was attributed to the mainly short-range character of both transitions~\cite{Menendez:2018JPS,Menendez:2017JPCS}.   These studies provide strong support to experimental programs to measure HIDCE reactions, through examples such as $^{12}$C($^{18}$O,$^{18}$Ne)$^{12}$Be~\cite{Takaki:2015} and $(^{11}$B, $^{11}$Li)~\cite{Takahisa:2017} at the Research Center of Nuclear Physics, Osaka University, and others in the NUMEN project at the Laboratori Nazionali del Sud (LNS), Istituto Nazionale di Fisica Nucleare (INFN)~\cite{Cappuzzello:2018}. The expectation is that the  cross-section in HIDCE reactions can place a constraint on the NMEs for $0\nu\beta\beta$ decay if the correlation exists and is universal. Santopinto {\em et al.}~\cite{Santopinto:2018} argued that it is possible to factorize the HIDCE cross section into reaction and nuclear parts. The latter can be further written as a product of the DGT NMEs for projectile and target nuclei.  This study showed that the  DGT NME is linearly correlated  with the total NME for $0\nu\beta\beta$ decay predicted by the interacting boson model (IBM). Recently, Brase {\em et al.}~\cite{Brase:2021} exploited  this correlation to determine the NMEs for $0\nu\beta\beta$ decay in heavier candidate nuclei with an effective field theory (EFT).  
 
 Ref.~\cite{Shimizu:2018PRL} shows, however, that the correlation does not appear in the results of the quasiparticle random-phase approximation (QRPA)~\cite{Simkovic:2018}, which has been extensively used to calculate $0\nu\beta\beta$-decay NMEs~\cite{Faessler:1998JPG,Rodin:2006,Simkovic:2013,Mustonen:2013,Terasaki:2015PRC,Fang:2018}. The contradiction between this method and others needs to be resolved to clarify the significance of HIDCE experiments ~\cite{Takahisa:2017,Cappuzzello:2018,Cappuzzello:2020}. In this paper, we use {\em ab initio} methods to address the issue.  So that we can compare our results with those obtained previously, we do not include the recently discovered contact transition operator~\cite{Cirigliano:2018,Cirigliano:2021PRL}, even though it might affect the $M^{0\nu\beta\beta}$ significantly within an {\em ab initio} framework~\cite{Wirth:2021}.

 The paper is organized as follows. In Sec.~\ref{sec:operators}, the formulas for the NMEs of both DGT transition and $0\nu\beta\beta$ decay are presented. The feature of neutrino potentials regularized with dipole form factors in coordinate space is discussed. In Sec.~\ref{sec:results}, we briefly introduce the many-body methods and nuclear Hamiltonians that are employed in the {\em ab initio} calculations. The results are also discussed in comparison with a scale-separation analysis in Sec.~\ref{sec:scales}.  The in-medium renormalization effect is discussed in comparison with conventional shell-model results in Sec.~\ref{sec:effect}. Our conclusions are summarized in Sec.~\ref{sec:summary}.

\section{The  DGT and $0\nu\beta\beta$ transitions } 
\label{sec:operators}

The spin-parity $J^\pi$ of the ground state of an even-even nucleus is $0^+$. A DGT transition connects this state to the final states of its neighboring even-even nucleus with spin-parity of $0^+$ or $2^+$. Here we only consider the  ground-state to ground-state DGT transition as its NME is expected to be closest to that of the $0\nu\beta\beta$ decay,  even though it might be just a small fraction (about $10^{-4}$) of the total DGT strength for the isospin-changing transitions~\cite{Haxton1984PPNP,Vogel:1988,Shimizu:2018PRL}.
The NME of the DGT transition is defined as
\begin{align}  
\label{eq:DGT}
M^{\rm DGT} &=\left\langle 0^+_f \left|\sum_{1,2}[\bm{\sigma}_1\otimes \bm{\sigma}_2]^0 \tau^{+}_1\tau^{+}_2\right| 0^+_i\right\rangle,
\end{align}  
where $\boldsymbol{\sigma}$ and $\tau$ are the spin and isospin operators, respectively.   The nonzero matrix element of the isospin-raising operator is $\bra{pp}  \tau^{+}_1\tau^{+}_2\ket{nn}=1$.  The ground-state wave functions of initial and final nuclei $|0^+_{i/f}\rangle$  also enter into the expression for the  NME  $M^{0\nu\beta\beta}$ of $0\nu\beta\beta$ decay,  which in the standard mechanism has the following form:
\begin{equation}
\label{eq:NME}
M^{0 \nu \beta \beta}  =\sum_{\alpha}\left\langle 0^+_f\left|  \sum_{1,2} h_{\alpha,K}(r_{12})   C^K_\alpha \cdot  S^K_\alpha  \tau^{+}_1\tau^{+}_2\right| 0^+_i\right\rangle.
\end{equation} 
The symbol $\alpha$ runs over Fermi, GT, and tensor terms. The spin-spatial part $C^K_\alpha \cdot  S^K_\alpha$ in \eqref{eq:NME} is the scalar product of two tensors with their expressions  given by
\bsub\begin{align}
C^0_{\rm F} &= 1, \quad S_{\rm F}^0=1, \\
C^0_{\rm GT}&=1,\quad S_{\rm GT}^0=\bm{\sigma}_1\cdot \bm{\sigma}_2, \\
C^2_{\rm T}&=\sqrt{\dfrac{24\pi}{5}}Y_2(\hat{\bm{r}}_{12}),\quad S_{\rm T}^2=\left[\boldsymbol{\sigma}_{1}\otimes \boldsymbol{\sigma}_{2}\right]^2.
\end{align}
\esub 

The coordinate-space neutrino potential is given by 
\begin{align}
\label{eq:neutrion_potential_r12}
 h_{\alpha,K}(r_{12}) 
 &=
\dfrac{2R_A}{\pi g^2_A} \int^\infty_0 \dd q \, q^2 \dfrac{ h_{\alpha,K} (\bm q^2)}{q(q+E_d)} j_K(qr_{12}), 
\end{align}
where $R_A=1.2\,A^{1/3}$ fm is introduced to make the matrix element $M^{0 \nu \beta \beta}$ dimensionless. The relative coordinate between the two decaying neutrons is defined as $\mathbf{r}_{12}=\mathbf{r}_1-\mathbf{r}_2$ and its magnitude $r_{12}=\lvert\vec{r}_{12}\rvert$ and direction vector $\hat{\vec{r}}_{12}=\vec{r}_{12}/\lvert\vec{r}_{12}\rvert$. The average excitation energy $E_d$ is chosen as $E_d=1.12\,A^{1/2}$ MeV~\cite{Haxton1984PPNP}. We make this choice to facilitate comparison with prior work. In an EFT framework, $E_d$ corresponds to a sub-leading correction~\cite{Cirigliano:2018PRC}. The function $j_K(qr_{12})$ is the spherical Bessel function of rank $K$, where $K=0$ for the Fermi and GT terms, and $K=2$ for the tensor term. The functions $h_{\alpha,K} (\bm q^2)$ are defined in terms of the vector ($g_V$), axial-vector ($g_A$), induced pseudoscalar ($g_P$) and weak-magnetism ($g_M$) coupling constants 
\bsub
\begin{align}
\label{eq:Fermi}
h_{\rm F,0}(\vec{q}^2) &= -g^2_V(\vec{q}^2),\\
\label{eq:GT}
h_{\rm GT,0}(\vec{q}^2) &= g^2_A(\vec{q}^2)-\dfrac{2}{3} \dfrac{\vec{q}^2}{2m_p} g_A(\vec{q}^2)g_{\rm P}(\vec{q}^2)\nonumber\\
&\hphantom{{}={}}+\dfrac{1}{3} \dfrac{\vec{q}^4}{4m^2_p} g^2_P(\vec{q}^2) +
\dfrac{2}{3} \dfrac{\vec{q}^2}{4m^2_p} g^2_M(\vec{q}^2),\\ 
\label{eq:Tensor}
 h_{\rm T,2}(\vec{q}^2) &=  \dfrac{2}{3} \dfrac{\vec{q}^2}{2m_p} g_A(\vec{q}^2)g_P(\vec{q}^2)
-\dfrac{1}{3} \dfrac{\vec{q}^4}{4m^2_p} g^2_P(\vec{q}^2) \nonumber\\
&\hphantom{{}={}}+
\dfrac{1}{3} \dfrac{\vec{q}^2}{4m^2_p} g^2_M(\vec{q}^2),
\end{align}
\esub 
where the coupling constants are regularized by the following dipole form factors,
\bsub\begin{align}
g_V(\vec{q}^2) &= g_V(0)\left(1+ \vec{q}^2/\Lambda^2_V\right)^{-2},\\
g_A(\vec{q}^2) &= g_A(0)\left(1+ \vec{q}^2/\Lambda^2_A\right)^{-2},\\
g_M(\vec{q}^2) &= g_V(\vec{q}^2)\left(1+\kappa_1\right),\\
g_P(\vec{q}^2) &= g_A(\vec{q}^2) \left(\dfrac{2m_p}{ \vec{q}^2+m^2_\pi}\right).
\end{align}
\esub
If not mentioned explicitly, we choose the vector and axial-vector coupling constants $g_V(0)=1$, $g_A(0)=1.27$, the anomalous nucleon iso\-vector magnetic moment $\kappa_1=\mu^{(a)}_n-\mu^{(a)}_p=3.7$, and the cutoff values $\Lambda_V=\SI{0.85}{\GeV}$ and 
$\Lambda_A=\SI{1.09}{\GeV}$, following Refs.~\cite{Simkovic:1999,Simkovic:2009PRC}.

\begin{figure}
\centering
\includegraphics[width=8cm]{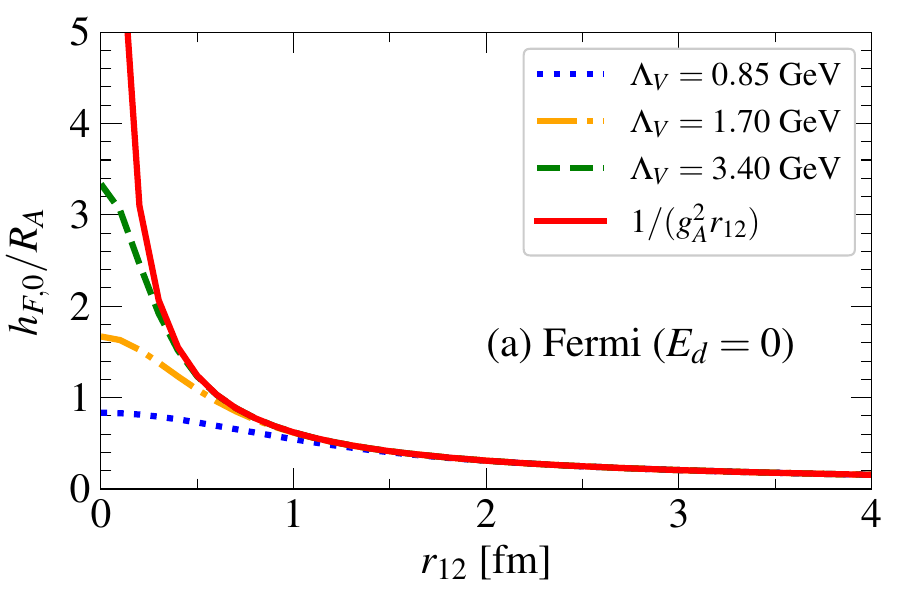}
\includegraphics[width=8cm]{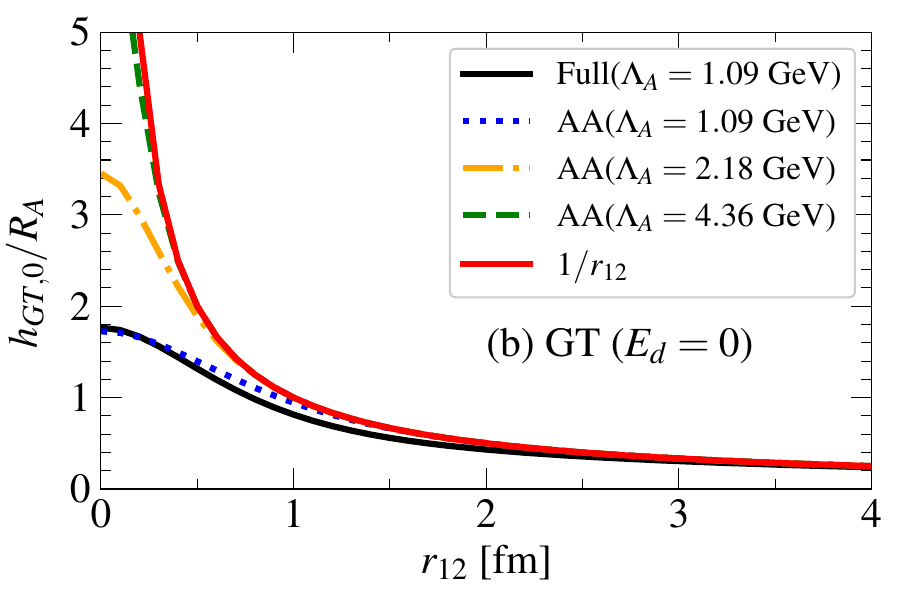}
\includegraphics[width=8cm]{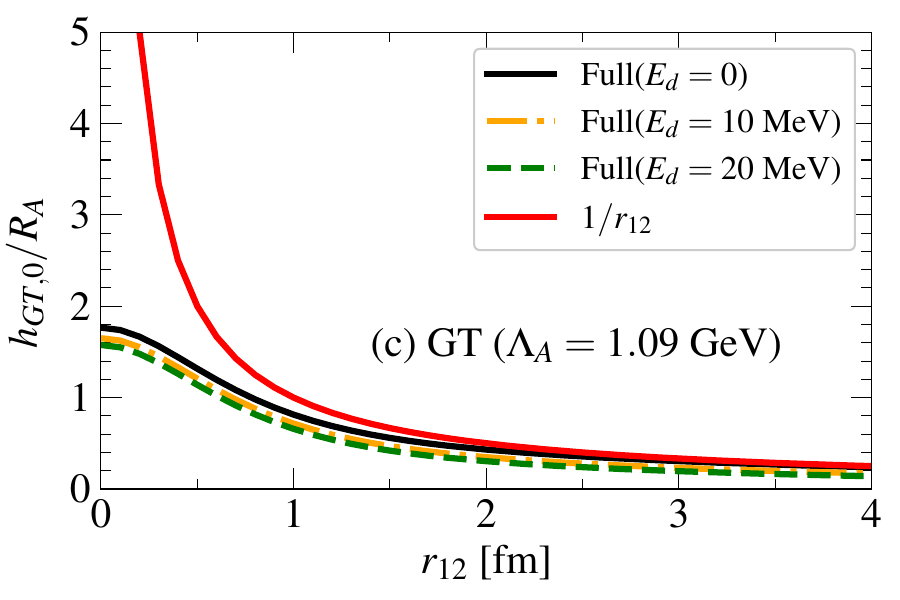}
\caption{The (a) Fermi and (b)(c) GT-type neutrino potentials \eqref{eq:neutrion_potential_r12} as a function of $r_{12}$. The radius parameter $R_A$ is excluded in the neutrino potentials to facilitate comparison with the Coulomb-like potential $1/r_{12}$. Different choices of cutoff values $\Lambda_V$ and $\Lambda_A$ are employed in the dipole form factors. In (c), a different value of the average excitation energy $E_d$ is used.  All the Fermi-type neutrino potentials are multiplied with a minus sign to facilitate comparison.  In the panel (b),  the potential $h^{AA}_{\rm GT,0}(\vec{q}^2) = g^2_A(\vec{q}^2)$ is compared to that of the full GT operator (\ref{eq:GT}) with the cutoff value change to $\Lambda_A=1.09$~GeV.}
\label{fig:potential_r12}
\end{figure}

\begin{figure}
\centering
\includegraphics[width=8.5cm]{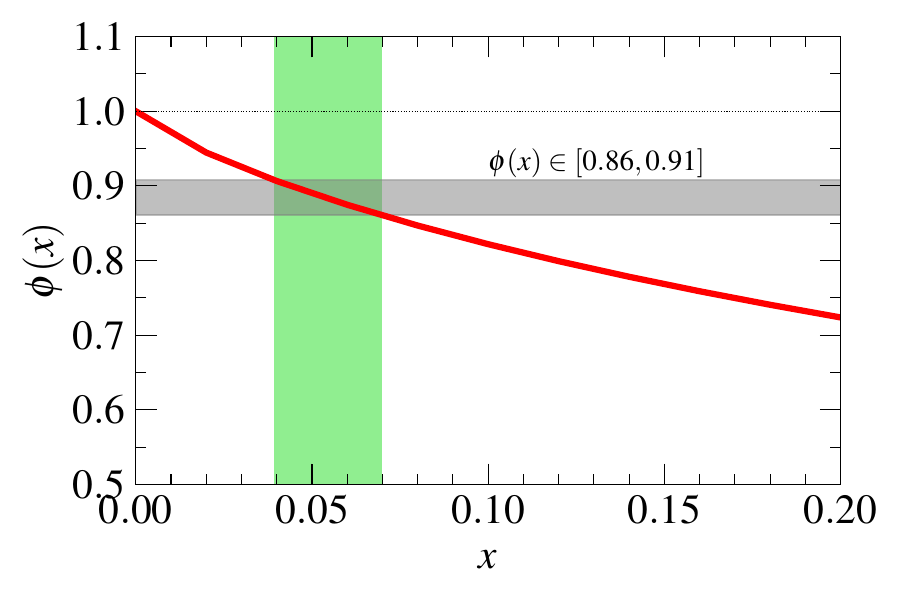}
\caption{The function $\phi(x)$ defined in Eq. \eqref{eq:phi} with the limits of $\phi(0)=1$ and $\phi(\infty)=0$. The shaded area indicates the interval where the average excitation energy takes $E_d\in[7.76,13.71]$ MeV and the relative distance between the two decaying nucleons $r_{12}=1.0$ fm.}
\label{fig:phi_r12}
\end{figure}

The dipole form factor regularizes the short-range behavior of the neutrino potentials. In the simplest case of $h_{\rm F, 0}(\bm q^2)=-g^2_V(0)$, and $h_{\rm GT, 0}(\bm q^2)=g^2_A(0)$, i.e., without dipole form factors and induced higher-order currents, the Fermi- and GT-type neutrino potentials in coordinate space can be derived analytically~\cite{Greuling:1960}:
\bsub\begin{align} 
\label{eq:potential_Ed0}
 h_{\rm F,0}(r_{12}) \to \tilde{h}_{\rm F,0}(r_{12}) 
 &=-\left(\frac{g^2_V}{g^2_A}\right)R_A\frac{\phi(E_d r_{12})}{r_{12}}, \\
 h_{\rm GT,0}(r_{12}) \to \tilde{h}_{\rm GT,0}(r_{12}) 
 &=R_A\frac{\phi(E_d r_{12})}{r_{12}}, 
\end{align}
\esub
where the function $\phi(x)$ is defined as~\cite{Greuling:1960,Haxton1984PPNP}
\begin{align} 
\label{eq:phi}
\phi(x)&=\frac{2}{\pi} \sin(x) \operatorname{Ci}(x)+\cos(x)\left[1-\frac{2}{\pi} \operatorname{Si}(x)\right],
\end{align} 
with the sine and cosine integrals
\begin{align} 
\operatorname{Si}(x) 
=\int_{0}^{x} \frac{\sin t}{t} d t,\quad \operatorname{Ci}(x) 
=-\int_{x}^{\infty} \frac{\cos t}{t} dt.  
\end{align} 

To see how the short-range behavior of the neutrino potential $h_{\alpha,0}$ is regularized by the dipole form factors, we choose different values for the cutoffs $\Lambda_V$ and $\Lambda_A$ in $g_V(\vec{q}^2)$ and $g_A(\vec{q}^2)$, respectively. The corresponding Fermi and GT neutrino potentials are displayed in Fig.~\ref{fig:potential_r12}(a) and (b), respectively. It is shown that the use of a smaller cutoff value $\Lambda_{V/A}$ leads to a larger modification on the neutrino potentials in the short-distance region. As expected, the neutrino potentials approach a Coulomb-like potential in the limit $\Lambda_{V/A}\to\infty$ and $E_d=0$, i.e., $\tilde{h}_{GT,0}(r_{12})\to R_A/r_{12}$ as $\phi(0)=1$.  Besides, we show in Fig.~\ref{fig:potential_r12}(c) how a nonzero value of $E_d$ reduces the entire neutrino potential.  For the $0\nu\beta\beta$-decay candidate nuclei  with mass number $48\leq A\leq150$, the empirical value of $E_d$ is $E_d\in[7.76,13.71]$ MeV~\cite{Haxton1984PPNP}, and $r_{12}\simeq 1.0$ fm, which gives $\phi(E_dr_{12})\in[0.86, 0.91]$, as shown in Fig.~\ref{fig:phi_r12}.

The NMEs of both DGT and $0\nu\beta\beta$ transitions can be conveniently rewritten as a function of the relative coordinate $r_{12}$ between decaying nucleons~\cite{Simkovic:2008},
\begin{equation}
\label{eq:NMEr}
 M^{\kappa}_A =\int dr_{12} C^\kappa_A(r_{12}),
\end{equation}
where $\kappa$ stands for either $0\nu\beta\beta$ or DGT. It was pointed out in Ref.~\cite{Brase:2021} that in conventional nuclear models, nuclear wave functions and neutrino potentials are represented in a harmonic oscillator basis with the oscillator length given by $b=\sqrt{\hbar/(M_N\omega)}$, where $M_N$ is nucleon mass and the frequency $\omega$ scales as $A^{-1/3}$. Thus, the NME of $M^{0\nu\beta\beta}_{\rm GT}$ (proportional to $R_A/b$) is expected to scale as $A^{1/6}$ and the DGT matrix element $M^{\rm DGT}$ is expected to be correlated with $M^{0 \nu\beta\beta}_{\rm GT}\cdot A^{-1/6}$ for all isotopes. From another point of view, if the $0\nu\beta\beta$ decay is dominated by the short-range contribution, namely, the long-range Coulomb-like decaying behavior is regularized by a faster decaying two-nucleon wave function, one may expect that  the DGT matrix element $M^{\rm DGT}$ is correlated with $M^{0 \nu\beta\beta}_{\rm GT}\cdot A^{-1/3}$, where the factor $A^{-1/3}$ is from the radius $R_A$ introduced to make the NME dimensionless, c.f.(\ref{eq:neutrion_potential_r12}). These two correlation relations will be discussed using the results from the calculations of both conventional nuclear models and {\em ab initio} methods in the next section.

\section{{\em Ab initio} many-body calculations}  
 \label{sec:results}

\begin{figure}
\centering  
\includegraphics[width=4.2cm]{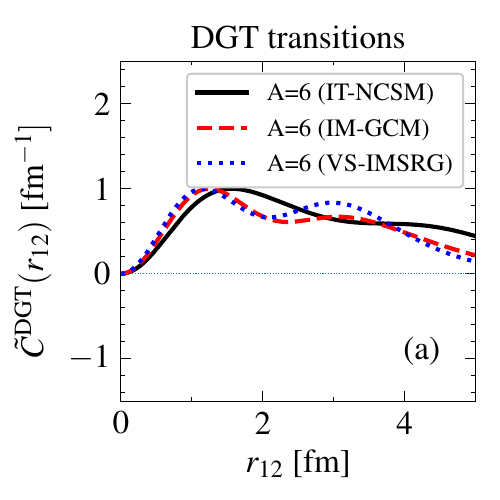} 
\includegraphics[width=4.2cm]{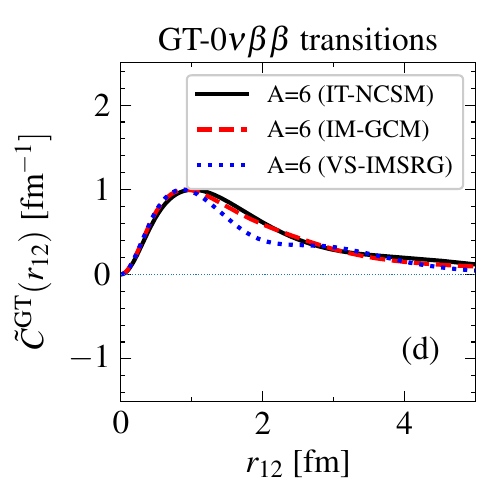}    
\includegraphics[width=4.2cm]{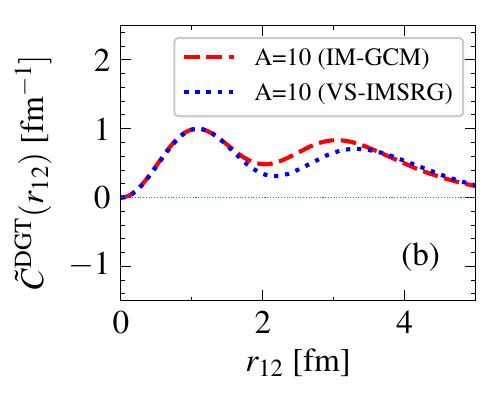} 
\includegraphics[width=4.2cm]{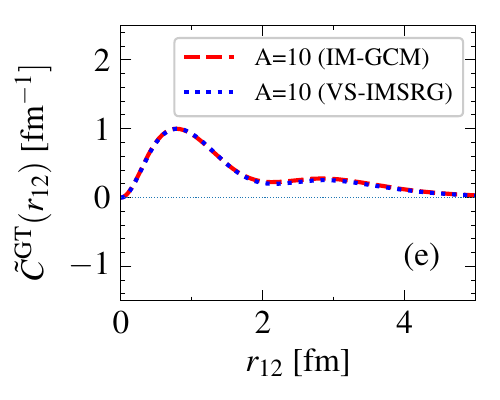}    
\includegraphics[width=4.2cm]{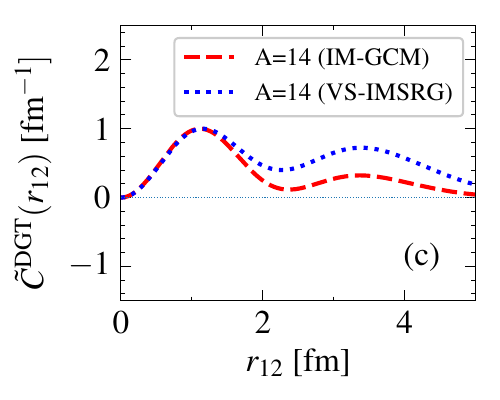} 
\includegraphics[width=4.2cm]{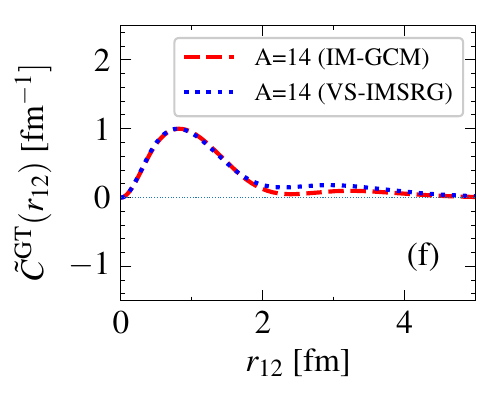}  
\caption{The transition densities  $\tilde{C}^\alpha(r_{12})$ for the isospin-conserving transitions of DGT (a-c) and  GT-$0\nu\beta\beta$ decay (d-f) from three {\em ab initio} calculations for  $\nuclide[6]{He}$,  $\nuclide[10]{Be}$, and  $\nuclide[14]{C}$ using the EM1.8/2.0 chiral force.  The tilde means that the transition densities are normalized to unity at the first peak position. }
\label{fig:abinitio_isospin_conserving}
\end{figure}

\begin{figure}
\centering  
\includegraphics[width=4.2cm]{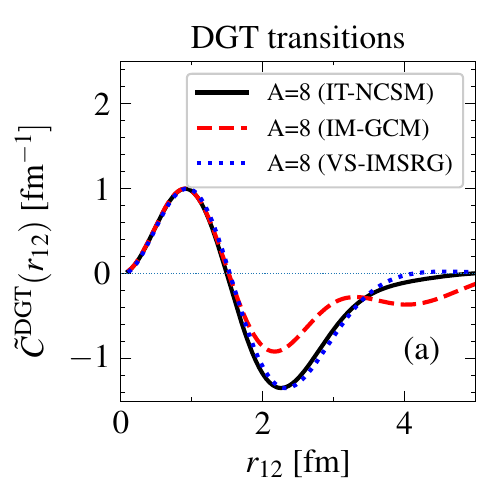} 
\includegraphics[width=4.2cm]{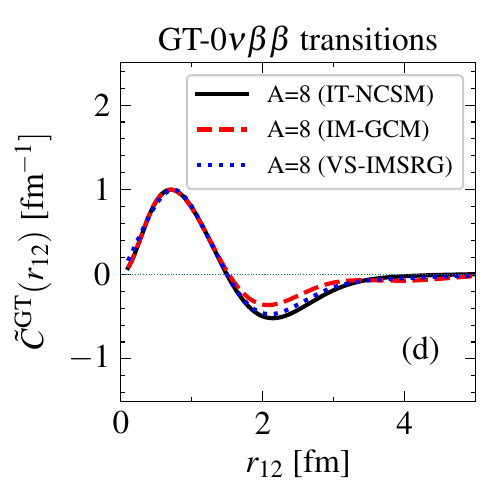}    
\includegraphics[width=4.2cm]{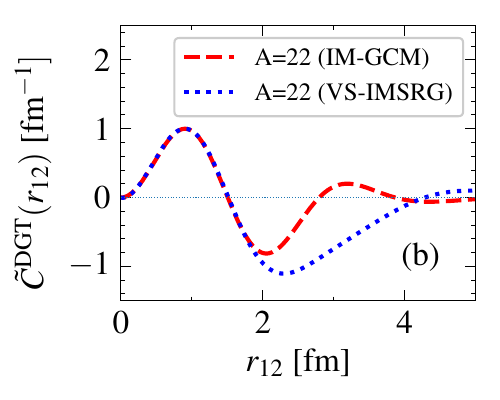} 
\includegraphics[width=4.2cm]{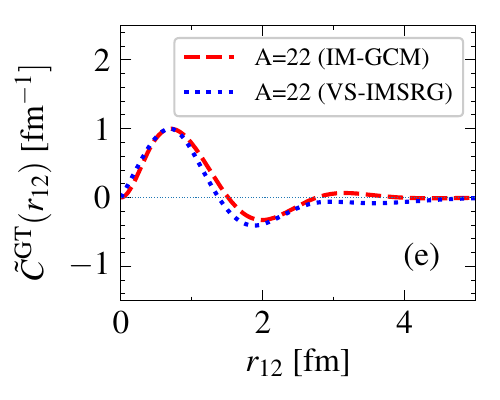} 
\includegraphics[width=4.2cm]{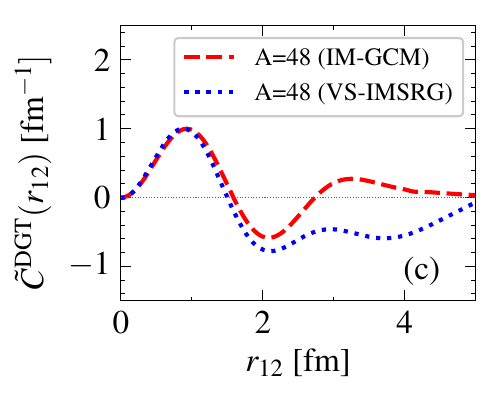} 
\includegraphics[width=4.2cm]{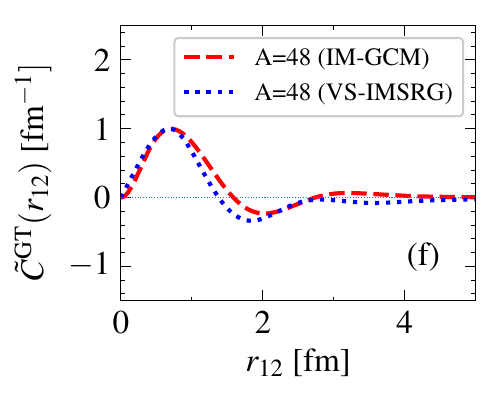} 
\caption{Same as Fig.~\ref{fig:abinitio_isospin_conserving}, but for  isospin-changing transitions of $\nuclide[8]{He}$,  $\nuclide[22]{O}$, and  $\nuclide[48]{Ca}$.  }
\label{fig:abinitio_isospin_changing}
\end{figure}

 \begin{figure*}
\centering
\includegraphics[width = \textwidth]{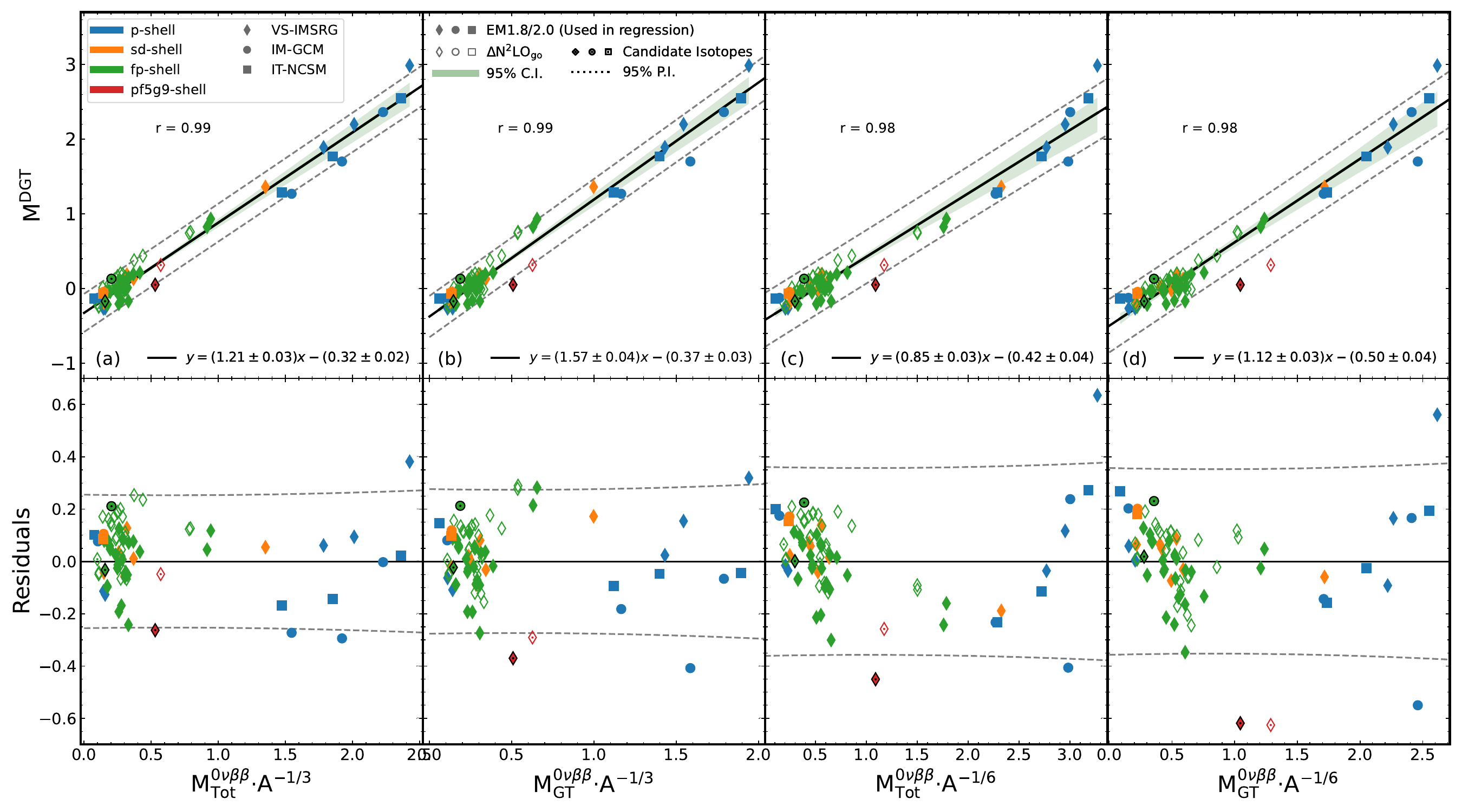}  
\caption{Correlation between the NMEs of DGT transitions ($M^{\rm DGT}$)  with those of GT-$0\nu\beta\beta$ decay ($M^{0 \nu\beta\beta}_{\rm GT}$) or the full NMEs of $0\nu \beta \beta$ decay ($M^{0 \nu\beta\beta}_{\rm Tot}$), scaled by a mass-number dependant factor of either A$^{-1/6}$ or A$^{-1/3}$, respectively. The NMEs are obtained from {\em ab initio} calculations with the chiral NN+3N interactions EM1.8/2.0 or $\Delta$N$^2$LO$_{\rm go}$ for isotopes ranging from $A = 6$ to $A = 76$. Only the results obtained with EM1.8/2.0 are used in the linear regression with residuals given in the bottom row. The shaded area indicates the confidence interval with a 95\% confidence level while the dashed line indicates the prediction interval at a 95\% confidence level. See text for details.  }
\label{fig:ab-initio-GT}
\end{figure*}
 
 In this section, we carry out {\em ab initio} nuclear many-body calculations with the  importance-truncated no-core shell model (IT-NCSM)~\cite{Roth:2009},  valence-space in-medium similarity renormalization group (VS-IMSRG)~\cite{Stroberg:2019}, and in-medium generator coordinate method (IM-GCM)~\cite{Yao:2018wq,Yao:2020PRL}  starting from chiral NN+3N interactions.  The latter two are different variants of IMSRG~\cite{Hergert:2016jk} which introduces a flow equation to gradually decouple the off-diagonal elements of the Hamiltonian that are connecting the  valence space and the excluded spaces,  or to decouple a preselected reference state  from all other states.  In the VS-IMSRG,  an effective Hamiltonian in a specific valence space is obtained, while in the IM-GCM~\cite{Yao:2018wq,Yao:2020PRL}, the reference state becomes a reasonable approximation to the ground state of the evolved Hamiltonian. The close-to-exact ground state is obtained with GCM by admixing other states that differ only in their collective parameters. A unitary transformation is defined via the flow equation and this transformation is consistently applied to all operators of interest.  We employ the chiral nuclear interaction (up to N\textsuperscript3LO)  by \citet{Entem:2003}, which we indicate by the label ``EM''.  We use the free-space SRG~\cite{Bogner:2010} to evolve the EM interaction to a resolution scale of $\lambda=\SI{1.8}{\per\fm}$. Following Refs.~\cite{Hebeler:2011,Nogga:2004il}, we construct the 3N interaction  directly, with a chiral cutoff of $\Lambda=\SI{2.0}{\per\fm}$. We refer to the resulting NN+3N Hamiltonian as EM$\lambda$/$\Lambda$, i.e., EM1.8/2.0 --- see Refs.~\cite{Hebeler:2011,Nogga:2004il} for  details. For comparison, we also employ the recently proposed chiral force $\Delta$N\textsuperscript2LO$_\text{GO}(394)$~\cite{Jiang2020},   a low-cutoff NN+3N interaction whose construction accounts for $\Delta$ isobars and whose parameters are constrained by $A\leq4$ few-body data as well as nuclear matter properties.
  For the 3N interaction, we discard all matrix elements involving states with $e_1+e_2+e_3>14$, where $e_i=2n_i+\ell_i$ is the number of oscillator quanta in state $i$. The frequency of the harmonic oscillator basis is chosen as $\hbar\omega=16$ MeV.

 \begin{figure}
     \centering
     \includegraphics[width = \columnwidth]{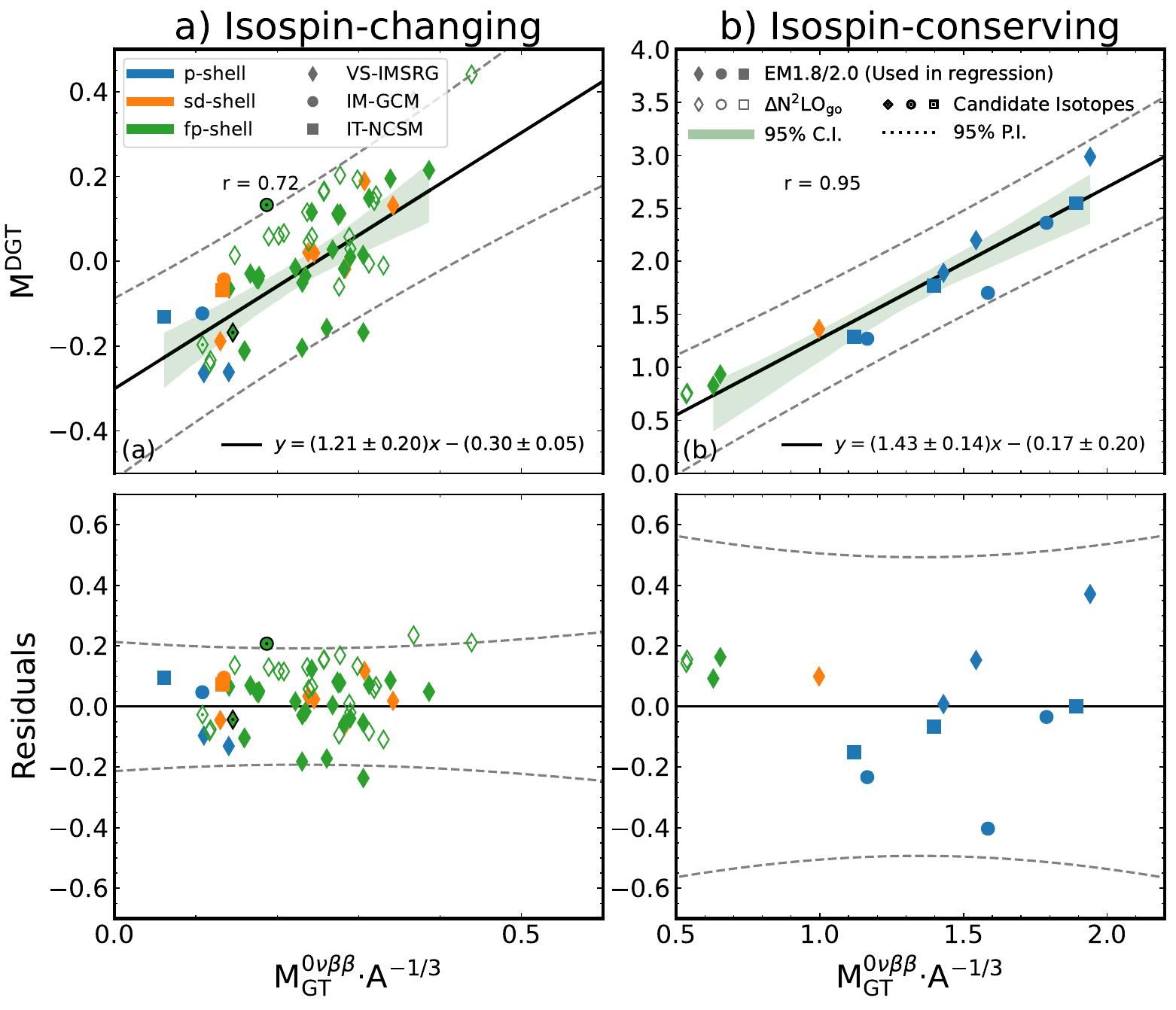}
     \caption{Correlation between $M^{\rm DGT}$ and $M^{0 \nu\beta\beta}_{\rm GT}\cdot A^{-1/3}$ when considering only isospin-changing transitions (left column) or only isospin-conserving transitions (right column). The green bands show the 95\% confidence interval while the dashed lines show the 95\% prediction interval.  The best fit lines from both cases agree with each other within $1\sigma$.}
     \label{fig:isospin-changing-conserving}
 \end{figure}

 Figures ~\ref{fig:abinitio_isospin_conserving} and ~\ref{fig:abinitio_isospin_changing} display the results of  three {\em ab initio} calculations for both isospin-conserving and isospin-changing DGT transitions and GT-$0\nu\beta\beta$ decay in a set of $p$, $sd$, and $fp$-shell nuclei, respectively,   where the same value of $e_{\rm Max}=8$ is used in the three methods for comparison. As discussed in the previous paper~\cite{Yao:2021PRC}, $N_{\rm Max}=8$ is usually employed in the IT-NCSM calculations, except for the transitions of $^{8}$He and $^{22}$O where the $N_{\rm Max}=6$ and $4$ is used, respectively.  In the two variants of IMSRG calculations, the transition density $C^\kappa(r_{12})$ is evaluated using the corresponding IMSRG-evolved transition operator at each value of the relative coordinate $r_{12}$ and the nuclear wave functions by the evolved Hamiltonian. One can see that the short-range parts of both transition densities  by all the three {\em ab initio} methods are on top of each other. The predictions for the long-range part of the isospin-changing DGT transitions differ because the long-range part of the transition densities is more sensitive to the way each method models many-body correlations. Due to the presence of the neutrino potential which decreases with $r_{12}$ approximately as $1/r_{12}$ (cf.~Fig.~\ref{fig:potential_r12}), the discrepancy in the long-range part of $\tilde C^{\rm GT}(r_{12})$ among the three methods is strongly suppressed. As a result, the short- and long-range parts of $0\nu\beta\beta$ decay of both isospin-conserving and isospin-changing types are consistently described in the three calculations. It is also seen that the transition densities of isospin-conserving transitions do not change sign as a function of $r_{12}$. In contrast, those of isospin-changing transitions oscillate with $r_{12}$, and thus the contributions of long- and short-range regions compensate for each other.
 Figs.~\ref{fig:abinitio_isospin_changing}(b) and (c) show that  the long-range contribution in the VS-IMSRG is generally larger than that in the IM-GCM. In the VS-IMSRG calculation, the long-range contribution to the DGT transition can be even larger than the short-range part, resulting in an inverted sign for the DGT NME. We note that varying the $e_{\rm Max}$ around the selected value  does not change the shape of the transition densities, but modifies slightly the height of the peaks.

 As discussed in Ref.~\cite{Shimizu:2018PRL} and in the next section, if the processes of both DGT transition and $0\nu\beta\beta$ decay  are dominated by the short-range contribution,  then these two types of matrix elements are expected to be correlated, irrespective of if the isospin is changing or  conserving in the process.  To verify this finding,  we show all NMEs of DGT transitions and  $0 \nu \beta \beta$ decay from the three {\em ab initio} calculations for the isotopes in different mass regions in Figure~\ref{fig:ab-initio-GT}. In these calculations, the value of $e_{\rm Max}$ is chosen to ensure the convergence of the NMEs, as discussed in Ref.~\cite{Yao:2021PRC}. In the VS-IMSRG calculation for heavier isotopes, $e_{\rm Max}=12$ is used.  The  NMEs $M^{\rm DGT}$ are plotted against those of  $0 \nu \beta \beta$ decay scaled as $M^{0 \nu \beta \beta}\cdot A^{-1/3}$ and $M^{0 \nu\beta \beta}\cdot A^{-1/6}$ in Fig.~\ref{fig:ab-initio-GT}(a,b) and Fig.~\ref{fig:ab-initio-GT}(c,d), respectively. The results are fitted to the following relation,
 \begin{equation}
 \label{eq:line_fit}
 M^{\rm DGT} = \alpha M^{0\nu\beta\beta}\cdot A^{\gamma} + \beta,    
\end{equation}
where the power parameter is fixed to be either $\gamma=-1/3$ or $-1/6$. The coefficient of correlation $r$  between the two quantities
\begin{align}  
\label{eq:r}
r &=\frac{\sum_i(x_i-\Bar{x})(y_i-\Bar{y})}{\sqrt{\sum_i(x_i-\Bar{x})^2\sum_i(y_i-\Bar{y})^2}}. 
\end{align} 
 is introduced as a statistical measure of the strength of the correlation relationship, where $x_i$ and $y_i$ are the samples on the $x$ and $y$ axis respectively, and $\Bar{x}$ and $\Bar{y}$ are the average of the samples on each axis. The value $r=1$ corresponds to a perfect positive linear correlation and $r=0$ indicates no linear relationship. The correlation relationships that we obtain are as follows,

 \bsub\begin{align}
 \label{eq: correlation13}
     M^{\rm DGT} &= 1.21(3) M^{0\nu\beta\beta}_{\rm Tot}\cdot A^{-1/3} -0.32(2),\quad r=0.99, \\
     M^{\rm DGT} &= 1.57(4) M^{0\nu\beta\beta}_{\rm GT}\cdot A^{-1/3} -0.37(3),\quad r=0.99, \\
 \label{eq: correlation16}
     M^{\rm DGT} &= 0.85(3) M^{0\nu\beta\beta}_{\rm Tot}\cdot A^{-1/6} -0.42(4),\quad r=0.98, \\
     M^{\rm DGT} &= 1.12(3) M^{0\nu\beta\beta}_{\rm GT}\cdot A^{-1/6} -0.50(4),\quad r=0.98.
 \end{align}
 \esub 
  The finite-sample-size error on the coefficient $r$ is evaluated using the formula\cite{Bowley1928} $\sigma_r\approx (1-r^2)/\sqrt{N-2}$ which gives $\sigma_r = 0.003$ for $r=0.99$ and  $\sigma_r = 0.006$ for $r=0.98$. One can see that the $M^{\rm DGT}$ is  correlated slightly stronger with the quantity $M^{0 \nu \beta \beta}\cdot A^{-1/3}$ than with  the quantity $M^{0 \nu \beta \beta}\cdot A^{-1/6}$, irrespective of if only the GT part or the total NME is considered for $M^{0 \nu \beta \beta}$. The Fermi part of $M^{0 \nu \beta \beta}$ is approximately one-third of the GT part, which explains why we still see a correlation when considering the total NME $M^{0 \nu \beta \beta}$. Besides, the residuals exhibit some clear pattern at low values of $M^{0 \nu \beta \beta}$ in all cases, more predominantly when scaling $M^{0 \nu \beta \beta}$ by $A^{-1/6}$. This finding not only further validates that using $A^{-1/3}$ is a better choice for the results of {\em ab initio} calculations, but it also suggests that the good correlation we find when looking at the whole data set is not representative of the whole situation.  We note that the results from  {\em ab initio} calculations using the $\Delta$N\textsuperscript2LO$_\text{GO}(394)$ interaction are consistent with the above correlation relation (\ref{eq: correlation13}).

\begin{figure}
\centering
\includegraphics[width= \columnwidth]{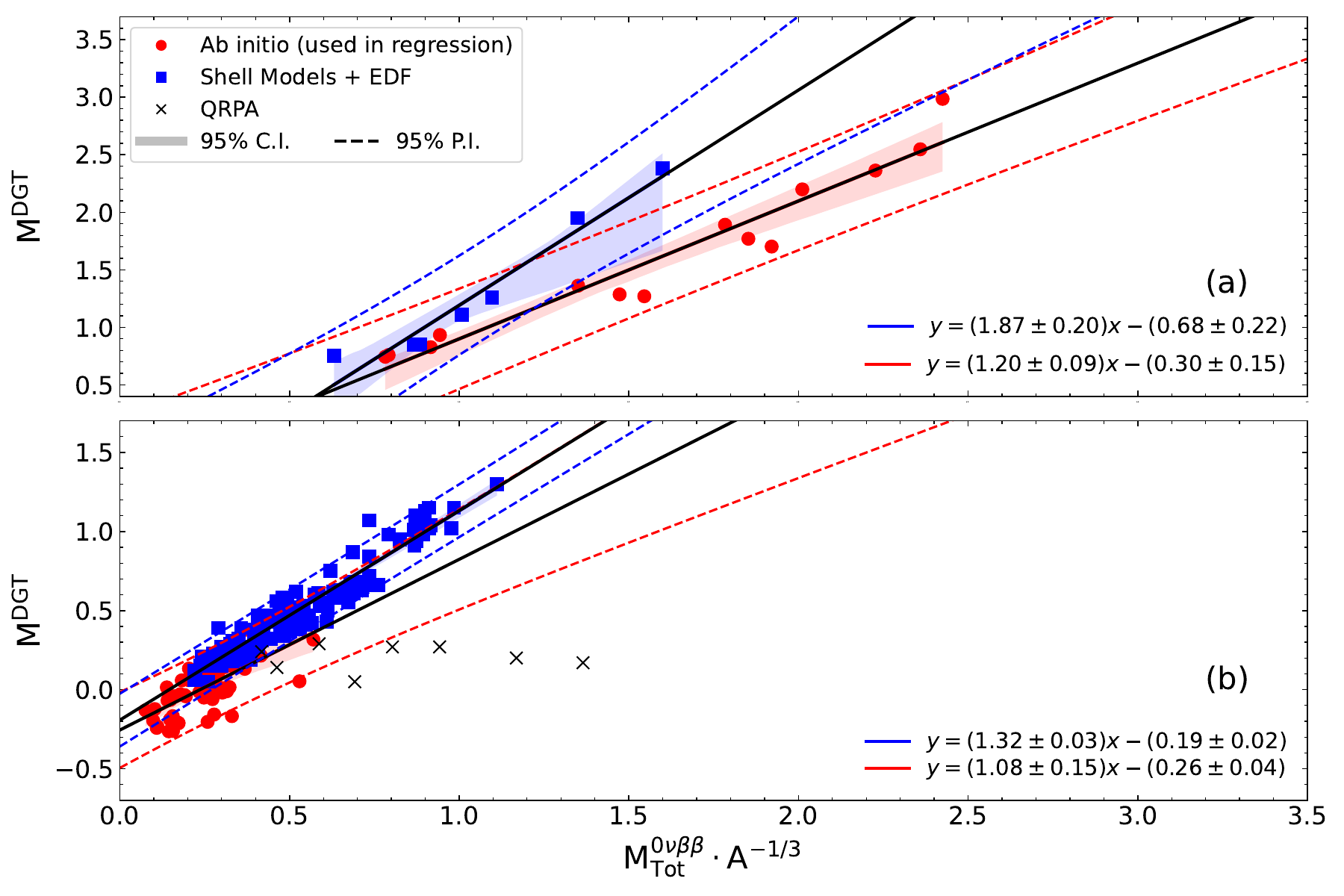}  
\caption{Correlation between $M^{\rm DGT}$ and $M^{0 \nu\beta\beta}_{\rm Tot}\cdot A^{-1/3}$  derived from the results of both  \emph{ab initio} (red) and conventional (blue) nuclear models~\cite{Rodriguez:2013PLB,Shimizu:2018PRL,Menendez:2018JPS,Menendez:2017JPCS,Simkovic:2018}, except for the results of QRPA calculations~\cite{Simkovic:2018} for isospin-conserving transitions in (a) and isospin-changing in (b).}
\label{fig:comparison-13}
\end{figure}

\begin{figure}
\centering
\includegraphics[width= \columnwidth]{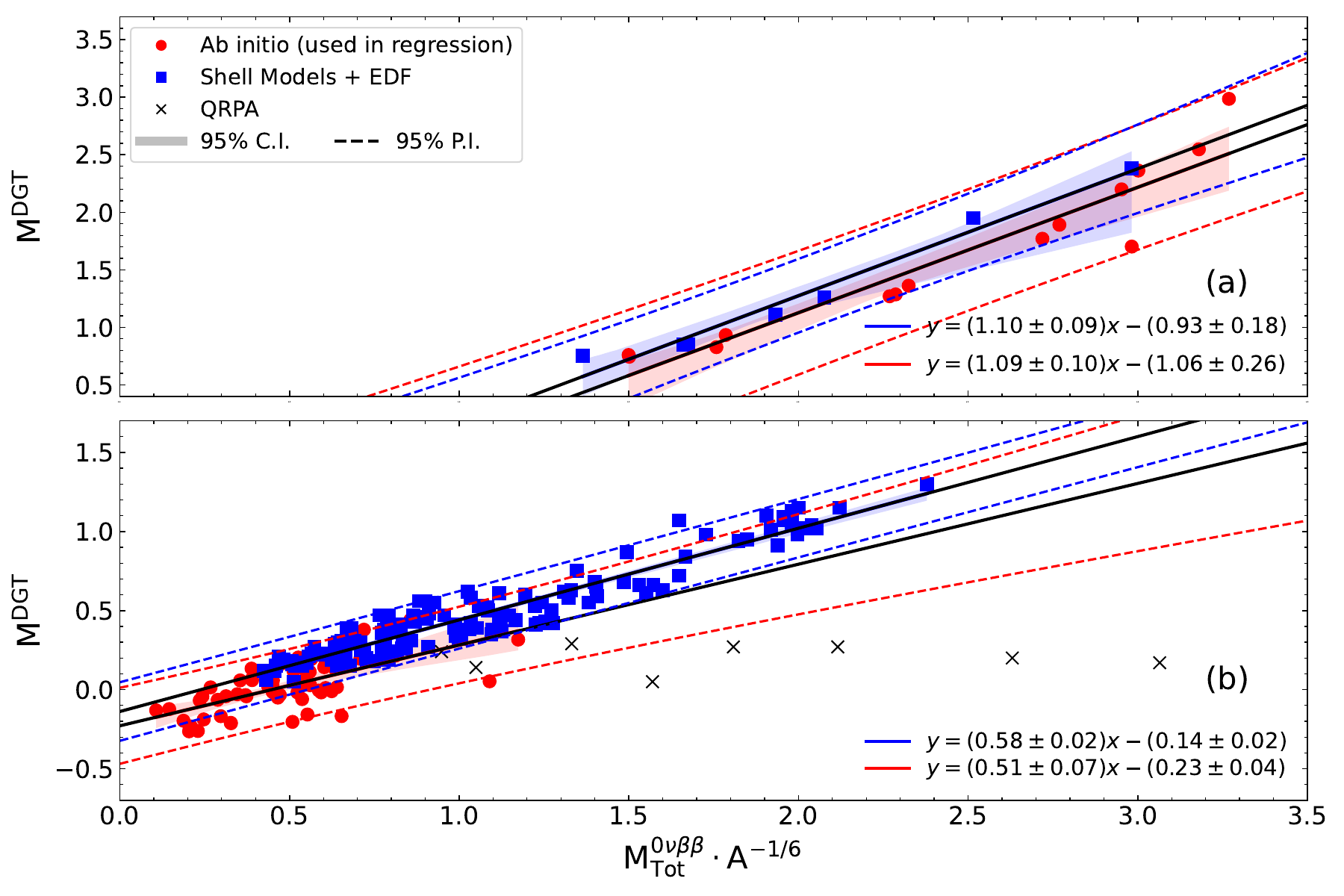}   
\caption{Same as Fig.~\ref{fig:comparison-13}, but for the correlation between $M^{\rm DGT}$ and $M^{0 \nu\beta\beta}_{\rm Tot}\cdot A^{-1/6}$ .}
\label{fig:comparison-16}
\end{figure}

 It is worth emphasizing that the matrix elements for isospin-changing transitions and isospin-conserving transitions are significantly different from each other. The relevant $0\nu\beta \beta$ decay is an isospin-changing transition, and thus its matrix element is generally smaller than those of isospin-conserving transitions for the same mass number $A$. Therefore, one may expect the correlation obtained from these two types of transitions to be different. With this consideration, we plot the NMEs of isospin-changing and isospin-conserving transitions separately in Fig.~\ref{fig:isospin-changing-conserving}(a) and (b), respectively. Two different linear regressions are carried out for these two types of transitions with the parameter $\gamma=-1/3$. We find the following correlation relationships
 \bsub\begin{align}
    \Delta T=2: M^{\rm DGT} &= 1.21(20) M^{0\nu\beta\beta}_{\rm GT}\cdot A^{-1/3} -0.30(5), \\
    \Delta T=0: M^{\rm DGT} &= 1.43(14) M^{0\nu\beta\beta}_{\rm GT}\cdot A^{-1/3} -0.17(20).
 \end{align}
 \esub 
 It is shown that the correlation coefficient ($r=0.72$, $\sigma_r=0.08$) of isospin-changing transitions ($ \Delta T=2$) is much smaller than that  ($r=0.95$, $\sigma_r=0.03$) of isospin-conserving ($\Delta T=0$) transitions. 
 In other words, the correlation relation is much weaker for the NMEs of isospin-changing transitions.

 Figures~\ref{fig:comparison-13} and ~\ref{fig:comparison-16} summarize the results from the calculations of  both \emph{ab initio}  and conventional nuclear models  (shell models~\cite{Shimizu:2018PRL,Menendez:2018JPS,Menendez:2017JPCS},  energy-density-functional (EDF)~\cite{Rodriguez:2013PLB} and QRPA~\cite{Simkovic:2018}).  Treating the matrix elements of   isospin-conserving and isospin-changing processes, separately, we find the following relation for the nuclear matrix elements from the {\em ab initio} calculations, 
 \bsub\begin{align}
    \Delta T=2: M^{\rm DGT} &= 1.08(15) M^{0\nu\beta\beta}_{\rm Tot}\cdot A^{-1/3} -0.26(4), \\
    \Delta T=0: M^{\rm DGT} &= 1.20(9) M^{0\nu\beta\beta}_{\rm Tot}\cdot A^{-1/3} -0.30(15),\\ 
    \Delta T=2: M^{\rm DGT} &= 0.51(7) M^{0\nu\beta\beta}_{\rm Tot}\cdot A^{-1/6} -0.23(4), \\
    \Delta T=0: M^{\rm DGT} &= 1.09(10) M^{0\nu\beta\beta}_{\rm Tot}\cdot A^{-1/6} -1.06(26),
 \end{align}
 \esub 
 and the following relation for the nuclear matrix elements from the conventional nuclear-model calculations, 
 \bsub\begin{align}
    \Delta T=2: M^{\rm DGT} & = 1.32(3) M^{0\nu\beta\beta}_{\rm Tot}\cdot A^{-1/3} -0.19(2), \\
    \Delta T=0: M^{\rm DGT} &= 1.87(20) M^{0\nu\beta\beta}_{\rm Tot}\cdot A^{-1/3} -0.68(22),\\
    \Delta T=2: M^{\rm DGT} & = 0.58(2) M^{0\nu\beta\beta}_{\rm Tot}\cdot A^{-1/6} -0.14(2), \\
    \Delta T=0: M^{\rm DGT} &= 1.10(9) M^{0\nu\beta\beta}_{\rm Tot}\cdot A^{-1/6} -0.93(18).
 \end{align}
 \esub 
 It is seen that the correlation relations are significantly different for the matrix elements of $ \Delta T=2$ and $\Delta T=0$ transitions. The slope for the $\Delta T=0$ transitions is generally larger than that for the $\Delta T=2$ transitions. Besides, the slope by the {\em ab initio} calculation is generally smaller than that by the conventional nuclear model calculations, because of a stronger cancellation between the long-range and short-range contributions in the DGT NMEs by  {\em ab initio} methods. Among all the methods concerned, the QRPA predicts the smallest value for the DGT NMEs, indicating the occurrence of the strongest cancellations~\cite{Simkovic:2018}. 
 
  To compare with the correlation relation in Refs.~\cite{Shimizu:2018PRL,Brase:2021}, we follow their way to derive the correlation relation based on the matrix elements of both isospin-conserving and isospin-changing processes  from the calculations of conventional nuclear models (excluding the results of QRPA), which reads
  \bsub\begin{align}
 \label{eq: correlation13_sm}
     M^{\rm DGT} &= 1.41(3) M^{0\nu\beta\beta}_{\rm Tot}\cdot A^{-1/3} -0.23(2), \quad r = 0.96, \\
 \label{eq: correlation16_sm}
     M^{\rm DGT} &= 0.65(2) M^{0\nu\beta\beta}_{\rm Tot}\cdot A^{-1/6} -0.20(2), \quad r = 0.94,
 \end{align}
 \esub 
 the latter of which is consistent with the interval of parameters $\alpha\in [0.447,0.699]$, $\beta\in [-0.18,-0.056]$, with the power parameter $\gamma=-1/6$ found in Ref.~\cite{Brase:2021}. The value of the coefficient $r$ indicates that the DGT matrix elements $M^{\rm DGT}$ are  slightly stronger correlated with $M^{0\nu\beta\beta}_{\rm Tot}\cdot A^{-1/3}$ than with $M^{0\nu\beta\beta}_{\rm Tot}\cdot A^{-1/6}$, as discussed in Fig.~\ref{fig:ab-initio-GT}.

 It is worth  mentioning that the NMEs of the ground-state to ground-state DGT transition  of isospin-changing transitions would be exactly zero if the spin-isospin SU(4) symmetry were conserved in atomic nuclei as in this case the initial and final states would belong to different irreducible representations of the SU(4) group.  Therefore, different values of the DGT NMEs predicted by different nuclear models indicate that  the SU(4) symmetry is broken to different extents in the ground-state wave functions. The studies on the breaking of the SU(4) symmetry in different nuclear models and its impact on the correlation relation  might be able to provide us with a more profound understanding, but these studies are beyond the scope of this work.

\section{A scale-separation analysis\label{sec:scales}}

\begin{figure}
    \centering
    \includegraphics[width = 7cm]{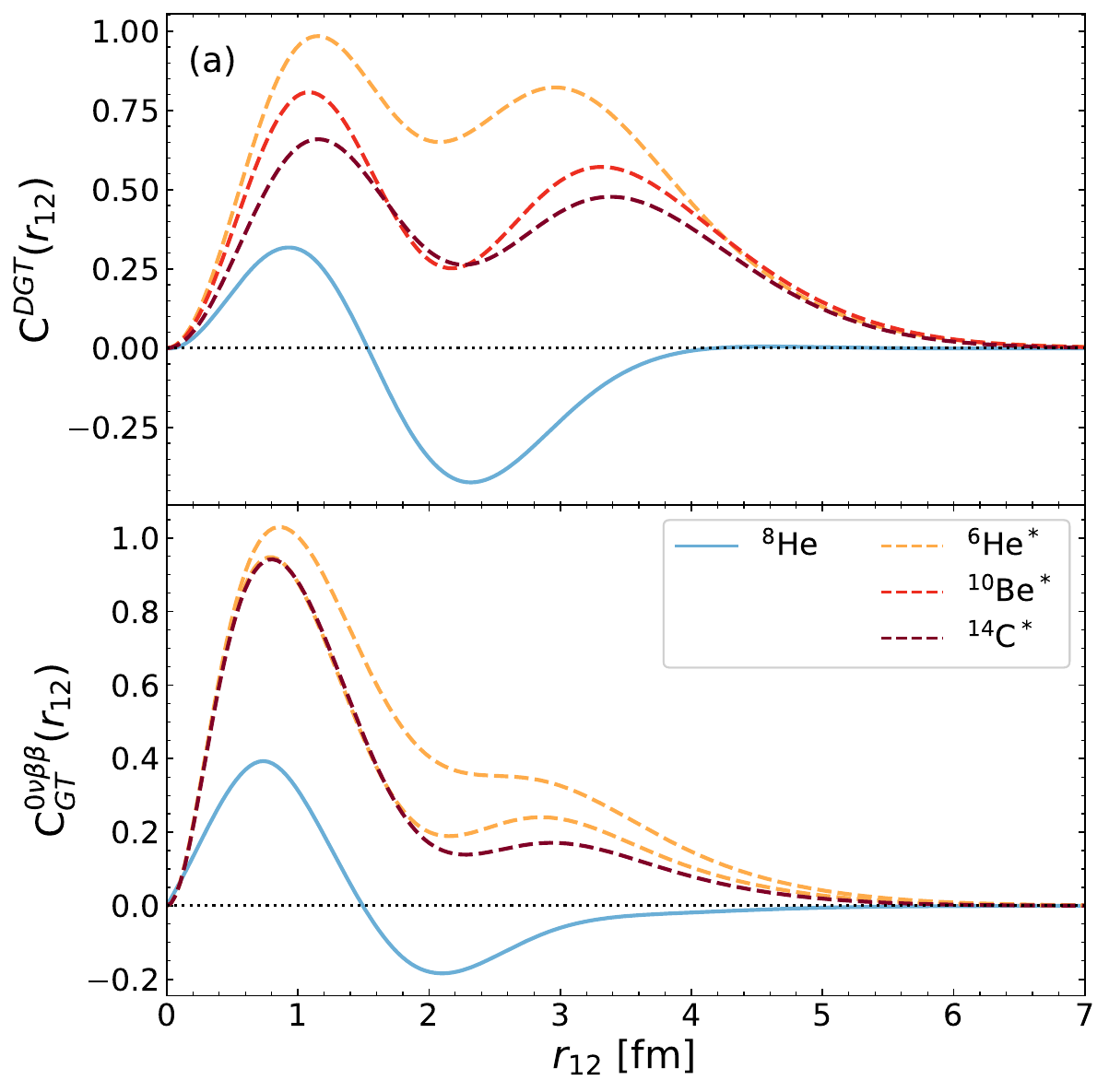}
    \includegraphics[width = 7cm]{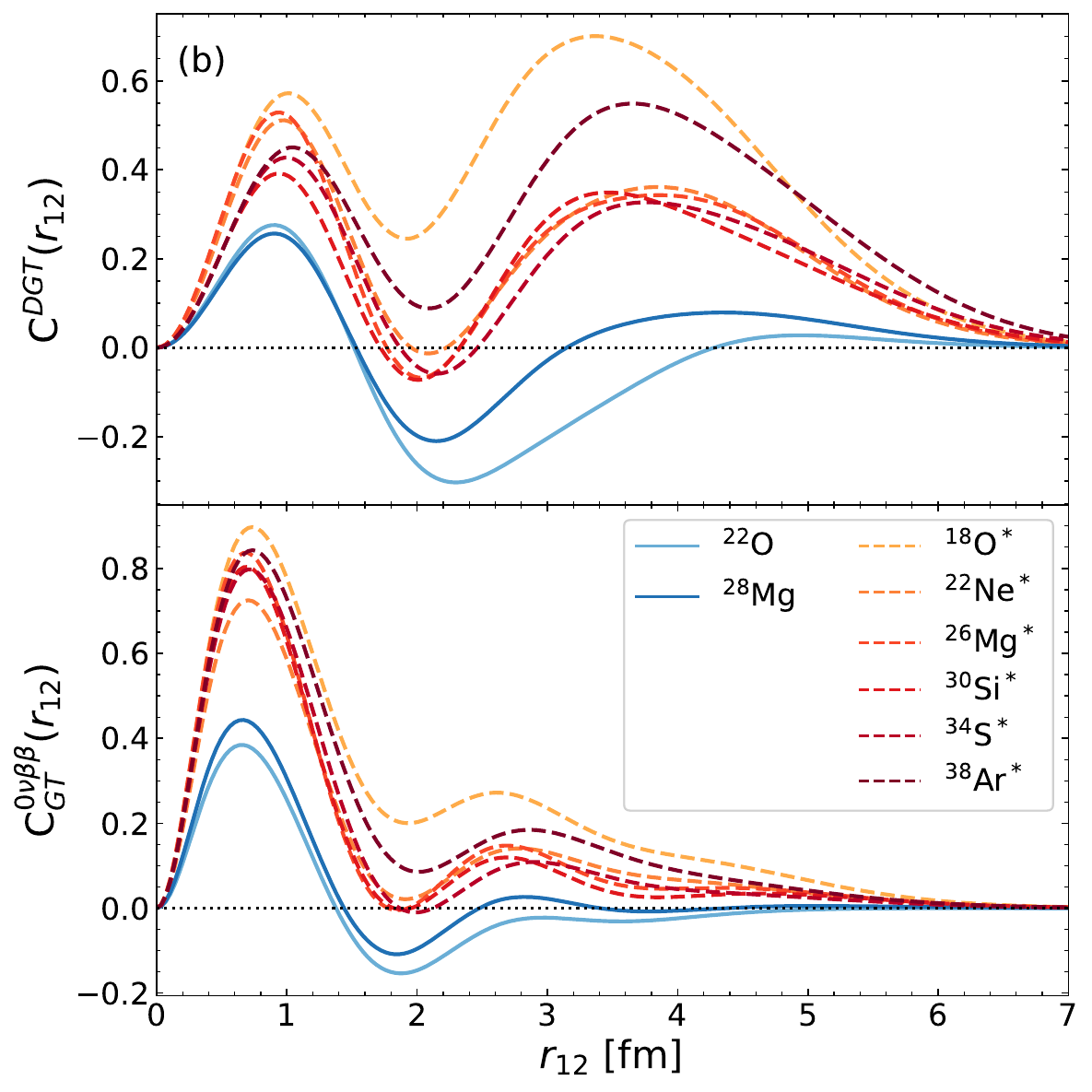}
    \includegraphics[width = 7cm]{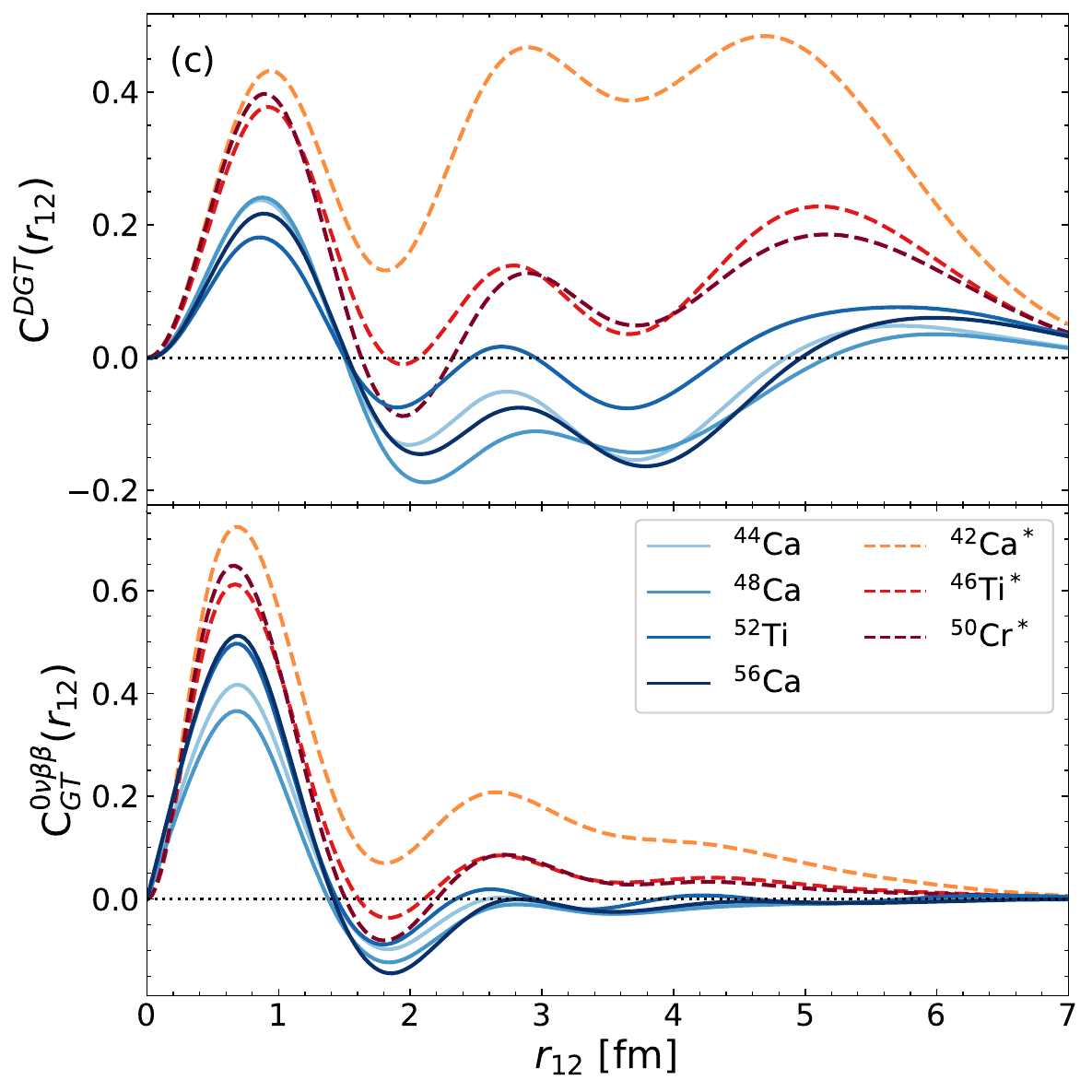}
    \caption{The transition densities (cf.~Eq.~\eqref{eq:NMEr}) of isotopes in $p$, $sd$- and $fp$ shells from the VS-IMSRG calculation with the EM1.8/2.0 interaction. Asterisks indicate isospin conserving transitions. }
    \label{fig:vsimsrg-transition-densities}
\end{figure}

\begin{figure}
    \centering
    \includegraphics[width = 7cm]{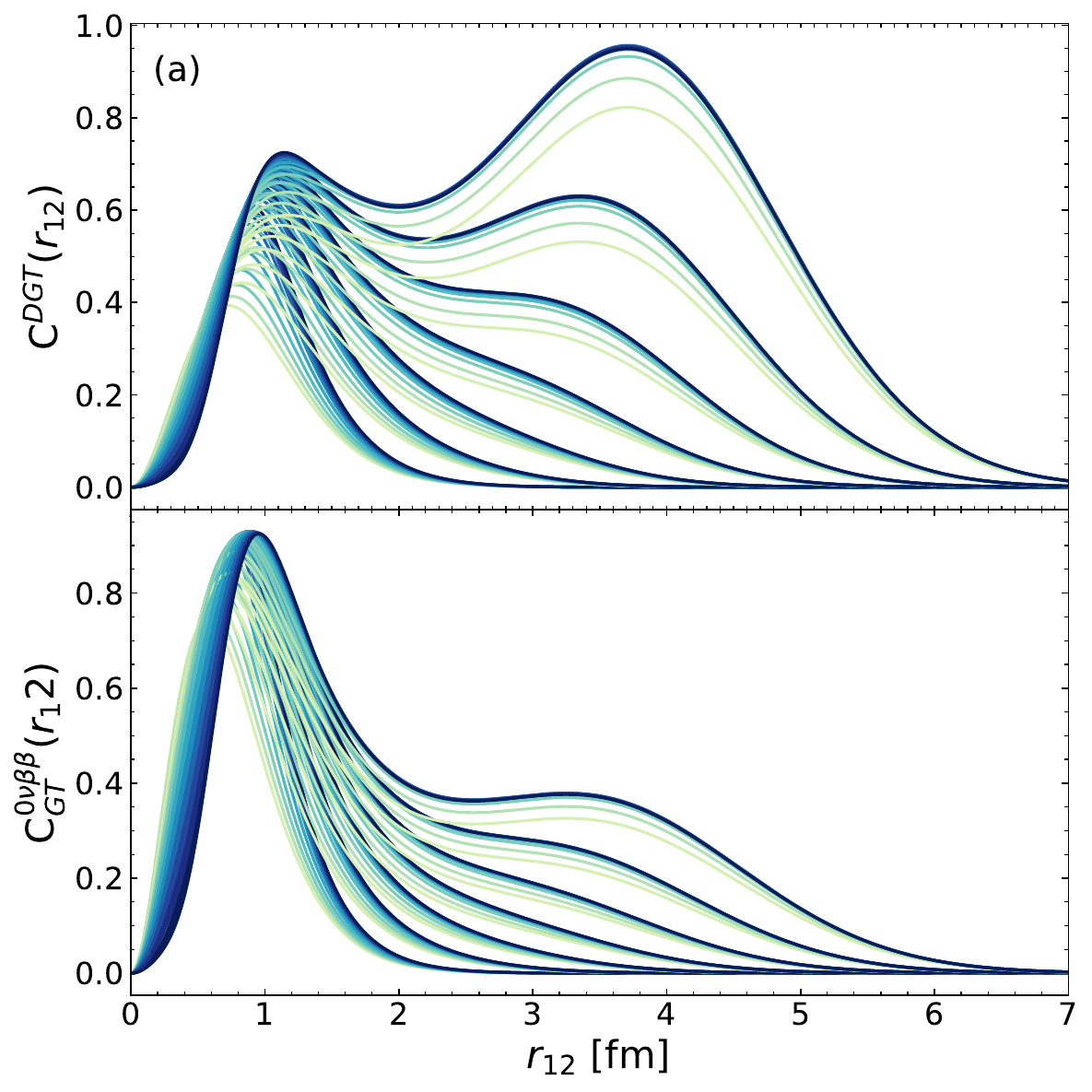}
    \includegraphics[width = 7cm]{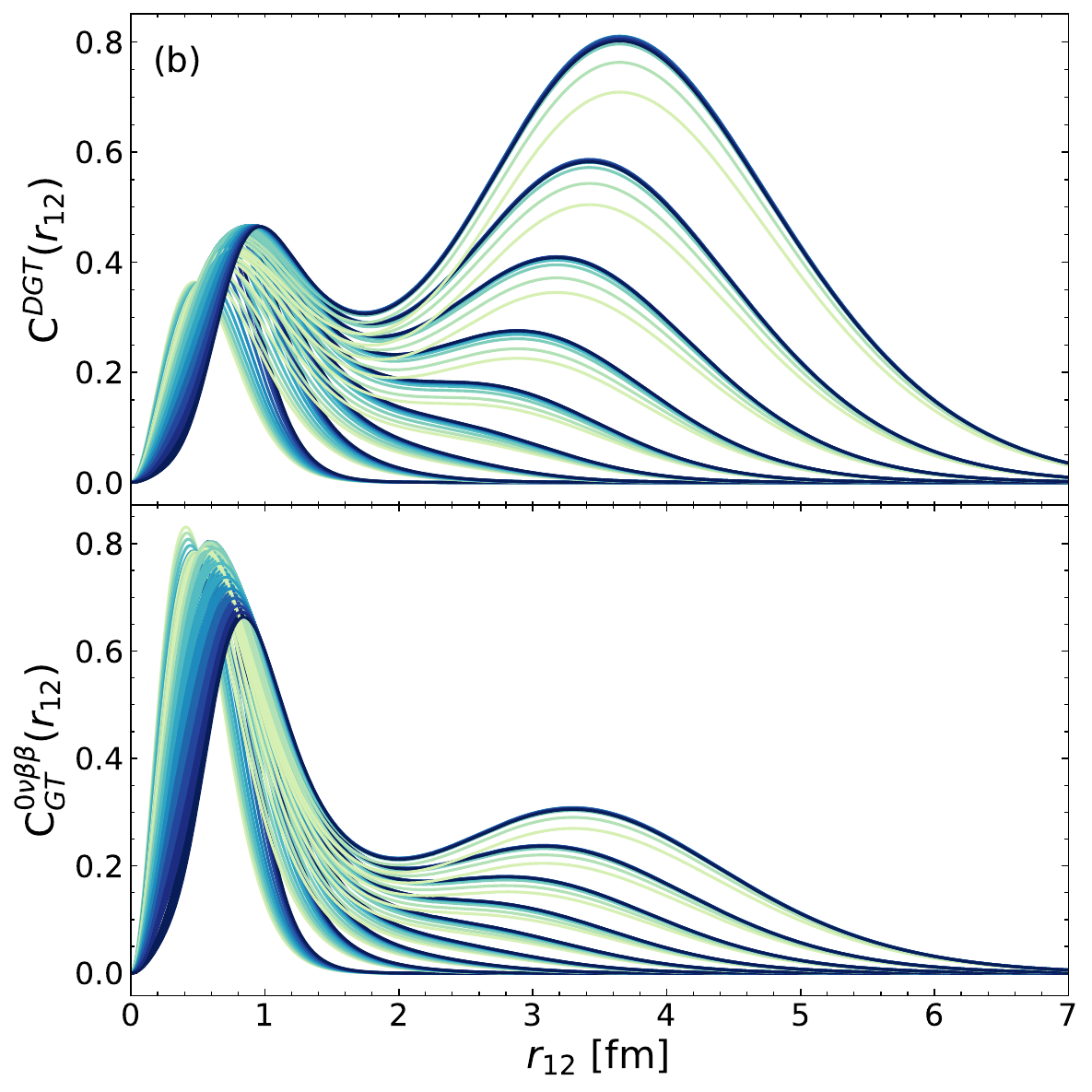}
    \includegraphics[width = 7cm]{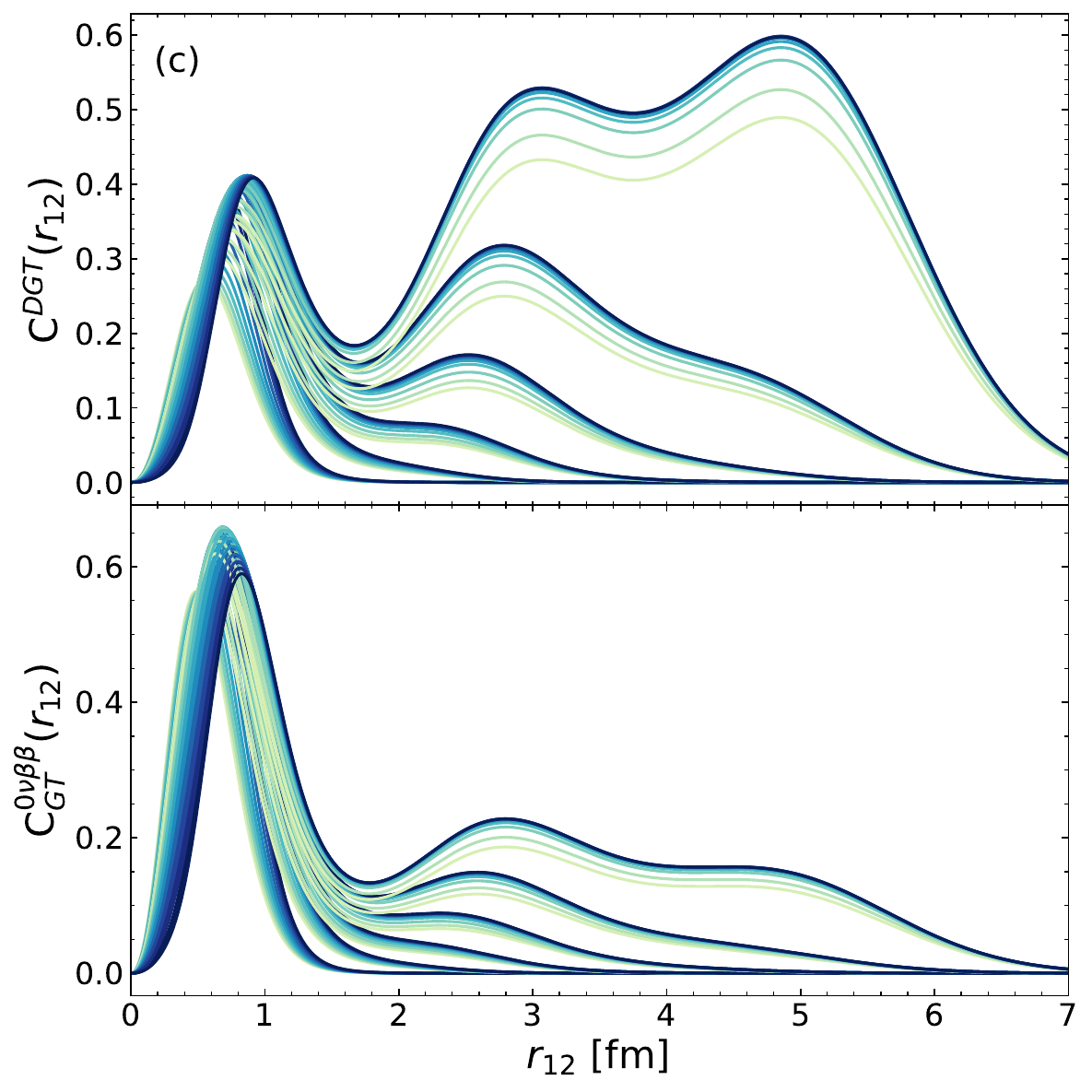}
    \caption{Sampled transition densities $\tilde{C}^{\kappa}_A(r_{12})$ with the parameters fitted to the results of VS-IMSRG calculations for the isospin-conserving transitions in $p$-, $sd$- and $fp$-shell nuclei, except for the parameters $c$ and $d$, which vary from 0 to 1 and from 0.5 fm$^2$ to 4.5 fm$^2$, respectively. }
    \label{fig:fit-isospin-conserving-vsimsrg}
\end{figure}

In this section, we carry out a scale-separation analysis of the correlation between the NMEs of DGT transitions and $0\nu\beta\beta$ decay based on the argument~\cite{Anderson:2010,Bogner:2012,Cruz-Torres:2021} that the long- and short-distance physics in atomic nuclei can be rather well separated. In this case, the nuclear many-body wave functions $\Psi_{i/f}(1,2, \ldots A)$ of initial and final nuclei in the transitions can be approximately factorized into the product of a universal short-distance two-body wave function $\phi_c(\mathbf{r}_{12})$  and a state-dependent long-range $A$-body wave function $\chi^{(i/f)}_c$~\cite{Weiss:2021}
\begin{equation}
\label{eq:wfs_r12to0}
\Psi_{i/f} \xrightarrow{r_{12}\to 0} \sum_c \phi_c(\mathbf{r}_{12}) \chi^{(i/f)}_c(\mathbf{R}_{12}; 3, \ldots, A), 
\end{equation}
 where $\mathbf{R}_{12}=(\mathbf{r}_1+\mathbf{r}_2)/2$, and $c$ distinguishes channels (spin-isospin) with different quantum numbers for the pair of nucleons.  For the NMEs of DGT and $0\nu\beta\beta$ transitions, the pair of neutrons in the s-wave state with total spin $S=0$, angular momentum $J=0$ and isospin $T=1$ converting into a pair of protons with the same quantum numbers provides the predominate contribution. If only this channel (labeled with $c=0$) is considered and the two-body wave function $\phi_c(r_{12})$ is assumed to be isospin-independent, one has $\phi_{nn}(r_{12})=\phi_{pp}(r_{12})=\phi_0(r_{12})$. Thus, it is reasonable to parametrize the transition densities defined in \eqref{eq:NMEr} into the following forms,
 \bsub
 \label{eq:transiton_density}
 \begin{align} 
  \tilde{C}^{\rm GT}_A(r_{12})&= -3 r^2_{12} h_{\rm GT,0}(r_{12})\rho_{nn}(r_{12}) C^0_{ppnn}(A_f,A_i,r_{12}),\\
 \tilde{C}^{\rm DGT}_A(r_{12})&=\sqrt{3}r^2_{12}  \rho_{nn}(r_{12}) C^0_{ppnn}(A_f,A_i,r_{12}),
 \end{align}
 \esub
 where the factor of $(-3)$ in the GT transition is from the spin operator. The DGT defined in \eqref{eq:DGT} brings an additional factor of $(-1/\sqrt{3})$. The two-nucleon density is defined as $\rho_{nn}(r_{12})=|\phi_0(r_{12})|^2$. The overlap function $C^0_{ppnn}(A_f,A_i, r_{12})$ is  determined by the $A$-body wave functions $\chi^{(i/f)}_0$ of initial and final nuclei multiplied by the number of pairs~\cite{Weiss:2021}.
 In the limit of $r_{12}\to 0$, the two-nucleon density $\rho_{nn}(r_{12})$ is expected to be universal for different nuclei but may depend on the employed nuclear force. The overlap function $C^0_{ppnn}(A_f,A_i,r_{12})$ in the short-distance region $r_{12}<1.0$~fm can be approximated by a constant ${\cal N}^A_{ppnn}(f,i)$ depending only on the nucleus and on the details of employed  nuclear force~\cite{Weiss:2018,Cruz-Torres:2021,Tropiano:2021}. The ratios ${\cal N}^A_{ppnn}(f,i)/{\cal N}^{A'}_{ppnn}(f,i)$ for any two nuclei with mass numbers $A$ and $A'$ are less model-dependent. Recently, this property has been exploited in the calculation of the NMEs of $0\nu\beta\beta$ candidate nuclei by combining quantum Monte Carlo and the nuclear shell model~\cite{Weiss:2021}.

 If the processes of both DGT transition and $0\nu\beta\beta$ decay are dominated by the short-range contribution, as pointed out in Ref.~\cite{Shimizu:2018PRL}, then one would have the ratio of these two matrix elements from Eq.(\ref{eq:transiton_density}),
\begin{equation}
\begin{aligned}
    \frac{M^{0\nu\beta\beta}_{\rm GT}}{M^{\rm DGT}} 
    \approx& \lim_{\epsilon \to 0}  \frac{-3\int_0^{\epsilon} dr_{12} r_{12}^2 \rho_{nn}(r_{12}) h_{\rm GT,0}(r_{12})}{\sqrt{3}\int_0^{\epsilon}dr_{12} r_{12}^2 \rho_{nn}(r_{12})}  \\ 
     =&\lim_{\epsilon \to 0} 
    -\sqrt{3} R_A \frac{\int_0^{\epsilon} dr_{12} r_{12} \rho_{nn}(r_{12}) }{\int_0^{\epsilon}dr_{12} r_{12}^2 \rho_{nn}(r_{12})}
    = \lim_{\epsilon \to 0}  C(\epsilon) A^{1/3},
    \end{aligned}
\end{equation}
where $C(\epsilon)$ is a constant depending on the short-range parameter $\epsilon$. This expression holds for both isospin-changing and isospin-conserving transitions, because the isospin effects are encoded in the overlap functions, which cancel in the ratio. Of course, in the realistic case, the validity of the approximations employed to derive the above relation varies with atomic nuclei, and it will be studied in combination with the results of {\em ab initio} calculations.

\begin{figure}
    \centering
    \includegraphics[width = 7cm]{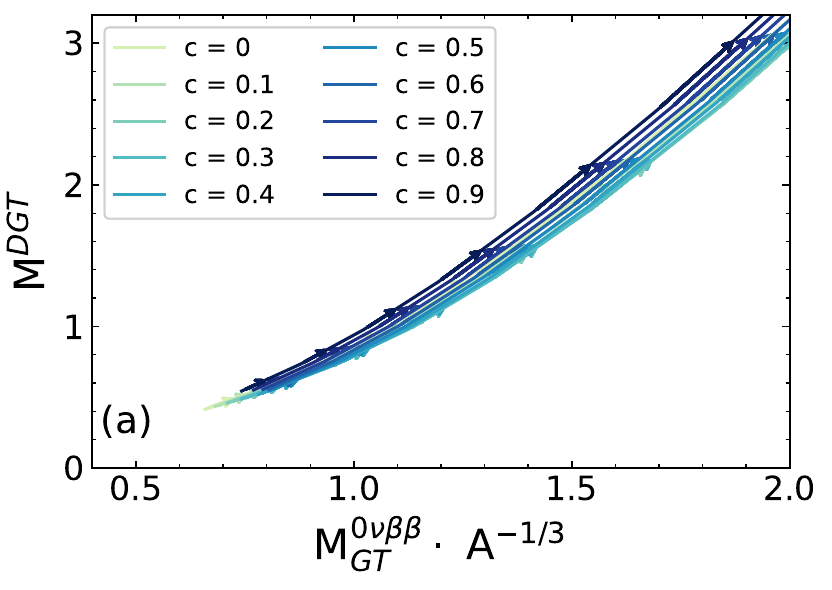}
    \includegraphics[width = 7cm]{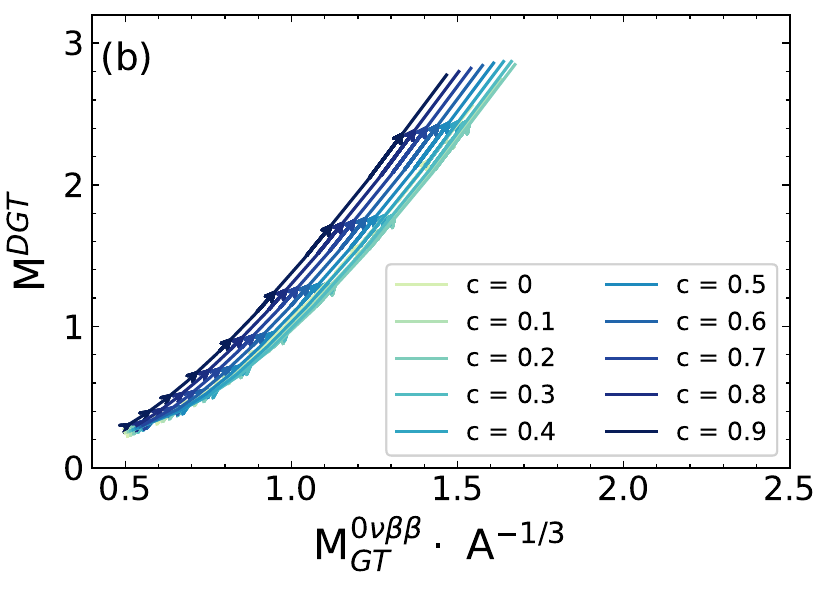}
    \includegraphics[width = 7cm]{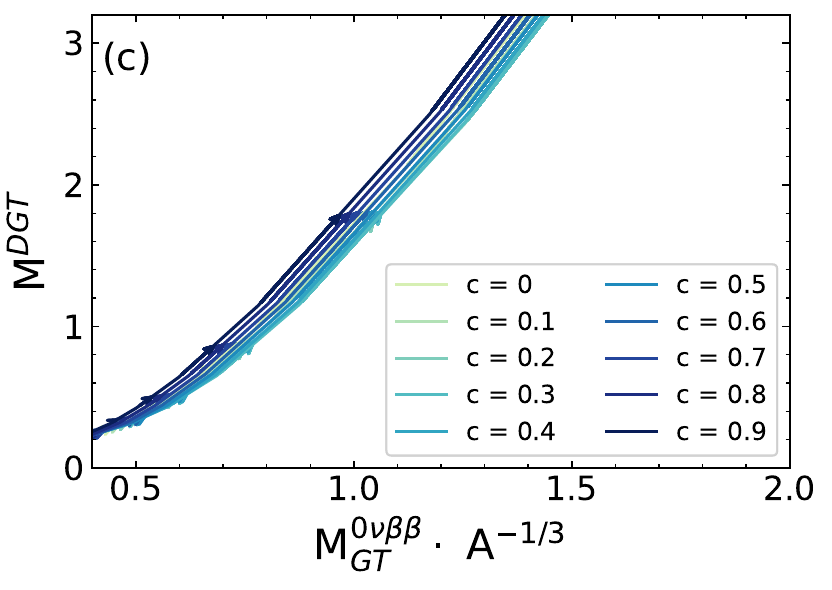}
    \caption{The NMEs from the integral of the sampled transition densities in Fig.~\ref{fig:fit-isospin-conserving-vsimsrg}.  Arrows show the direction of increasing the $d$ value.}
    \label{fig:isospin-conserving-vsimsrg-correlation}
\end{figure}

\begin{figure}
    \centering
    \includegraphics[width=7cm]{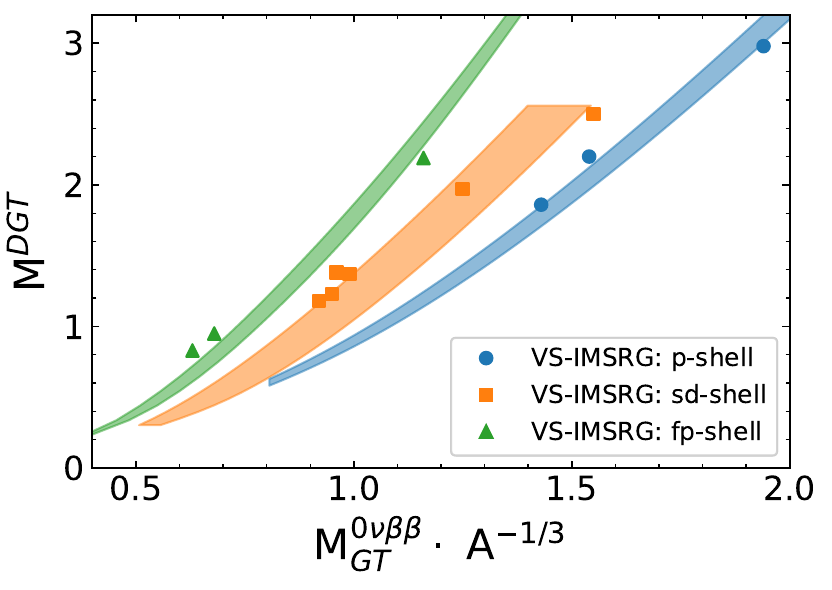}
    \caption{The correlation relations are derived from the sampled transition densities for isospin-conserving transitions. The area is originated from the variants of the parameters $(c,d)$ which control the short-range and long-range behavior, respectively. The NMEs from the VS-IMSRG calculations are added for comparison.}
    \label{fig:isospin-conserving-vsimsrg-correlation-comparison}
\end{figure}

 Let us first parametrize the two-nucleon density $\rho_{nn}(r)$  with the following function form 
 \begin{equation}
  \label{eq:rho_nn_paramer}
    \rho_{nn}(r)=F_{\rm SRC}(r)e^{-r^2/d},
 \end{equation}
 where the parameter $d$ controls the long-range decay behavior. The  function $F_{\rm SRC}(r)$ is introduced to mimic the effect of short-range correlation (SRC)~\cite{Cruz:2018PLB},
 \begin{equation}
  \label{eq:SRC}
  F_{\rm SRC}(r)=1 -ce^{-ar^2}\Bigg(\gamma+\sum^3_{i=1}b_i r^{i+1}\Bigg).
  \end{equation}
 The parameters $a=3.17~{\rm fm^{-2}}$, $\gamma=0.995$, $c=1$, $b_1=1.81~{\rm fm^{-2}}$, $b_2=5.90~{\rm fm^{-3}}$, $b_3=-9.87~{\rm fm^{-4}}$ 
 were obtained in Ref.~\cite{Cruz:2018PLB} by fitting to the results  from cluster variational Monte Carlo calculations for the proton-proton/neutron-neutron correlation functions.  Here, we vary the values of the parameters ($c, d$) within an interval producing reasonable  transition densities to simulate the impacts from the nuclear-force dependence and nucleus dependence~\cite{Cruz-Torres:2021} on the correlation relation between the NMEs of DGT and GT-$0\nu\beta\beta$ transitions. Specifically, we vary the parameter $c$ between $0$ and $1$, and the $d$ between $0.5$ and $5$ to examine the sensitivity of the correlation relation to these two parameters.

 As shown in Figs.~\ref{fig:abinitio_isospin_conserving} and \ref{fig:abinitio_isospin_changing}, the transition densities $\tilde C^\kappa(r_{12})$  of isospin-conserving  and isospin-changing transitions have different dependence on $r_{12}$, so we treat these two transitions separately. Besides, it has been found in previous many-body calculations~\cite{Simkovic:2008,Menendez:2009,Yao:2020PRL} that the transition densities $C^\kappa(r_{12})$ may possess node structures varying in detail with nuclear models and the mass regions of isotopes. We note that this structure is generally similar for the same type of transitions in the isotopes of the same mass region for a given nuclear model. To reproduce this node structure, we approximate the overlap function $C^0_{ppnn}(A_f,A_i,r_{12})$ with the following simple form,
 \begin{equation}
 \label{eq:overlap_function_approx}
     C^0_{ppnn}(A_f,A_i,r_{12})  \simeq p(r_{12})q(r_{12}){\cal N}^A_{ppnn}(f,i),
 \end{equation}
 where $p(r)$ and $q(r)$ are polynomial functions $c_0\big[ 1+ \sum_{i=1}^N c_{2i-1}(r-c_{2i})^{2i}\big]$ with different values of $c_i$ for isospin-changing and isospin-conserving transitions. The parameter $N$ is determined by the node structure of the transition densities and the parameters $c_i$ need to be fitted to transition densities of each nuclear model. In our case, as shown later, we fit to the results of a few transitions in each valence space and take the average values given for the $c_i$'s from those fits.  We also imposed  $p(r) = q(r)$ for the case of isospin-conserving transitions to ensure isospin symmetry.  
 Fig.~\ref{fig:vsimsrg-transition-densities} displays the transition densities of isotopes in $p$, $sd$- and $fp$-shells from the VS-IMSRG calculation with the EM1.8/2.0 interaction. One can see that the distribution of the transition density could have a complex  structure with more than two peaks. In particular, long-range contribution to the DGT transition could be even more significant than the short-range contribution, and it could enhance (quench) the isospin-conserving (isospin-changing) transitions.

 To examine the correlation between the NMEs of DGT and $0\nu\beta\beta$ decay in a more general way, we use the transition densities in Fig.~\ref{fig:vsimsrg-transition-densities} to optimize the parameters $(c,d)$ in the two-nucleon density and $c_i$s in the function $p(r_{12})$ and $q(r_{12})$. Once these parameters are determined, we vary the parameters $(c, d)$ around their optimal values which allows to generate more transition densities    to look at the correlation relation. The sampled transition densities $\tilde{C}^{\kappa}_A(r_{12})$ are displayed in Fig.~\ref{fig:fit-isospin-conserving-vsimsrg}. One can see that the function form \eqref{eq:transiton_density} for the transition density, together with the two-nucleon density $\rho_{nn}(r)$ in \eqref{eq:rho_nn_paramer} and the polynomial function for the overlap $C^{0}_{ppnn}(A_f,A_i,r_{12})$ in (
 \ref{eq:overlap_function_approx}) can reproduce nicely the main structure of both isospin-conserving and isospin-changing transition densities in each mass regions from the VS-IMSRG calculations.

 The sampled transitions are then integrated over the coordinate $r_{12}$ which leads to the NME $\tilde M_{\rm DGT}$ and $\tilde{M}^{0\nu\beta\beta}_{\rm GT}$,  
 \begin{align}
     \tilde{M}^{\kappa}=\int dr_{12}\tilde{C}^{\kappa}_A(r_{12}),
 \end{align} 
 where $\tilde{C}^\kappa_A(r_{12})$ has been defined in \eqref{eq:transiton_density} with the overlap function  \eqref{eq:overlap_function_approx}.  The NME $\tilde{M}^\kappa$ might differ from the actual value of the NME $M^{\kappa}$ because of the unknown constant ${\cal N}^A_{ppnn}(f,i)$, but it does not impact the analysis of the correlation relation between them, except for the intercept parameter.  
 \begin{figure}
    \centering
    \includegraphics[width = 7cm]{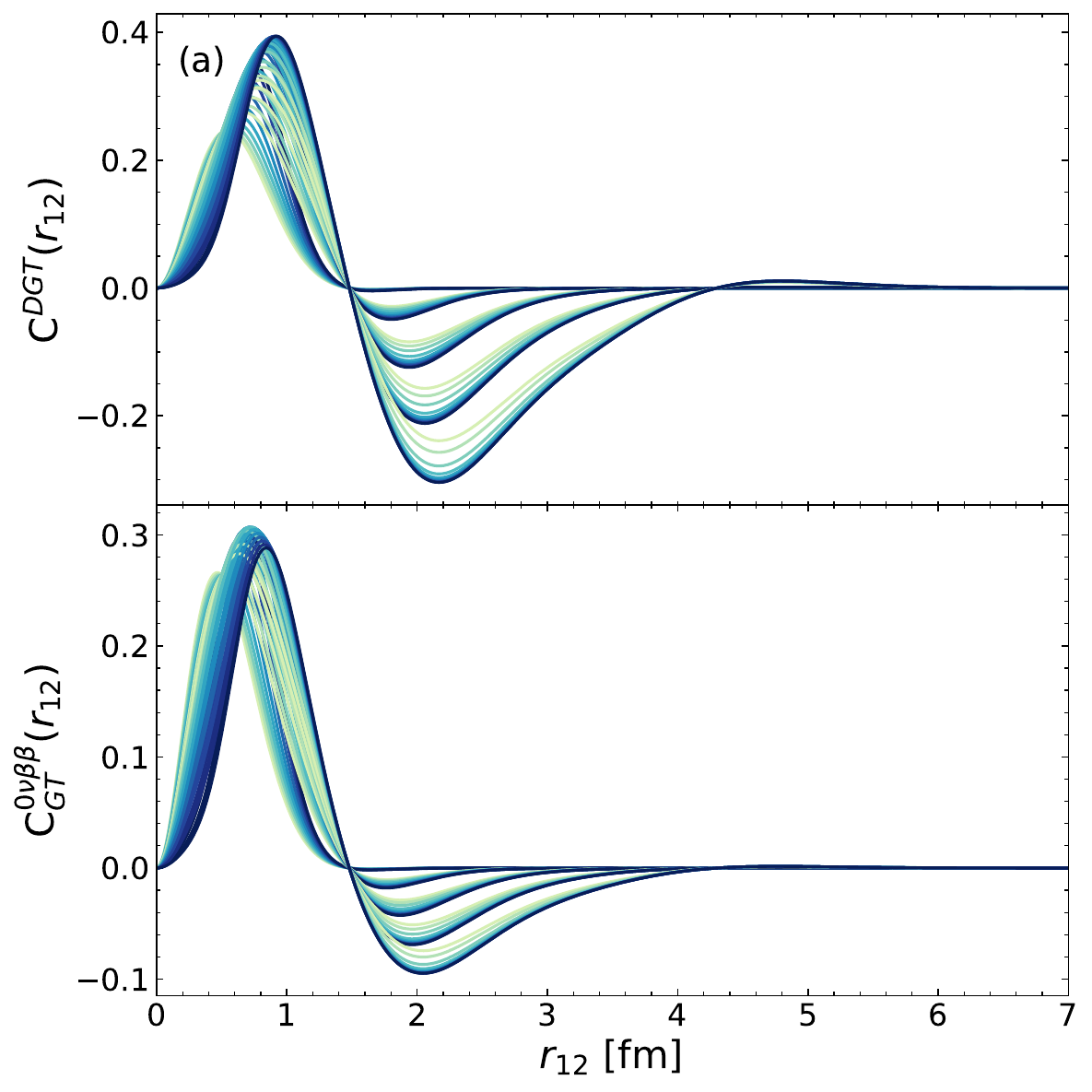}
    \includegraphics[width = 7cm]{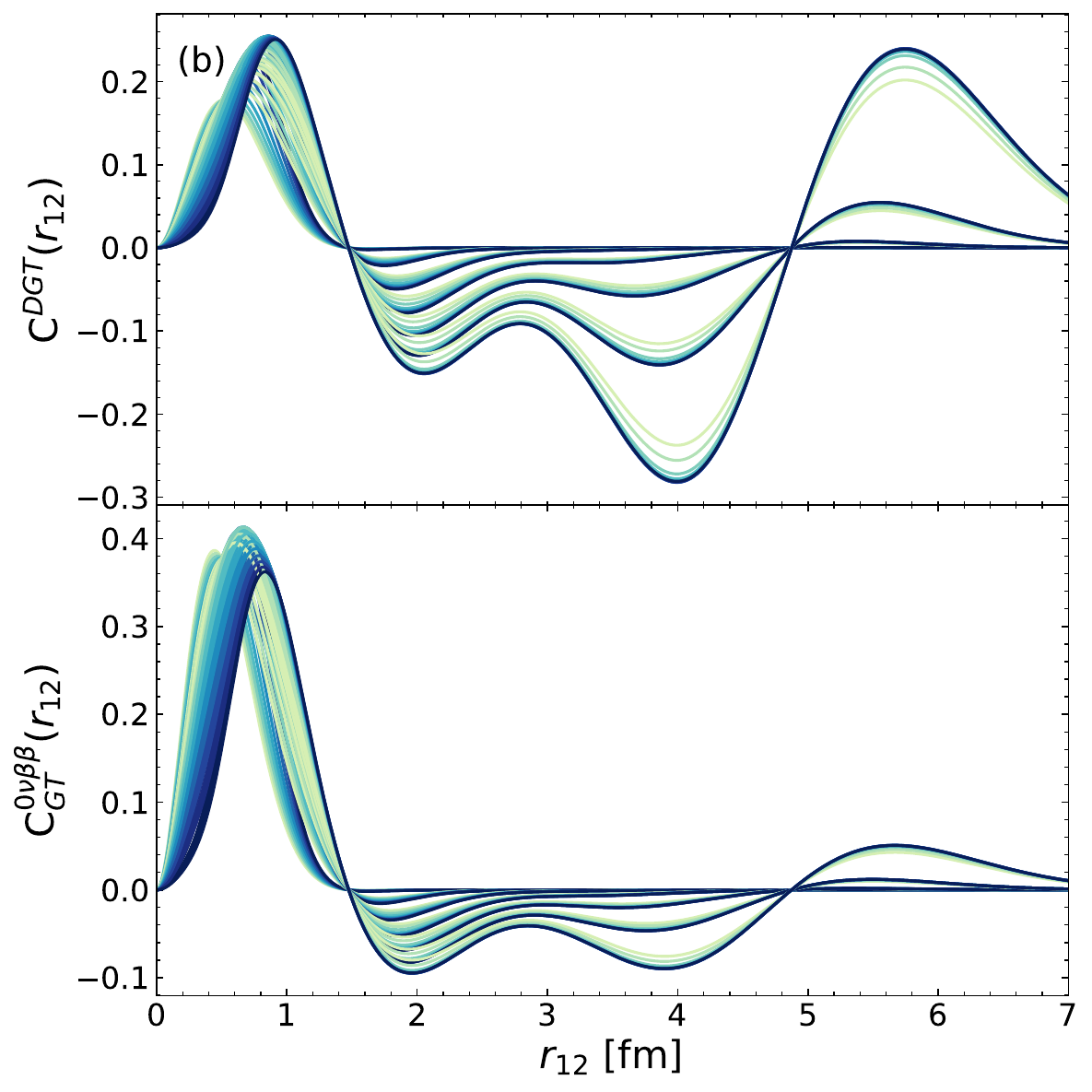}
    \caption{Sampled transition densities  with parameters fitted to the VS-IMSRG results for isospin-changing transitions in the (a) $sd$- and (b) $fp$-shell nuclei, with the parameters $c$ and $d$ increasing from 0 to 1 and from 0.5 fm$^2$ to 4.5 fm$^2$, respectively.  }
    \label{fig:fit-isospin-changing-vsimsrg}
\end{figure}

\begin{figure}
    \centering
    \includegraphics[width=7cm]{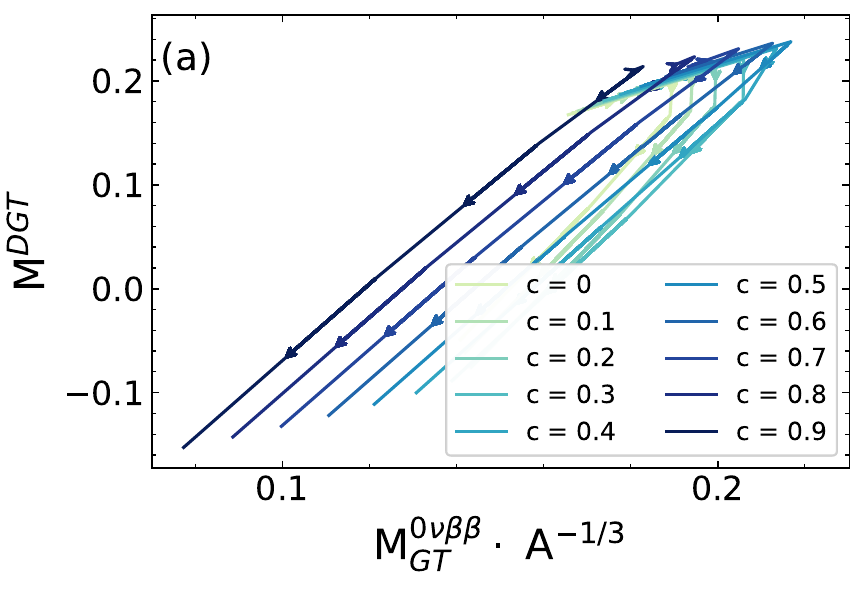}
    \includegraphics[width=7cm]{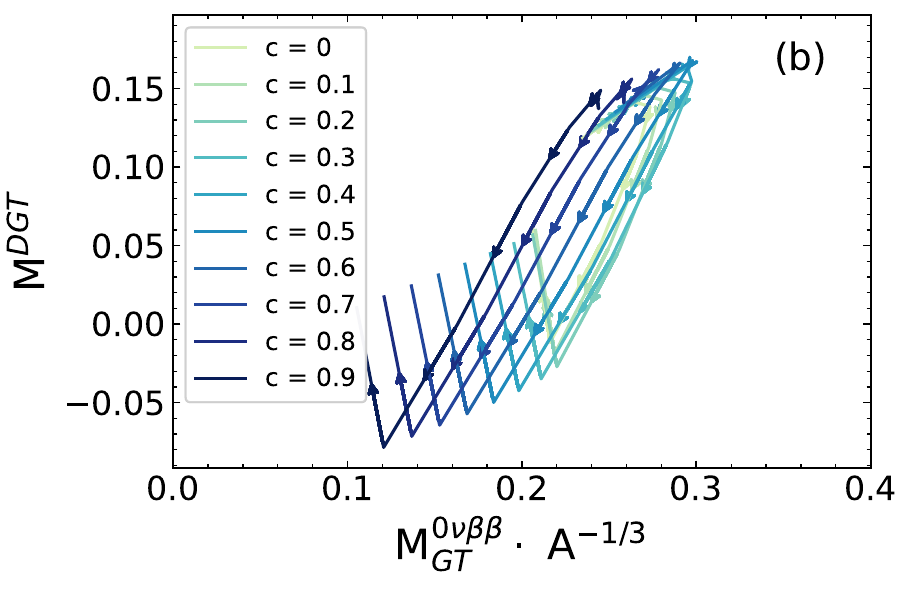}
    \caption{The NMEs from the integral of the sampled transition densities in Fig.~\ref{fig:fit-isospin-changing-vsimsrg}.  (a) for $sd$-shell nuclei and (b) for $fp$-shell nuclei.  Arrows show the direction of increasing the $d$ value. }
    \label{fig:isospin-changing-vsimsrg-correlation}
\end{figure}

\begin{figure} 
    \includegraphics[width = 6cm]{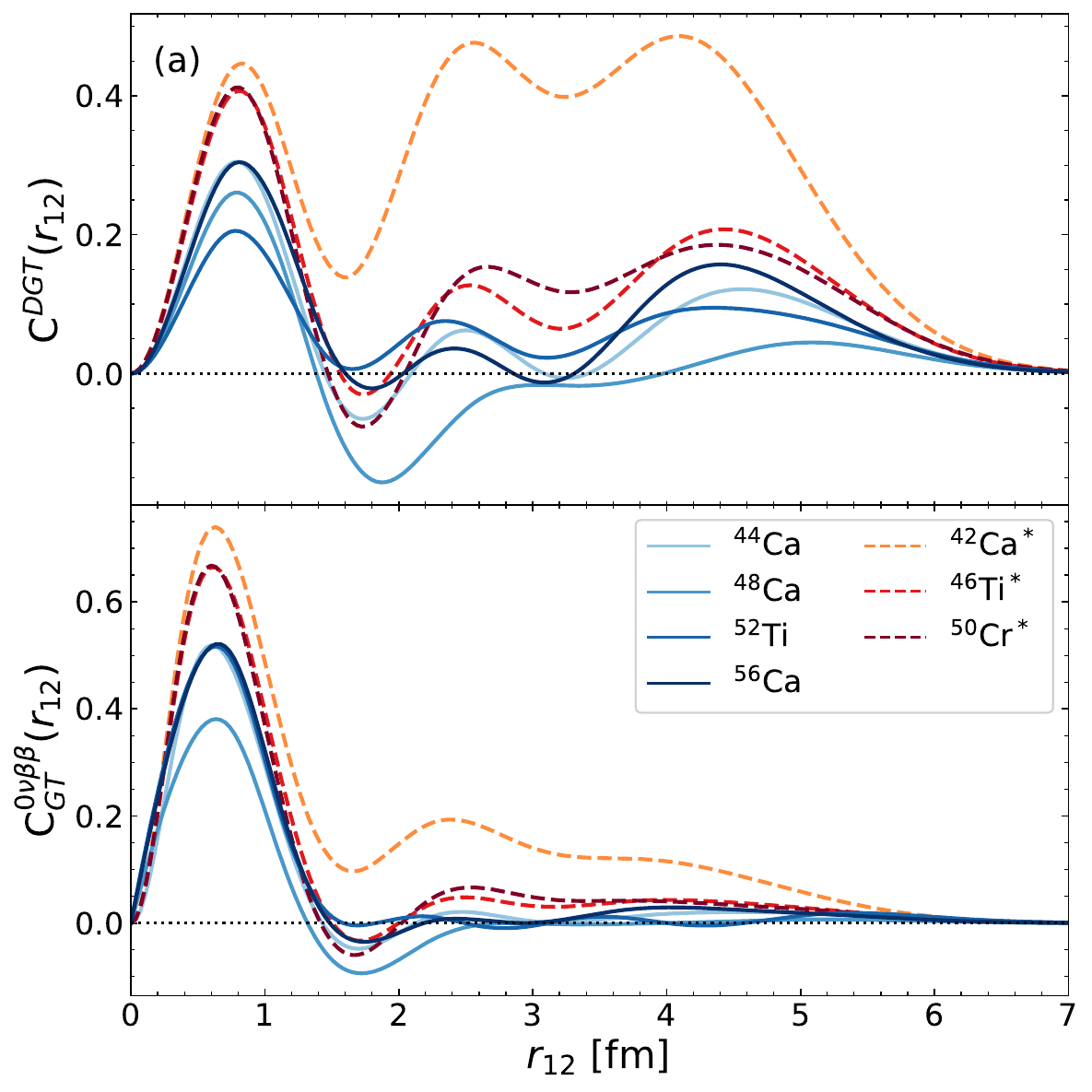}
    \includegraphics[width = 6cm]{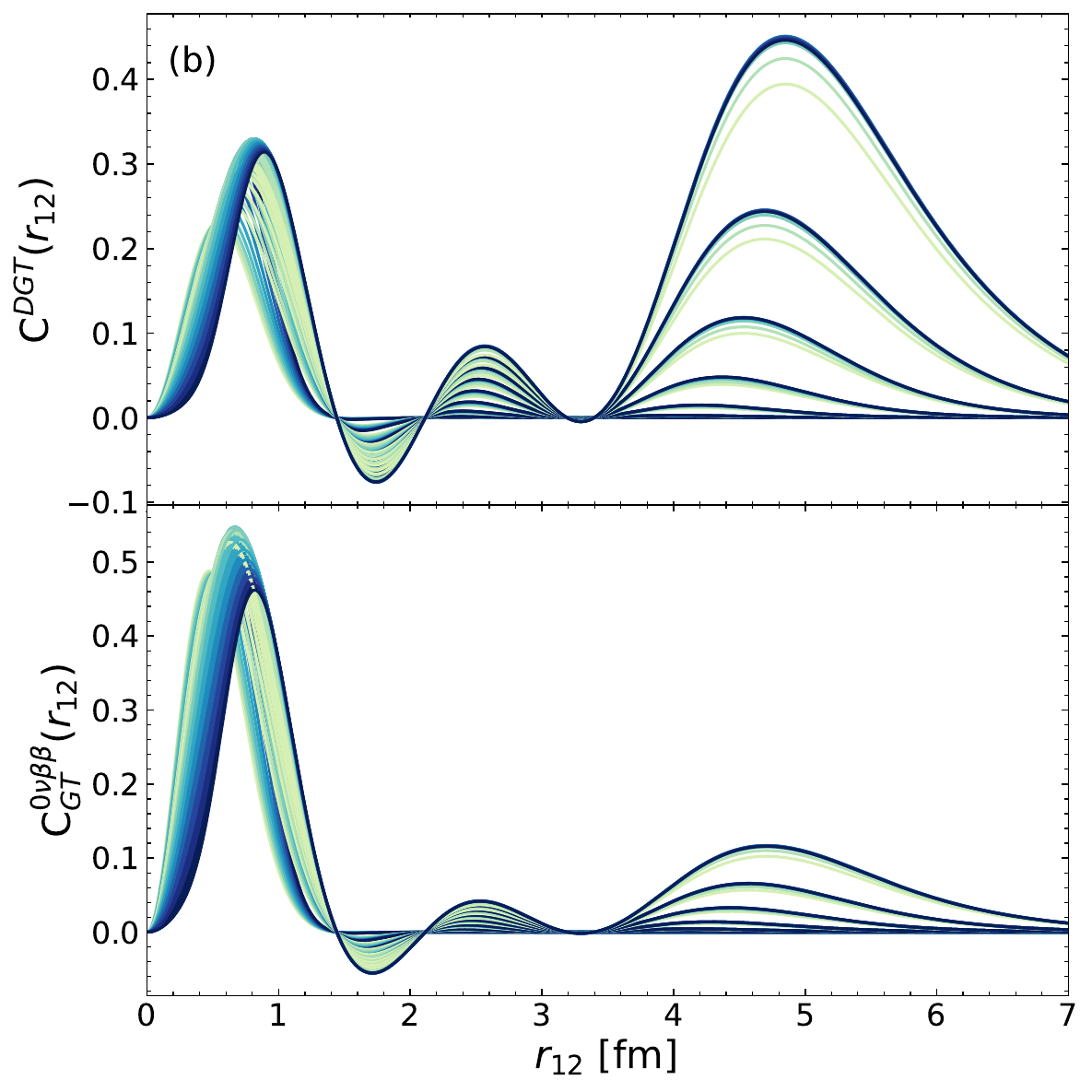}
    \includegraphics[width = 6cm]{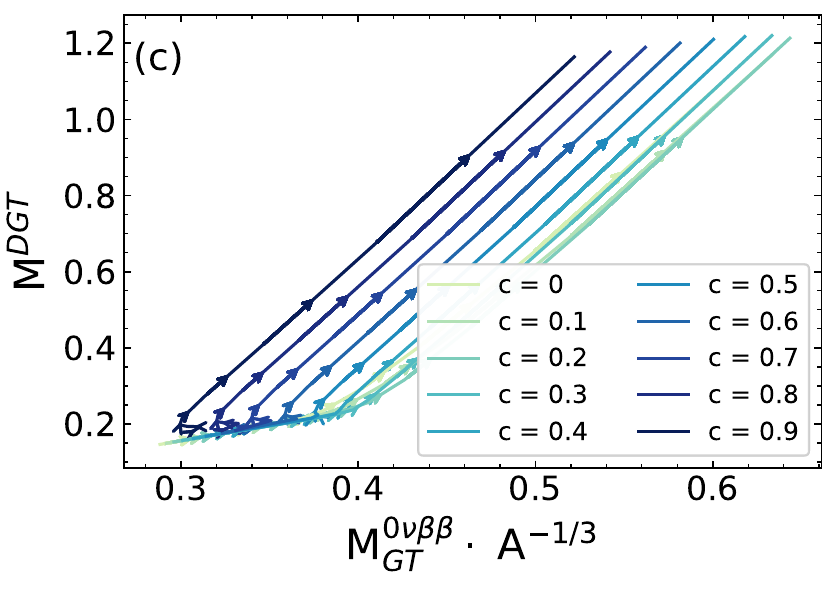}
    \caption{(a) The transition densities from conventional  shell-model calculations; (b) Sampled transition densities with parameters fitted to the isospin-changing transition densities in (a). (c) The NMEs from the integral of the sampled transition densities in (b). }
    \label{fig:isospin-changing-sm-correlation}
\end{figure}

 The NMEs $\tilde M_{\rm DGT}$ and $\tilde{M}^{0\nu\beta\beta}_{\rm GT}$ for isospin-conserving transitions are displayed in Fig.~\ref{fig:isospin-conserving-vsimsrg-correlation}. It is shown that these two types of NMEs are correlated in some way. With the transition densities generated by varying the  parameters ($c, d$) around  those values from the VS-IMSRG calculations for the $p$- and $sd$-shell isotopes, the resultant $M^{\rm DGT}$ is increasing with $M^{0\nu\beta\beta}_{\rm GT}$ in a parabolic form approximately, while that derived from the $fp$-shell isotopes is in a linear form. These correlation relations are summarized in Fig.~\ref{fig:isospin-conserving-vsimsrg-correlation-comparison}, where the NMEs from the VS-IMSRG calculations are added for comparison. It is shown that the location of the VS-IMSRG calculations is generally within the area of the correlation relation for the isotopes in each mass region. However, the correlation relations for the isotopes in different mass regions are offset from each other. It implies that there are probably different correlation relations for the isospin-conserving transitions of isotopes in different mass regions.

The correlation relationship between the NMEs of isospin-changing transitions is more complicated. The corresponding sampled transition densities are displayed in Fig.~\ref{fig:fit-isospin-changing-vsimsrg} and the obtained NMEs are shown in Fig.~\ref{fig:isospin-changing-vsimsrg-correlation}.
Again, one can see that the main structure of the transition densities exhibited in those by the VS-IMSRG calculation is reproduced in the sampled ones. Due to the strong cancellation between long-range and short-range contributions, the final DGT NME is significantly quenched. As a result, the value of the DGT NME varies from a small negative value to a small positive value with the parameters ($c, d$). There is a kind of weak correlation between the NMEs of DGT transitions and $0\nu\beta\beta$ decay, depending much on the mass region of the isotopes and the values of ($c,d$).

For comparison, we perform a similar analysis based on the transition densities from conventional shell-model calculations for isotopes in $sd$- and $fp$-shells. The results are shown in Fig.~\ref{fig:isospin-changing-sm-correlation}. In contrast to the results from the VS-IMSRG calculations, the linear correlation relation between the NMEs is very robust. Varying the parameters $(c,d)$ seems only change the intercept parameter of the linear correlation relation. The main difference between the sampled densities in Fig.~\ref{fig:fit-isospin-changing-vsimsrg} and Fig.~\ref{fig:isospin-changing-sm-correlation} is the contributions from the intermediate- and long-range regions to the NME. A strong cancellation is shown in transition densities derived from the VS-IMSRG calculation, but not in those from the conventional shell-model calculations.

\begin{figure*}[]
\centering
\includegraphics[width=\textwidth]{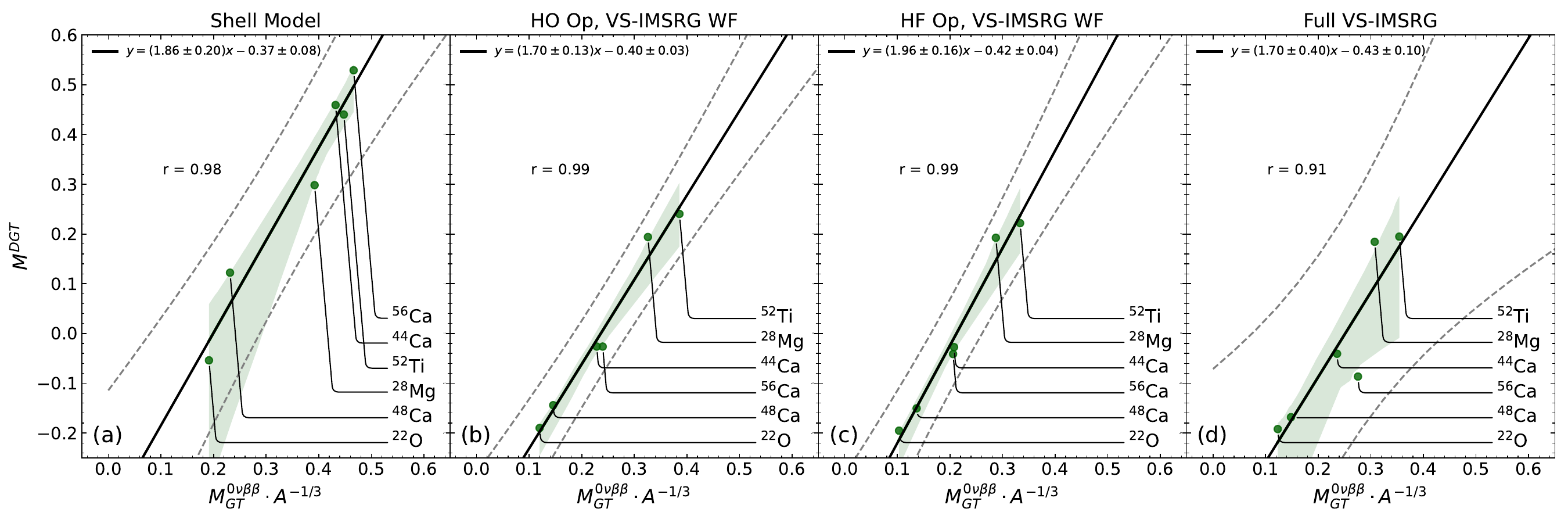}  
\caption{Correlation between the NMEs $M^{\rm DGT}$ of DGT transitions and the $M^{0 \nu\beta\beta}_{\rm GT}$ scaled by $A^{-1/3}$ using a subset of isospin-changing transitions for the $sd$- and $fp$-shells nuclei from the conventional shell-model calculation as well as the results obtained with the VS-IMSRG wave functions and the transition operators in the harmonic oscillator (HO) basis, Hartree-Fock basis (HF), or fully (full) evolved basis. See text for details.  
}
\label{fig:comparison_sm_vsimsrg_correlation_subset}
\end{figure*}
 
\begin{figure}[]
\centering 
\includegraphics[width=8.4cm]{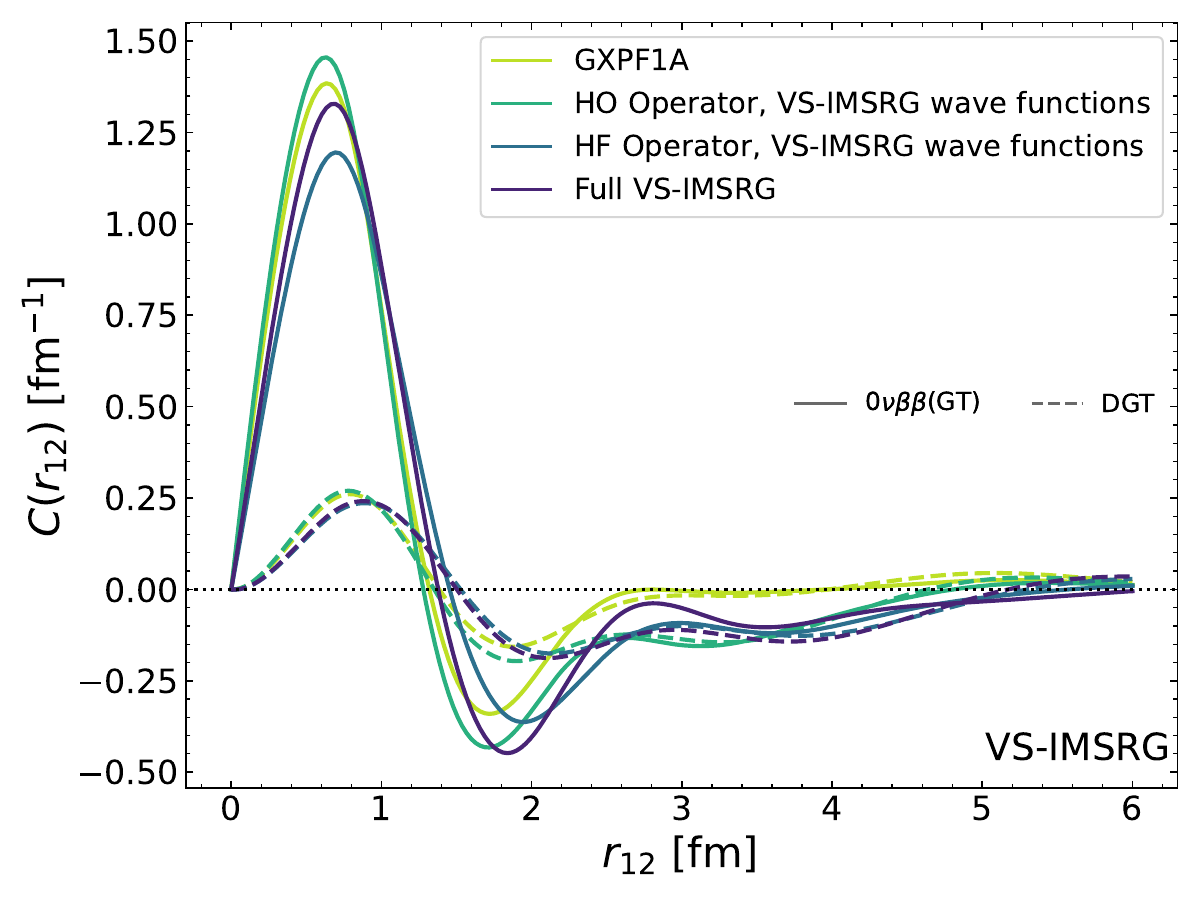} 
\caption{The transition densities of both DGT and the GT part of the $0\nu\beta\beta$ decay from $^{48}$Ca to $^{48}$Ti from the conventional nuclear shell model calculation with the GXPF1A interaction and the VS-IMSRG calculation using the transition operators in three types of basis: HO basis, HF basis, and the fully evolved basis. 
}
\label{fig:Compare_shellmodel_distribution}
\end{figure}

\begin{figure}[]
\centering
\includegraphics[width=7cm]{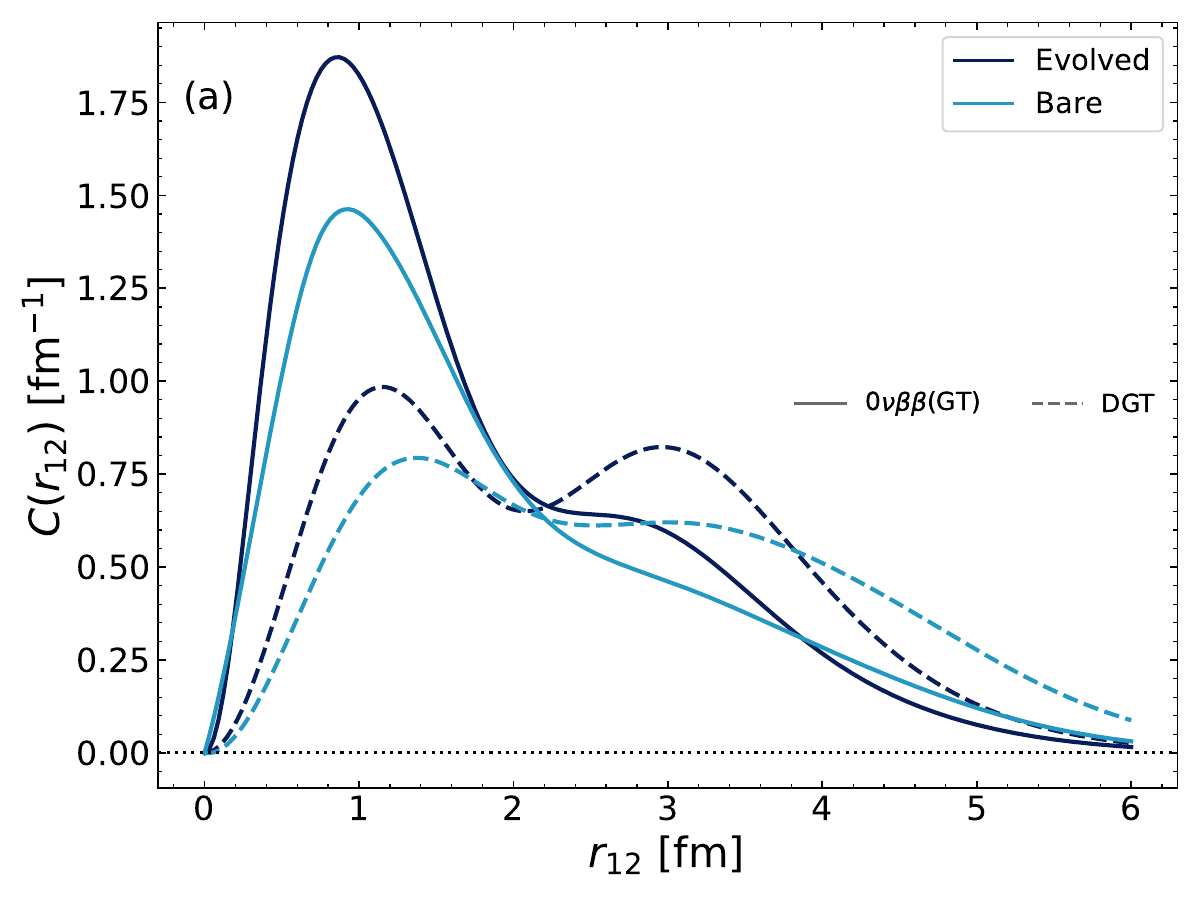} 
\includegraphics[width=7cm]{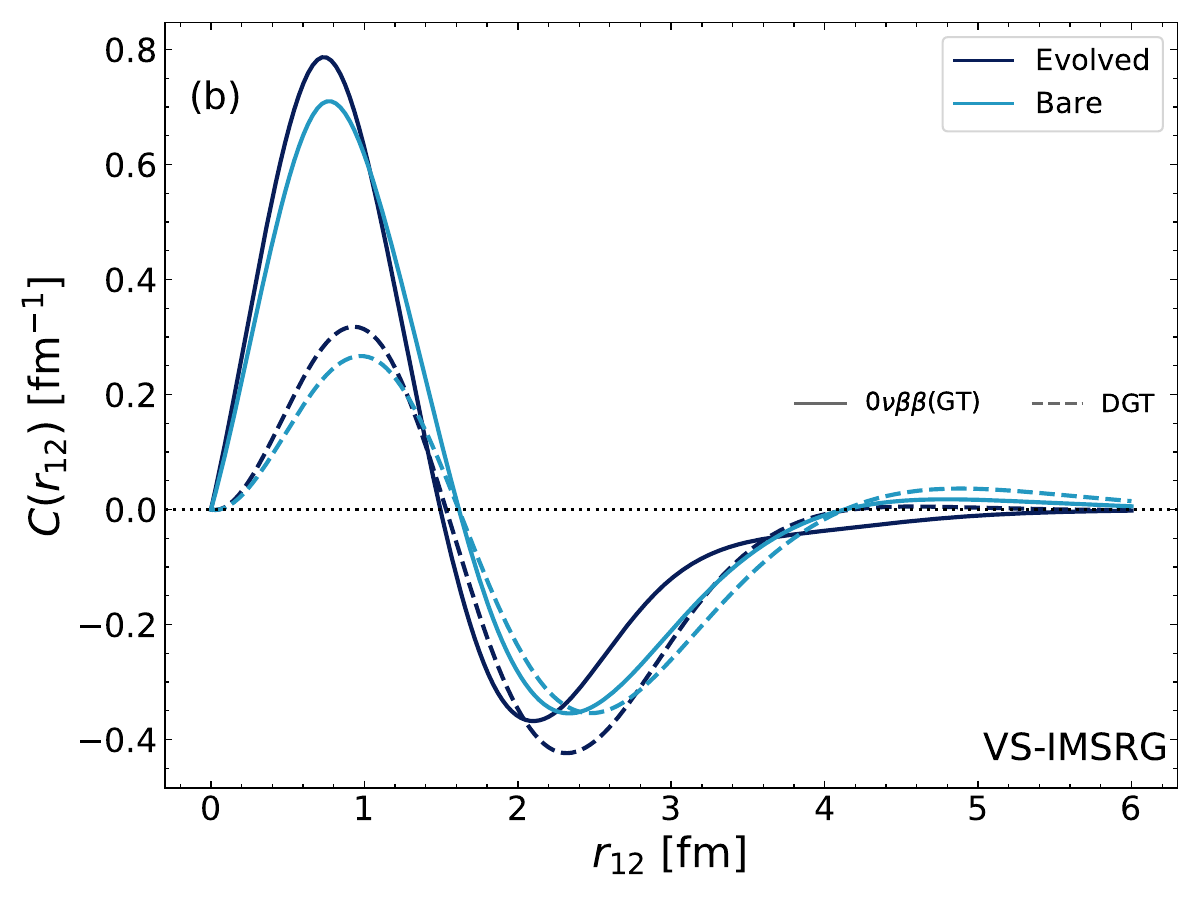} 
\caption{Evolution of the transition distribution of the Gamow-Teller (GT) part of the $0\nu\beta\beta$ decay and DGT transitions from the  VS-ISMSRG in (a) $^{6}$He and (b) $^8$He.}
\label{fig:IMSRG_isospin_conserving_evolution}
\end{figure}

\begin{figure}[]
\centering
\includegraphics[width=7cm]{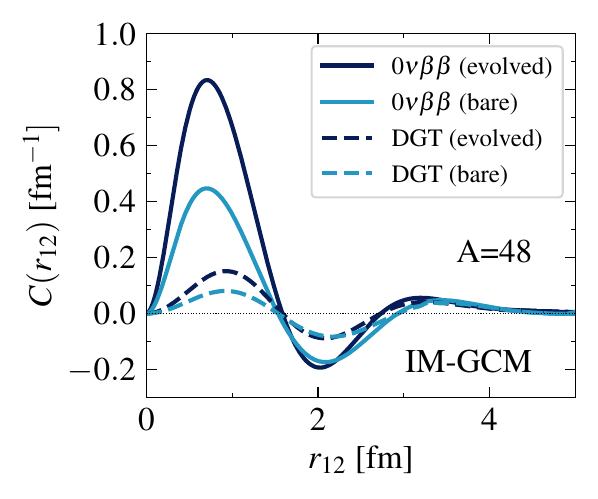}  
\caption{The transition densities of the DGT and GT part of the $0\nu\beta\beta$ decay in $^{48}$Ca from the IM-GCM calculation using either the bare or evolved transition operator, where the same nuclear wave functions but different transition operators are used, respectively. }
\label{fig:IMSRG_Ca2Ti_IMGCM}
\end{figure}
 
The previous analysis starts from (but is not limited to) the assumption that the transition matrix elements are dominated by the short-ranged part of the operator. As can be seen in Fig.~\ref{fig:abinitio_isospin_changing}, the DGT operator satisfies this requirement rather poorly. It is worth considering whether limiting this operator to shorter distances would enhance the scale separation, and thus improve the correlation with the $0\nu\beta\beta$ amplitude. Such a restriction to short distances could conceivably be motivated by the light-ion induced double charge exchange reaction mechanism being surface peaked, and therefore requiring both exchanged nucleons to be relatively localized. Our aim here is not to model the reaction process realistically, but to test whether requiring the DGT operator to be short-range improves the correlation with the NME of $0\nu\beta\beta$ decay.  To this end, we define the NME of the surface localized DGT as
 \begin{align}  
\label{eq:DGT_surf}
M^{\rm DGT}_{\rm surf} &=\left\langle 0^+_f\left|\sum_{1,2} f(r_{12}) g(r_{\rm CM})[\bm{\sigma}_1\otimes \bm{\sigma}_2]^0 \tau^{+}_1\tau^{+}_2\right| 0^+_i\right\rangle,
\end{align} 
where 
\begin{align}
    f(r_{12}) &= \exp\Bigg(-r_{12}^2/2\Bigg),\\ 
    g(R_{12}) &= 1-\exp\bigg(-\frac{R_{12}^2}{2R_A^2}\bigg)
\end{align}
with $R_{12}=|(\mathbf{r}_{1}+\mathbf{r}_{2})/2|$ representing the position of the center of mass of the two particles. The functions $f(r_{12})$ and $g(R_{12})$ ensure that the two particles are close to each other and on the surface of the nucleus, respectively. We find that the correlation of the NME by this operator with $M^{0\nu\beta\beta}\cdot A^{-1/3}$ is worse than the standard DGT operator in (\ref{eq:DGT}).

 \section{The in-medium renormalization effect}
\label{sec:effect}

In this section, we try to understand the origin of the discrepancy between the results of the conventional nuclear shell model and VS-IMSRG as these two methods are comparable. To further understand why a correlation is found in the nuclear shell-model calculation for isospin-changing transitions, we select a subset of transitions that shows a very strong correlation. This subset consists of the transitions $^{22}{\rm O} \to^{22}$Ne, $^{28}{\rm Mg} \to^{28}$Si, $^{44}{\rm Ca} \to^{44}$Ti, $^{48}{\rm Ca} \to^{48}$Ti, $^{56}{\rm Ca} \to^{56}$Ti and $^{52}{\rm Ti} \to^{52}$Cr.
In the conventional shell-model calculations, the USDB interaction \cite{Brown:2006PRC} is used for the $sd$-shell  nuclei, and the  GXPF1A interaction \cite{Honma2005}  for the $fp$-shell nuclei. The results are shown in
Fig.~\ref{fig:comparison_sm_vsimsrg_correlation_subset}. Based on the results within this subset, one finds the following correlation relationship
 \begin{equation}
     M^{\rm DGT} = 1.86(20) M^{0\nu\beta\beta}_{\rm GT}\cdot A^{-1/3} -0.37(8),\quad r=0.98.
 \end{equation}

  The correlation coefficient $r=0.98$ indicates that the two NMEs are strongly correlated, as shown in Fig.~\ref{fig:comparison_sm_vsimsrg_correlation_subset}(a). This result is consistent with the previous shell-model study~\cite{Shimizu:2018PRL}, as expected. In contrast, Fig.~\ref{fig:comparison_sm_vsimsrg_correlation_subset}(d) shows that the correlation relation from the full VS-IMSRG calculation is weakened with the correlation coefficient $r=0.91$.
  We note that the $0\nu\beta\beta$ NMEs of these nuclei by the  the VS-IMSRG are generally smaller than  those by the conventional shell-model calculations, while the DGT NMEs are much smaller and even with opposite sign. This is mainly due to the configuration mixing in nuclear wave functions predicted differently in the calculations using the conventional shell-model interactions and those derived from VS-IMSRG.

  To better understand how the in-medium renormalization effect from the VS-IMSRG evolution on the transition operator affects the correlation,  we provide two intermediate results in Fig.~\ref{fig:comparison_sm_vsimsrg_correlation_subset}(b) and (c), where the nuclear wave functions are from the VS-IMSRG calculation, while the transition operator in the harmonic oscillator (HO) basis or in the Hartree-Fock (HF) basis is used, respectively.  In these two intermediate results, the transition operator is not consistently evolved.  One can see that the correlation in the results of calculations with either the HO transition operator or the HF operator is even stronger than that of the conventional shell model in which the operator is represented in the harmonic oscillator basis.

  Taking $^{48}$Ca as an example, we illustrate how the transition density distribution looks like in the four types of calculations shown in Fig.~\ref{fig:comparison_sm_vsimsrg_correlation_subset}. The transition densities are displayed in Fig.~\ref{fig:Compare_shellmodel_distribution}. 
 One can see that the use of the transition operator from the HO one to the IMSRG evolved one modifies the transition densities slightly in both short-range ($\simeq 0.6$ fm) and long-range ($\simeq2.0-3.0$ fm) regions. Quantitatively, this modification is slightly different for different isotopes.  Figure~\ref{fig:IMSRG_isospin_conserving_evolution}   shows 
 the transition densities for the isospin-conserving transition $^{6}{\rm He}\to ^{6}$Be and for the isospin-changing transition $^{8}{\rm He}\to ^{8}$Be obtained with and without VS-IMSRG evolution.
 The evolution enhances the short-range part of the transition densities for the isospin-conserving transitions, leading to an overall enhancement of the NME. This behavior is found in all the isospin-conserving cases. 
For the isospin-changing transition of $^{8}$He, the evolution increases the magnitude of the peaks of the distributions but, due to cancellations, does not necessarily increase the final NMEs.
This behavior is found in all isospin-changing cases. In particular, one finds that the renormalization effect on the short-range part of the transition density is more significant in the isospin-conserving transitions than in the isospin-changing transitions. 
In short, the effect of VS-IMSRG evolution on isospin-changing transitions is generally small, but the details of the cancellation between short- and long-range components depend on the nucleus and the operator, which finally degrades the correlation.

To help assess the method-dependence of these conclusions,  Fig.~\ref{fig:IMSRG_Ca2Ti_IMGCM} displays the  transition densities of the DGT and GT part of the $0\nu\beta\beta$ decay in $^{48}$Ca from the IM-GCM calculation using either the bare or evolved transition operator, where the renormalization effect  enlarges significantly the short-range contribution for both transitions.
It has been discussed in Ref.~\cite{Yao:2020PRL} that the multi-reference IMSRG flow incorporates the effects of pairing in high-energy orbitals, greatly enhancing the contribution of the $J=0$ pair of nucleons to the NMEs.
We note that one should not make a direct comparison of the operator renormalization effects in the VS-IMSRG and IM-GCM calculations, as the operator renormalization and many-body correlations are partitioned in a different way.  It is, however, meaningful to compare the final matrix elements obtained with renormalized operators; in this case the discrepancy between the results of these two calculations reflects the error due to the missing of higher-body operators in both methods.  As shown in Ref.~\cite{Yao:2021PRC} the inclusion of an induced three-body transition operator helps reduce the discrepancy between them. Therefore, these two variants of IMSRG provide a complementary description of NMEs of neutrinoless double-beta decay.

\section{Conclusion} 
 \label{sec:summary}
 
In this work, we have explored the possible correlation between  the NMEs $M^{0\nu\beta\beta}$ of ground-state to ground-state $0\nu\beta\beta$ decay and those $M^{\rm DGT}$ of  DGT transitions in a set of nuclei in different mass regions with three {\em ab initio} methods starting from the same NN+3N chiral interactions. We have found that the obtained $M^{\rm DGT}$ is correlated with the quantity $M^{0\nu\beta\beta} A^{-1/3}$ slightly stronger than with $M^{0\nu\beta\beta} A^{-1/6}$ for isospin-conserving transitions, where the long-range and short-range contributions add coherently, leading to  large values of both transition matrix elements. 
However, the correlation relation turns out to be much weaker in  isospin-changing transitions where the long-range and short-range contributions compensate each other, leading to small values of $M^{\rm DGT}$. This conclusion has been confirmed with a scale-separation analysis in which we have sampled a set of transition densities for both isospin-conserving and isospin-changing transitions by changing the short-range and long-range behavior around the results from the VS-IMSRG calculation.  

We have also explored the origin of the discrepancy between the NMEs from conventional shell-model calculations and VS-IMSRG calculations. Our studies have shown that apart from the discrepancy mainly due to the configuration mixing predicted differently in the calculations using the conventional shell-model interactions and those derived from VS-IMSRG,  the in-medium renormalization effect from the VS-IMSRG evolution on the transition operator  varies with isotopes, which spoils somewhat the correlation relation.

Combining the NMEs of both isospin-conserving and isospin-changing transitions from the three {\em ab initio} calculations, we observe a strong correlation. However, the correlation is considerably weaker for the experimentally relevant isospin-changing transitions.
In other words, a large uncertainty will likely still exist in the $M^{0\nu\beta\beta}$ even if the ground-state to ground-state DGT transition of the candidate nucleus is precisely measured. 
It is worth mentioning that our current analysis is mainly based on the results of calculations with the chiral interaction EM1.8/2.0. A comprehensive way to examine the correlation relation can be carried out by computing the NMEs with a set of chiral nuclear forces with the low-energy constants varying within acceptable regions~\cite{Hu:2021}. Any data on the transition NME of DGT transition 
may provide a constraint on the chiral interaction, and finally on the predicted  $M^{0\nu\beta\beta}$. Besides, we note that recently Romeo {\em et al.}~\cite{Romeo:2021}  found a good linear correlation  between the $0\nu\beta\beta$ decays  and the double gamma transitions, suggesting another potential way to constrain the NMEs of $0\nu\beta\beta$ decay. The present study can be extended straightforwardly to examine that linear correlation relation as well.  

\section*{Acknowledgments}

 We thank J. Menéndez for sending us the results from the calculations of conventional nuclear models and for his careful reading of this manuscript and fruitful discussions. J.M.Y. also thanks C.F. Jiao, J. Meng, and Y.F. Niu for extensive discussions. This work is supported in part by the National Natural Science Foundation of China (Grant No. 12141501) and the Fundamental Research Funds for the Central Universities, Sun Yat-sen University,  the U.S.\ Department of Energy, Office of Science, Office of Nuclear Physics under Awards No.\ DE-SC0017887,  No.\ DE-SC0018083 (NUCLEI SciDAC-4 Collaboration), DE-FG02-97ER41019, DE-AC02-06CH11357,  and DE-SC0015376 (the DBD Topical Theory Collaboration), NSERC, the Arthur B.\ McDonald Canadian Astroparticle Physics Research Institute, the Canadian Institute for Nuclear Physics, the U.S.~Department of Energy (DOE) under contract DE-FG02-97ER41014, and the Deutsche Forschungsgemeinschaft (DFG, German Research Foundation) -- Project-ID 279384907 -- SFB 1245. TRIUMF receives funding via a contribution through the National Research Council of Canada. The IM-GCM and IT-NCSM calculations were carried out using the computing resources provided by the Institute for Cyber-Enabled Research at Michigan State University, and the U.S.~National Energy Research Scientific Computing Center (NERSC), a DOE Office of Science User Facility supported by the Office of Science of the U.S.~Department of Energy under Contract No.\ DE-AC02-05CH11231.   The VS-IMSRG computations were performed with an allocation of computing resources on Cedar at WestGrid and Compute Canada, and on the Oak Cluster at TRIUMF managed by the University of British Columbia Department of Advanced Research Computing (ARC). 
 

\begin{thebibliography}{93}%
\makeatletter
\providecommand \@ifxundefined [1]{%
 \@ifx{#1\undefined}
}%
\providecommand \@ifnum [1]{%
 \ifnum #1\expandafter \@firstoftwo
 \else \expandafter \@secondoftwo
 \fi
}%
\providecommand \@ifx [1]{%
 \ifx #1\expandafter \@firstoftwo
 \else \expandafter \@secondoftwo
 \fi
}%
\providecommand \natexlab [1]{#1}%
\providecommand \enquote  [1]{``#1''}%
\providecommand \bibnamefont  [1]{#1}%
\providecommand \bibfnamefont [1]{#1}%
\providecommand \citenamefont [1]{#1}%
\providecommand \href@noop [0]{\@secondoftwo}%
\providecommand \href [0]{\begingroup \@sanitize@url \@href}%
\providecommand \@href[1]{\@@startlink{#1}\@@href}%
\providecommand \@@href[1]{\endgroup#1\@@endlink}%
\providecommand \@sanitize@url [0]{\catcode `\\12\catcode `\$12\catcode
  `\&12\catcode `\#12\catcode `\^12\catcode `\_12\catcode `\%12\relax}%
\providecommand \@@startlink[1]{}%
\providecommand \@@endlink[0]{}%
\providecommand \url  [0]{\begingroup\@sanitize@url \@url }%
\providecommand \@url [1]{\endgroup\@href {#1}{\urlprefix }}%
\providecommand \urlprefix  [0]{URL }%
\providecommand \Eprint [0]{\href }%
\providecommand \doibase [0]{http://dx.doi.org/}%
\providecommand \selectlanguage [0]{\@gobble}%
\providecommand \bibinfo  [0]{\@secondoftwo}%
\providecommand \bibfield  [0]{\@secondoftwo}%
\providecommand \translation [1]{[#1]}%
\providecommand \BibitemOpen [0]{}%
\providecommand \bibitemStop [0]{}%
\providecommand \bibitemNoStop [0]{.\EOS\space}%
\providecommand \EOS [0]{\spacefactor3000\relax}%
\providecommand \BibitemShut  [1]{\csname bibitem#1\endcsname}%
\let\auto@bib@innerbib\@empty
\bibitem [{\citenamefont {Furry}(1939)}]{Furry:1939}%
  \BibitemOpen
  \bibfield  {author} {\bibinfo {author} {\bibfnamefont {W.~H.}\ \bibnamefont
  {Furry}},\ }\href {\doibase 10.1103/PhysRev.56.1184} {\bibfield  {journal}
  {\bibinfo  {journal} {Phys. Rev.}\ }\textbf {\bibinfo {volume} {56}},\
  \bibinfo {pages} {1184} (\bibinfo {year} {1939})}\BibitemShut {NoStop}%
\bibitem [{\citenamefont {Tanabashi}\ and\ \citenamefont {the
  others}(2018)\citenamefont {Tanabashi} \emph {et~al.}}]{PDG:2018}%
  \BibitemOpen
  \bibfield  {author} {\bibinfo {author} {\bibfnamefont {M.}~\bibnamefont
  {Tanabashi}} \emph {et~al.} (\bibinfo {collaboration} {Particle Data
  Group}),\ }\href {\doibase 10.1103/PhysRevD.98.030001} {\bibfield  {journal}
  {\bibinfo  {journal} {Phys. Rev. D}\ }\textbf {\bibinfo {volume} {98}},\
  \bibinfo {pages} {030001} (\bibinfo {year} {2018})}\BibitemShut {NoStop}%
\bibitem [{\citenamefont {Anton}\ \emph {et~al.}(2019)\citenamefont {Anton},
  \citenamefont {Badhrees}, \citenamefont {Barbeau}, \citenamefont {Beck},
  \citenamefont {Belov}, \citenamefont {Bhatta}, \citenamefont {Breidenbach},
  \citenamefont {Brunner}, \citenamefont {Cao}, \citenamefont {Cen},
  \citenamefont {Chambers}, \citenamefont {Cleveland}, \citenamefont {Coon},
  \citenamefont {Craycraft}, \citenamefont {Daniels}, \citenamefont {Danilov},
  \citenamefont {Darroch}, \citenamefont {Daugherty}, \citenamefont {Davis},
  \citenamefont {Delaquis}, \citenamefont {Der Mesrobian-Kabakian},
  \citenamefont {DeVoe}, \citenamefont {Dilling}, \citenamefont {Dolgolenko},
  \citenamefont {Dolinski}, \citenamefont {Echevers}, \citenamefont {Fairbank},
  \citenamefont {Fairbank}, \citenamefont {Farine}, \citenamefont {Feyzbakhsh},
  \citenamefont {Fierlinger}, \citenamefont {Fudenberg}, \citenamefont
  {Gautam}, \citenamefont {Gornea}, \citenamefont {Gratta}, \citenamefont
  {Hall}, \citenamefont {Hansen}, \citenamefont {Hoessl}, \citenamefont
  {Hufschmidt}, \citenamefont {Hughes}, \citenamefont {Iverson}, \citenamefont
  {Jamil}, \citenamefont {Jessiman}, \citenamefont {Jewell}, \citenamefont
  {Johnson}, \citenamefont {Karelin}, \citenamefont {Kaufman}, \citenamefont
  {Koffas}, \citenamefont {Kr\"ucken}, \citenamefont {Kuchenkov}, \citenamefont
  {Kumar}, \citenamefont {Lan}, \citenamefont {Larson}, \citenamefont
  {Lenardo}, \citenamefont {Leonard}, \citenamefont {Li}, \citenamefont {Li},
  \citenamefont {Li}, \citenamefont {Licciardi}, \citenamefont {Lin},
  \citenamefont {MacLellan}, \citenamefont {McElroy}, \citenamefont {Michel},
  \citenamefont {Mong}, \citenamefont {Moore}, \citenamefont {Murray},
  \citenamefont {Njoya}, \citenamefont {Nusair}, \citenamefont {Odian},
  \citenamefont {Ostrovskiy}, \citenamefont {Piepke}, \citenamefont {Pocar},
  \citenamefont {Reti\`ere}, \citenamefont {Robinson}, \citenamefont {Rowson},
  \citenamefont {Ruddell}, \citenamefont {Runge}, \citenamefont {Schmidt},
  \citenamefont {Sinclair}, \citenamefont {Soma}, \citenamefont {Stekhanov},
  \citenamefont {Tarka}, \citenamefont {Todd}, \citenamefont {Tolba},
  \citenamefont {Totev}, \citenamefont {Veenstra}, \citenamefont
  {Veeraraghavan}, \citenamefont {Vogel}, \citenamefont {Vuilleumier},
  \citenamefont {Wagenpfeil}, \citenamefont {Watkins}, \citenamefont {Weber},
  \citenamefont {Wen}, \citenamefont {Wichoski}, \citenamefont {Wrede},
  \citenamefont {Wu}, \citenamefont {Xia}, \citenamefont {Yahne}, \citenamefont
  {Yang}, \citenamefont {Yen}, \citenamefont {Zeldovich},\ and\ \citenamefont
  {Ziegler}}]{EXO-200:2019}%
  \BibitemOpen
  \bibfield  {author} {\bibinfo {author} {\bibfnamefont {G.}~\bibnamefont
  {Anton}}, \bibinfo {author} {\bibfnamefont {I.}~\bibnamefont {Badhrees}},
  \bibinfo {author} {\bibfnamefont {P.~S.}\ \bibnamefont {Barbeau}}, \bibinfo
  {author} {\bibfnamefont {D.}~\bibnamefont {Beck}}, \bibinfo {author}
  {\bibfnamefont {V.}~\bibnamefont {Belov}}, \bibinfo {author} {\bibfnamefont
  {T.}~\bibnamefont {Bhatta}}, \bibinfo {author} {\bibfnamefont
  {M.}~\bibnamefont {Breidenbach}}, \bibinfo {author} {\bibfnamefont
  {T.}~\bibnamefont {Brunner}}, \bibinfo {author} {\bibfnamefont {G.~F.}\
  \bibnamefont {Cao}}, \bibinfo {author} {\bibfnamefont {W.~R.}\ \bibnamefont
  {Cen}}, \bibinfo {author} {\bibfnamefont {C.}~\bibnamefont {Chambers}},
  \bibinfo {author} {\bibfnamefont {B.}~\bibnamefont {Cleveland}}, \bibinfo
  {author} {\bibfnamefont {M.}~\bibnamefont {Coon}}, \bibinfo {author}
  {\bibfnamefont {A.}~\bibnamefont {Craycraft}}, \bibinfo {author}
  {\bibfnamefont {T.}~\bibnamefont {Daniels}}, \bibinfo {author} {\bibfnamefont
  {M.}~\bibnamefont {Danilov}}, \bibinfo {author} {\bibfnamefont
  {L.}~\bibnamefont {Darroch}}, \bibinfo {author} {\bibfnamefont {S.~J.}\
  \bibnamefont {Daugherty}}, \bibinfo {author} {\bibfnamefont {J.}~\bibnamefont
  {Davis}}, \bibinfo {author} {\bibfnamefont {S.}~\bibnamefont {Delaquis}},
  \bibinfo {author} {\bibfnamefont {A.}~\bibnamefont {Der Mesrobian-Kabakian}},
  \bibinfo {author} {\bibfnamefont {R.}~\bibnamefont {DeVoe}}, \bibinfo
  {author} {\bibfnamefont {J.}~\bibnamefont {Dilling}}, \bibinfo {author}
  {\bibfnamefont {A.}~\bibnamefont {Dolgolenko}}, \bibinfo {author}
  {\bibfnamefont {M.~J.}\ \bibnamefont {Dolinski}}, \bibinfo {author}
  {\bibfnamefont {J.}~\bibnamefont {Echevers}}, \bibinfo {author}
  {\bibfnamefont {W.}~\bibnamefont {Fairbank}}, \bibinfo {author}
  {\bibfnamefont {D.}~\bibnamefont {Fairbank}}, \bibinfo {author}
  {\bibfnamefont {J.}~\bibnamefont {Farine}}, \bibinfo {author} {\bibfnamefont
  {S.}~\bibnamefont {Feyzbakhsh}}, \bibinfo {author} {\bibfnamefont
  {P.}~\bibnamefont {Fierlinger}}, \bibinfo {author} {\bibfnamefont
  {D.}~\bibnamefont {Fudenberg}}, \bibinfo {author} {\bibfnamefont
  {P.}~\bibnamefont {Gautam}}, \bibinfo {author} {\bibfnamefont
  {R.}~\bibnamefont {Gornea}}, \bibinfo {author} {\bibfnamefont
  {G.}~\bibnamefont {Gratta}}, \bibinfo {author} {\bibfnamefont
  {C.}~\bibnamefont {Hall}}, \bibinfo {author} {\bibfnamefont {E.~V.}\
  \bibnamefont {Hansen}}, \bibinfo {author} {\bibfnamefont {J.}~\bibnamefont
  {Hoessl}}, \bibinfo {author} {\bibfnamefont {P.}~\bibnamefont {Hufschmidt}},
  \bibinfo {author} {\bibfnamefont {M.}~\bibnamefont {Hughes}}, \bibinfo
  {author} {\bibfnamefont {A.}~\bibnamefont {Iverson}}, \bibinfo {author}
  {\bibfnamefont {A.}~\bibnamefont {Jamil}}, \bibinfo {author} {\bibfnamefont
  {C.}~\bibnamefont {Jessiman}}, \bibinfo {author} {\bibfnamefont {M.~J.}\
  \bibnamefont {Jewell}}, \bibinfo {author} {\bibfnamefont {A.}~\bibnamefont
  {Johnson}}, \bibinfo {author} {\bibfnamefont {A.}~\bibnamefont {Karelin}},
  \bibinfo {author} {\bibfnamefont {L.~J.}\ \bibnamefont {Kaufman}}, \bibinfo
  {author} {\bibfnamefont {T.}~\bibnamefont {Koffas}}, \bibinfo {author}
  {\bibfnamefont {R.}~\bibnamefont {Kr\"ucken}}, \bibinfo {author}
  {\bibfnamefont {A.}~\bibnamefont {Kuchenkov}}, \bibinfo {author}
  {\bibfnamefont {K.~S.}\ \bibnamefont {Kumar}}, \bibinfo {author}
  {\bibfnamefont {Y.}~\bibnamefont {Lan}}, \bibinfo {author} {\bibfnamefont
  {A.}~\bibnamefont {Larson}}, \bibinfo {author} {\bibfnamefont {B.~G.}\
  \bibnamefont {Lenardo}}, \bibinfo {author} {\bibfnamefont {D.~S.}\
  \bibnamefont {Leonard}}, \bibinfo {author} {\bibfnamefont {G.~S.}\
  \bibnamefont {Li}}, \bibinfo {author} {\bibfnamefont {S.}~\bibnamefont {Li}},
  \bibinfo {author} {\bibfnamefont {Z.}~\bibnamefont {Li}}, \bibinfo {author}
  {\bibfnamefont {C.}~\bibnamefont {Licciardi}}, \bibinfo {author}
  {\bibfnamefont {Y.~H.}\ \bibnamefont {Lin}}, \bibinfo {author} {\bibfnamefont
  {R.}~\bibnamefont {MacLellan}}, \bibinfo {author} {\bibfnamefont
  {T.}~\bibnamefont {McElroy}}, \bibinfo {author} {\bibfnamefont
  {T.}~\bibnamefont {Michel}}, \bibinfo {author} {\bibfnamefont
  {B.}~\bibnamefont {Mong}}, \bibinfo {author} {\bibfnamefont {D.~C.}\
  \bibnamefont {Moore}}, \bibinfo {author} {\bibfnamefont {K.}~\bibnamefont
  {Murray}}, \bibinfo {author} {\bibfnamefont {O.}~\bibnamefont {Njoya}},
  \bibinfo {author} {\bibfnamefont {O.}~\bibnamefont {Nusair}}, \bibinfo
  {author} {\bibfnamefont {A.}~\bibnamefont {Odian}}, \bibinfo {author}
  {\bibfnamefont {I.}~\bibnamefont {Ostrovskiy}}, \bibinfo {author}
  {\bibfnamefont {A.}~\bibnamefont {Piepke}}, \bibinfo {author} {\bibfnamefont
  {A.}~\bibnamefont {Pocar}}, \bibinfo {author} {\bibfnamefont
  {F.}~\bibnamefont {Reti\`ere}}, \bibinfo {author} {\bibfnamefont {A.~L.}\
  \bibnamefont {Robinson}}, \bibinfo {author} {\bibfnamefont {P.~C.}\
  \bibnamefont {Rowson}}, \bibinfo {author} {\bibfnamefont {D.}~\bibnamefont
  {Ruddell}}, \bibinfo {author} {\bibfnamefont {J.}~\bibnamefont {Runge}},
  \bibinfo {author} {\bibfnamefont {S.}~\bibnamefont {Schmidt}}, \bibinfo
  {author} {\bibfnamefont {D.}~\bibnamefont {Sinclair}}, \bibinfo {author}
  {\bibfnamefont {A.~K.}\ \bibnamefont {Soma}}, \bibinfo {author}
  {\bibfnamefont {V.}~\bibnamefont {Stekhanov}}, \bibinfo {author}
  {\bibfnamefont {M.}~\bibnamefont {Tarka}}, \bibinfo {author} {\bibfnamefont
  {J.}~\bibnamefont {Todd}}, \bibinfo {author} {\bibfnamefont {T.}~\bibnamefont
  {Tolba}}, \bibinfo {author} {\bibfnamefont {T.~I.}\ \bibnamefont {Totev}},
  \bibinfo {author} {\bibfnamefont {B.}~\bibnamefont {Veenstra}}, \bibinfo
  {author} {\bibfnamefont {V.}~\bibnamefont {Veeraraghavan}}, \bibinfo {author}
  {\bibfnamefont {P.}~\bibnamefont {Vogel}}, \bibinfo {author} {\bibfnamefont
  {J.-L.}\ \bibnamefont {Vuilleumier}}, \bibinfo {author} {\bibfnamefont
  {M.}~\bibnamefont {Wagenpfeil}}, \bibinfo {author} {\bibfnamefont
  {J.}~\bibnamefont {Watkins}}, \bibinfo {author} {\bibfnamefont
  {M.}~\bibnamefont {Weber}}, \bibinfo {author} {\bibfnamefont {L.~J.}\
  \bibnamefont {Wen}}, \bibinfo {author} {\bibfnamefont {U.}~\bibnamefont
  {Wichoski}}, \bibinfo {author} {\bibfnamefont {G.}~\bibnamefont {Wrede}},
  \bibinfo {author} {\bibfnamefont {S.~X.}\ \bibnamefont {Wu}}, \bibinfo
  {author} {\bibfnamefont {Q.}~\bibnamefont {Xia}}, \bibinfo {author}
  {\bibfnamefont {D.~R.}\ \bibnamefont {Yahne}}, \bibinfo {author}
  {\bibfnamefont {L.}~\bibnamefont {Yang}}, \bibinfo {author} {\bibfnamefont
  {Y.-R.}\ \bibnamefont {Yen}}, \bibinfo {author} {\bibfnamefont {O.~Y.}\
  \bibnamefont {Zeldovich}}, \ and\ \bibinfo {author} {\bibfnamefont
  {T.}~\bibnamefont {Ziegler}} (\bibinfo {collaboration} {EXO-200
  Collaboration}),\ }\href {\doibase 10.1103/PhysRevLett.123.161802} {\bibfield
   {journal} {\bibinfo  {journal} {Phys. Rev. Lett.}\ }\textbf {\bibinfo
  {volume} {123}},\ \bibinfo {pages} {161802} (\bibinfo {year}
  {2019})}\BibitemShut {NoStop}%
\bibitem [{\citenamefont {Adams}\ \emph {et~al.}(2020)\citenamefont {Adams}
  \emph {et~al.}}]{CUORE:2019}%
  \BibitemOpen
  \bibfield  {author} {\bibinfo {author} {\bibfnamefont {D.~Q.}\ \bibnamefont
  {Adams}} \emph {et~al.} (\bibinfo {collaboration} {CUORE}),\ }\href {\doibase
  10.1103/PhysRevLett.124.122501} {\bibfield  {journal} {\bibinfo  {journal}
  {Phys. Rev. Lett.}\ }\textbf {\bibinfo {volume} {124}},\ \bibinfo {pages}
  {122501} (\bibinfo {year} {2020})},\ \Eprint
  {http://arxiv.org/abs/1912.10966} {arXiv:1912.10966 [nucl-ex]} \BibitemShut
  {NoStop}%
\bibitem [{\citenamefont {Agostini}\ \emph {et~al.}(2020)\citenamefont
  {Agostini} \emph {et~al.}}]{GERDA:2020}%
  \BibitemOpen
  \bibfield  {author} {\bibinfo {author} {\bibfnamefont {M.}~\bibnamefont
  {Agostini}} \emph {et~al.} (\bibinfo {collaboration} {GERDA}),\ }\href
  {\doibase 10.1103/PhysRevLett.125.252502} {\bibfield  {journal} {\bibinfo
  {journal} {Phys. Rev. Lett.}\ }\textbf {\bibinfo {volume} {125}},\ \bibinfo
  {pages} {252502} (\bibinfo {year} {2020})},\ \Eprint
  {http://arxiv.org/abs/2009.06079} {arXiv:2009.06079 [nucl-ex]} \BibitemShut
  {NoStop}%
\bibitem [{\citenamefont {Abe}\ \emph {et~al.}(2022)\citenamefont {Abe} \emph
  {et~al.}}]{KamLAND-Zen:2022}%
  \BibitemOpen
  \bibfield  {author} {\bibinfo {author} {\bibfnamefont {S.}~\bibnamefont
  {Abe}} \emph {et~al.} (\bibinfo {collaboration} {KamLAND-Zen}),\ }\href@noop
  {} {\  (\bibinfo {year} {2022})},\ \Eprint {http://arxiv.org/abs/2203.02139}
  {arXiv:2203.02139 [hep-ex]} \BibitemShut {NoStop}%
\bibitem [{\citenamefont {Xie}\ \emph {et~al.}(2021)\citenamefont {Xie},
  \citenamefont {Ni}, \citenamefont {Han},\ and\ \citenamefont
  {Wang}}]{Xie:2020}%
  \BibitemOpen
  \bibfield  {author} {\bibinfo {author} {\bibfnamefont {C.}~\bibnamefont
  {Xie}}, \bibinfo {author} {\bibfnamefont {K.}~\bibnamefont {Ni}}, \bibinfo
  {author} {\bibfnamefont {K.}~\bibnamefont {Han}}, \ and\ \bibinfo {author}
  {\bibfnamefont {S.}~\bibnamefont {Wang}},\ }\href {\doibase
  10.1007/s11433-020-1693-6} {\bibfield  {journal} {\bibinfo  {journal} {Sci.
  China Phys. Mech. Astron.}\ }\textbf {\bibinfo {volume} {64}},\ \bibinfo
  {pages} {261011} (\bibinfo {year} {2021})},\ \Eprint
  {http://arxiv.org/abs/2012.04552} {arXiv:2012.04552 [nucl-ex]} \BibitemShut
  {NoStop}%
\bibitem [{\citenamefont {Abgrall}\ \emph {et~al.}(2021)\citenamefont {Abgrall}
  \emph {et~al.}}]{LEGEND:2021}%
  \BibitemOpen
  \bibfield  {author} {\bibinfo {author} {\bibfnamefont {N.}~\bibnamefont
  {Abgrall}} \emph {et~al.} (\bibinfo {collaboration} {LEGEND}),\ }\href@noop
  {} {\  (\bibinfo {year} {2021})},\ \Eprint {http://arxiv.org/abs/2107.11462}
  {arXiv:2107.11462 [physics.ins-det]} \BibitemShut {NoStop}%
\bibitem [{\citenamefont {Armatol}\ \emph {et~al.}(2022)\citenamefont {Armatol}
  \emph {et~al.}}]{CUPID:2022}%
  \BibitemOpen
  \bibfield  {author} {\bibinfo {author} {\bibfnamefont {A.}~\bibnamefont
  {Armatol}} \emph {et~al.} (\bibinfo {collaboration} {CUPID}),\ }in\
  \href@noop {} {\emph {\bibinfo {booktitle} {{2022 Snowmass Summer Study}}}}\
  (\bibinfo {year} {2022})\ \Eprint {http://arxiv.org/abs/2203.08386}
  {arXiv:2203.08386 [nucl-ex]} \BibitemShut {NoStop}%
\bibitem [{\citenamefont {Cirigliano}\ \emph {et~al.}(2022)\citenamefont
  {Cirigliano} \emph {et~al.}}]{Cirigliano:2022}%
  \BibitemOpen
  \bibfield  {author} {\bibinfo {author} {\bibfnamefont {V.}~\bibnamefont
  {Cirigliano}} \emph {et~al.},\ }\href@noop {} {\  (\bibinfo {year} {2022})},\
  \Eprint {http://arxiv.org/abs/2203.12169} {arXiv:2203.12169 [hep-ph]}
  \BibitemShut {NoStop}%
\bibitem [{\citenamefont {Men\'endez}\ \emph {et~al.}(2009)\citenamefont
  {Men\'endez}, \citenamefont {Poves}, \citenamefont {Caurier},\ and\
  \citenamefont {Nowacki}}]{Menendez:2009}%
  \BibitemOpen
  \bibfield  {author} {\bibinfo {author} {\bibfnamefont {J.}~\bibnamefont
  {Men\'endez}}, \bibinfo {author} {\bibfnamefont {A.}~\bibnamefont {Poves}},
  \bibinfo {author} {\bibfnamefont {E.}~\bibnamefont {Caurier}}, \ and\
  \bibinfo {author} {\bibfnamefont {F.}~\bibnamefont {Nowacki}},\ }\href
  {\doibase 10.1016/j.nuclphysa.2008.12.005} {\bibfield  {journal} {\bibinfo
  {journal} {Nuclear Physics A}\ }\textbf {\bibinfo {volume} {818}},\ \bibinfo
  {pages} {139 } (\bibinfo {year} {2009})}\BibitemShut {NoStop}%
\bibitem [{\citenamefont {Rodr\'{\i}guez}\ and\ \citenamefont
  {Mart\'{\i}nez-Pinedo}(2010)}]{Rodriguez:2010}%
  \BibitemOpen
  \bibfield  {author} {\bibinfo {author} {\bibfnamefont {T.~R.}\ \bibnamefont
  {Rodr\'{\i}guez}}\ and\ \bibinfo {author} {\bibfnamefont {G.}~\bibnamefont
  {Mart\'{\i}nez-Pinedo}},\ }\href {\doibase 10.1103/PhysRevLett.105.252503}
  {\bibfield  {journal} {\bibinfo  {journal} {Phys. Rev. Lett.}\ }\textbf
  {\bibinfo {volume} {105}},\ \bibinfo {pages} {252503} (\bibinfo {year}
  {2010})}\BibitemShut {NoStop}%
\bibitem [{\citenamefont {Barea}\ \emph {et~al.}(2013)\citenamefont {Barea},
  \citenamefont {Kotila},\ and\ \citenamefont {Iachello}}]{Barea:2013}%
  \BibitemOpen
  \bibfield  {author} {\bibinfo {author} {\bibfnamefont {J.}~\bibnamefont
  {Barea}}, \bibinfo {author} {\bibfnamefont {J.}~\bibnamefont {Kotila}}, \
  and\ \bibinfo {author} {\bibfnamefont {F.}~\bibnamefont {Iachello}},\ }\href
  {\doibase 10.1103/PhysRevC.87.014315} {\bibfield  {journal} {\bibinfo
  {journal} {Phys. Rev. C}\ }\textbf {\bibinfo {volume} {87}},\ \bibinfo
  {pages} {014315} (\bibinfo {year} {2013})}\BibitemShut {NoStop}%
\bibitem [{\citenamefont {Mustonen}\ and\ \citenamefont
  {Engel}(2013)}]{Mustonen:2013}%
  \BibitemOpen
  \bibfield  {author} {\bibinfo {author} {\bibfnamefont {M.~T.}\ \bibnamefont
  {Mustonen}}\ and\ \bibinfo {author} {\bibfnamefont {J.}~\bibnamefont
  {Engel}},\ }\href {\doibase 10.1103/PhysRevC.87.064302} {\bibfield  {journal}
  {\bibinfo  {journal} {Phys. Rev. C}\ }\textbf {\bibinfo {volume} {87}},\
  \bibinfo {pages} {064302} (\bibinfo {year} {2013})}\BibitemShut {NoStop}%
\bibitem [{\citenamefont {Holt}\ and\ \citenamefont {Engel}(2013)}]{Holt:2013}%
  \BibitemOpen
  \bibfield  {author} {\bibinfo {author} {\bibfnamefont {J.~D.}\ \bibnamefont
  {Holt}}\ and\ \bibinfo {author} {\bibfnamefont {J.}~\bibnamefont {Engel}},\
  }\href {\doibase 10.1103/PhysRevC.87.064315} {\bibfield  {journal} {\bibinfo
  {journal} {Phys. Rev. C}\ }\textbf {\bibinfo {volume} {87}},\ \bibinfo
  {pages} {064315} (\bibinfo {year} {2013})}\BibitemShut {NoStop}%
\bibitem [{\citenamefont {Kwiatkowski}\ \emph {et~al.}(2014)\citenamefont
  {Kwiatkowski}, \citenamefont {Brunner}, \citenamefont {Holt}, \citenamefont
  {Chaudhuri}, \citenamefont {Chowdhury}, \citenamefont {Eibach}, \citenamefont
  {Engel}, \citenamefont {Gallant}, \citenamefont {Grossheim}, \citenamefont
  {Horoi}, \citenamefont {Lennarz}, \citenamefont {Macdonald}, \citenamefont
  {Pearson}, \citenamefont {Schultz}, \citenamefont {Simon}, \citenamefont
  {Senkov}, \citenamefont {Simon}, \citenamefont {Zuber},\ and\ \citenamefont
  {Dilling}}]{Kwiatkowski:2014}%
  \BibitemOpen
  \bibfield  {author} {\bibinfo {author} {\bibfnamefont {A.~A.}\ \bibnamefont
  {Kwiatkowski}}, \bibinfo {author} {\bibfnamefont {T.}~\bibnamefont
  {Brunner}}, \bibinfo {author} {\bibfnamefont {J.~D.}\ \bibnamefont {Holt}},
  \bibinfo {author} {\bibfnamefont {A.}~\bibnamefont {Chaudhuri}}, \bibinfo
  {author} {\bibfnamefont {U.}~\bibnamefont {Chowdhury}}, \bibinfo {author}
  {\bibfnamefont {M.}~\bibnamefont {Eibach}}, \bibinfo {author} {\bibfnamefont
  {J.}~\bibnamefont {Engel}}, \bibinfo {author} {\bibfnamefont {A.~T.}\
  \bibnamefont {Gallant}}, \bibinfo {author} {\bibfnamefont {A.}~\bibnamefont
  {Grossheim}}, \bibinfo {author} {\bibfnamefont {M.}~\bibnamefont {Horoi}},
  \bibinfo {author} {\bibfnamefont {A.}~\bibnamefont {Lennarz}}, \bibinfo
  {author} {\bibfnamefont {T.~D.}\ \bibnamefont {Macdonald}}, \bibinfo {author}
  {\bibfnamefont {M.~R.}\ \bibnamefont {Pearson}}, \bibinfo {author}
  {\bibfnamefont {B.~E.}\ \bibnamefont {Schultz}}, \bibinfo {author}
  {\bibfnamefont {M.~C.}\ \bibnamefont {Simon}}, \bibinfo {author}
  {\bibfnamefont {R.~A.}\ \bibnamefont {Senkov}}, \bibinfo {author}
  {\bibfnamefont {V.~V.}\ \bibnamefont {Simon}}, \bibinfo {author}
  {\bibfnamefont {K.}~\bibnamefont {Zuber}}, \ and\ \bibinfo {author}
  {\bibfnamefont {J.}~\bibnamefont {Dilling}},\ }\href {\doibase
  10.1103/PhysRevC.89.045502} {\bibfield  {journal} {\bibinfo  {journal} {Phys.
  Rev. C}\ }\textbf {\bibinfo {volume} {89}},\ \bibinfo {pages} {045502}
  (\bibinfo {year} {2014})}\BibitemShut {NoStop}%
\bibitem [{\citenamefont {Song}\ \emph {et~al.}(2014)\citenamefont {Song},
  \citenamefont {Yao}, \citenamefont {Ring},\ and\ \citenamefont
  {Meng}}]{Song:2014}%
  \BibitemOpen
  \bibfield  {author} {\bibinfo {author} {\bibfnamefont {L.~S.}\ \bibnamefont
  {Song}}, \bibinfo {author} {\bibfnamefont {J.~M.}\ \bibnamefont {Yao}},
  \bibinfo {author} {\bibfnamefont {P.}~\bibnamefont {Ring}}, \ and\ \bibinfo
  {author} {\bibfnamefont {J.}~\bibnamefont {Meng}},\ }\href {\doibase
  https://doi.org/10.1103/PhysRevC.90.054309} {\bibfield  {journal} {\bibinfo
  {journal} {Phys. Rev. C}\ }\textbf {\bibinfo {volume} {90}},\ \bibinfo
  {pages} {054309} (\bibinfo {year} {2014})}\BibitemShut {NoStop}%
\bibitem [{\citenamefont {Yao}\ \emph {et~al.}(2015)\citenamefont {Yao},
  \citenamefont {Song}, \citenamefont {Hagino}, \citenamefont {Ring},\ and\
  \citenamefont {Meng}}]{Yao:2015}%
  \BibitemOpen
  \bibfield  {author} {\bibinfo {author} {\bibfnamefont {J.~M.}\ \bibnamefont
  {Yao}}, \bibinfo {author} {\bibfnamefont {L.~S.}\ \bibnamefont {Song}},
  \bibinfo {author} {\bibfnamefont {K.}~\bibnamefont {Hagino}}, \bibinfo
  {author} {\bibfnamefont {P.}~\bibnamefont {Ring}}, \ and\ \bibinfo {author}
  {\bibfnamefont {J.}~\bibnamefont {Meng}},\ }\href {\doibase
  10.1103/PhysRevC.91.024316} {\bibfield  {journal} {\bibinfo  {journal} {Phys.
  Rev. C}\ }\textbf {\bibinfo {volume} {91}},\ \bibinfo {pages} {024316}
  (\bibinfo {year} {2015})}\BibitemShut {NoStop}%
\bibitem [{\citenamefont {Hyv\"arinen}\ and\ \citenamefont
  {Suhonen}(2015)}]{Hyvarinen:2015}%
  \BibitemOpen
  \bibfield  {author} {\bibinfo {author} {\bibfnamefont {J.}~\bibnamefont
  {Hyv\"arinen}}\ and\ \bibinfo {author} {\bibfnamefont {J.}~\bibnamefont
  {Suhonen}},\ }\href {\doibase 10.1103/PhysRevC.91.024613} {\bibfield
  {journal} {\bibinfo  {journal} {Phys. Rev. C}\ }\textbf {\bibinfo {volume}
  {91}},\ \bibinfo {pages} {024613} (\bibinfo {year} {2015})}\BibitemShut
  {NoStop}%
\bibitem [{\citenamefont {Horoi}\ and\ \citenamefont
  {Neacsu}(2016)}]{Horoi:2016}%
  \BibitemOpen
  \bibfield  {author} {\bibinfo {author} {\bibfnamefont {M.}~\bibnamefont
  {Horoi}}\ and\ \bibinfo {author} {\bibfnamefont {A.}~\bibnamefont {Neacsu}},\
  }\href {\doibase 10.1103/PhysRevC.93.024308} {\bibfield  {journal} {\bibinfo
  {journal} {Phys. Rev. C}\ }\textbf {\bibinfo {volume} {93}},\ \bibinfo
  {pages} {024308} (\bibinfo {year} {2016})}\BibitemShut {NoStop}%
\bibitem [{\citenamefont {Song}\ \emph {et~al.}(2017)\citenamefont {Song},
  \citenamefont {Yao}, \citenamefont {Ring},\ and\ \citenamefont
  {Meng}}]{Song:2017}%
  \BibitemOpen
  \bibfield  {author} {\bibinfo {author} {\bibfnamefont {L.~S.}\ \bibnamefont
  {Song}}, \bibinfo {author} {\bibfnamefont {J.~M.}\ \bibnamefont {Yao}},
  \bibinfo {author} {\bibfnamefont {P.}~\bibnamefont {Ring}}, \ and\ \bibinfo
  {author} {\bibfnamefont {J.}~\bibnamefont {Meng}},\ }\href {\doibase
  10.1103/PhysRevC.95.024305} {\bibfield  {journal} {\bibinfo  {journal} {Phys.
  Rev. C}\ }\textbf {\bibinfo {volume} {95}},\ \bibinfo {pages} {024305}
  (\bibinfo {year} {2017})}\BibitemShut {NoStop}%
\bibitem [{\citenamefont {Jiao}\ \emph {et~al.}(2017)\citenamefont {Jiao},
  \citenamefont {Engel},\ and\ \citenamefont {Holt}}]{Jiao:2017}%
  \BibitemOpen
  \bibfield  {author} {\bibinfo {author} {\bibfnamefont {C.~F.}\ \bibnamefont
  {Jiao}}, \bibinfo {author} {\bibfnamefont {J.}~\bibnamefont {Engel}}, \ and\
  \bibinfo {author} {\bibfnamefont {J.~D.}\ \bibnamefont {Holt}},\ }\href
  {\doibase 10.1103/PhysRevC.96.054310} {\bibfield  {journal} {\bibinfo
  {journal} {Phys. Rev. C}\ }\textbf {\bibinfo {volume} {96}},\ \bibinfo
  {pages} {054310} (\bibinfo {year} {2017})}\BibitemShut {NoStop}%
\bibitem [{\citenamefont {Yoshinaga}\ \emph {et~al.}(2018)\citenamefont
  {Yoshinaga}, \citenamefont {Yanase}, \citenamefont {Higashiyama},
  \citenamefont {Teruya},\ and\ \citenamefont {Taguchi}}]{Yoshinaga:2018}%
  \BibitemOpen
  \bibfield  {author} {\bibinfo {author} {\bibfnamefont {N.}~\bibnamefont
  {Yoshinaga}}, \bibinfo {author} {\bibfnamefont {K.}~\bibnamefont {Yanase}},
  \bibinfo {author} {\bibfnamefont {K.}~\bibnamefont {Higashiyama}}, \bibinfo
  {author} {\bibfnamefont {E.}~\bibnamefont {Teruya}}, \ and\ \bibinfo {author}
  {\bibfnamefont {D.}~\bibnamefont {Taguchi}},\ }\href
  {https://doi.org/10.1093/ptep/ptx174} {\bibfield  {journal} {\bibinfo
  {journal} {Prog. Theor. Exp. Phys.}\ }\textbf {\bibinfo {volume} {2018}},\
  \bibinfo {pages} {023D02} (\bibinfo {year} {2018})}\BibitemShut {NoStop}%
\bibitem [{\citenamefont {Fang}\ \emph {et~al.}(2018)\citenamefont {Fang},
  \citenamefont {Faessler},\ and\ \citenamefont {\ifmmode~\check{S}\else
  \v{S}\fi{}imkovic}}]{Fang:2018}%
  \BibitemOpen
  \bibfield  {author} {\bibinfo {author} {\bibfnamefont {D.-L.}\ \bibnamefont
  {Fang}}, \bibinfo {author} {\bibfnamefont {A.}~\bibnamefont {Faessler}}, \
  and\ \bibinfo {author} {\bibfnamefont {F.}~\bibnamefont
  {\ifmmode~\check{S}\else \v{S}\fi{}imkovic}},\ }\href {\doibase
  10.1103/PhysRevC.97.045503} {\bibfield  {journal} {\bibinfo  {journal} {Phys.
  Rev. C}\ }\textbf {\bibinfo {volume} {97}},\ \bibinfo {pages} {045503}
  (\bibinfo {year} {2018})}\BibitemShut {NoStop}%
\bibitem [{\citenamefont {Rath}\ \emph {et~al.}(2019)\citenamefont {Rath},
  \citenamefont {Chandra}, \citenamefont {Chaturvedi},\ and\ \citenamefont
  {Raina}}]{Rath:2019}%
  \BibitemOpen
  \bibfield  {author} {\bibinfo {author} {\bibfnamefont {P.~K.}\ \bibnamefont
  {Rath}}, \bibinfo {author} {\bibfnamefont {R.}~\bibnamefont {Chandra}},
  \bibinfo {author} {\bibfnamefont {K.}~\bibnamefont {Chaturvedi}}, \ and\
  \bibinfo {author} {\bibfnamefont {P.~K.}\ \bibnamefont {Raina}},\ }\href
  {\doibase 10.3389/fphy.2019.00064} {\bibfield  {journal} {\bibinfo  {journal}
  {Frontiers in Physics}\ }\textbf {\bibinfo {volume} {7}},\ \bibinfo {pages}
  {64} (\bibinfo {year} {2019})}\BibitemShut {NoStop}%
\bibitem [{\citenamefont {Terasaki}\ and\ \citenamefont
  {Iwata}(2019)}]{Terasaki:2019}%
  \BibitemOpen
  \bibfield  {author} {\bibinfo {author} {\bibfnamefont {J.}~\bibnamefont
  {Terasaki}}\ and\ \bibinfo {author} {\bibfnamefont {Y.}~\bibnamefont
  {Iwata}},\ }\href {\doibase 10.1103/PhysRevC.100.034325} {\bibfield
  {journal} {\bibinfo  {journal} {Phys. Rev. C}\ }\textbf {\bibinfo {volume}
  {100}},\ \bibinfo {pages} {034325} (\bibinfo {year} {2019})}\BibitemShut
  {NoStop}%
\bibitem [{\citenamefont {Coraggio}\ \emph {et~al.}(2020)\citenamefont
  {Coraggio}, \citenamefont {Gargano}, \citenamefont {Itaco}, \citenamefont
  {Mancino},\ and\ \citenamefont {Nowacki}}]{Coraggio:2020}%
  \BibitemOpen
  \bibfield  {author} {\bibinfo {author} {\bibfnamefont {L.}~\bibnamefont
  {Coraggio}}, \bibinfo {author} {\bibfnamefont {A.}~\bibnamefont {Gargano}},
  \bibinfo {author} {\bibfnamefont {N.}~\bibnamefont {Itaco}}, \bibinfo
  {author} {\bibfnamefont {R.}~\bibnamefont {Mancino}}, \ and\ \bibinfo
  {author} {\bibfnamefont {F.}~\bibnamefont {Nowacki}},\ }\href {\doibase
  10.1103/PhysRevC.101.044315} {\bibfield  {journal} {\bibinfo  {journal}
  {Phys. Rev. C}\ }\textbf {\bibinfo {volume} {101}},\ \bibinfo {pages}
  {044315} (\bibinfo {year} {2020})}\BibitemShut {NoStop}%
\bibitem [{\citenamefont {Deppisch}\ \emph {et~al.}(2020)\citenamefont
  {Deppisch}, \citenamefont {Graf}, \citenamefont {Iachello},\ and\
  \citenamefont {Kotila}}]{Deppisch:2020ztt}%
  \BibitemOpen
  \bibfield  {author} {\bibinfo {author} {\bibfnamefont {F.~F.}\ \bibnamefont
  {Deppisch}}, \bibinfo {author} {\bibfnamefont {L.}~\bibnamefont {Graf}},
  \bibinfo {author} {\bibfnamefont {F.}~\bibnamefont {Iachello}}, \ and\
  \bibinfo {author} {\bibfnamefont {J.}~\bibnamefont {Kotila}},\ }\href
  {\doibase 10.1103/PhysRevD.102.095016} {\bibfield  {journal} {\bibinfo
  {journal} {Phys. Rev. D}\ }\textbf {\bibinfo {volume} {102}},\ \bibinfo
  {pages} {095016} (\bibinfo {year} {2020})}\BibitemShut {NoStop}%
\bibitem [{\citenamefont {Wang}\ \emph {et~al.}(2021)\citenamefont {Wang},
  \citenamefont {Zhao},\ and\ \citenamefont {Meng}}]{Wang:2021}%
  \BibitemOpen
  \bibfield  {author} {\bibinfo {author} {\bibfnamefont {Y.~K.}\ \bibnamefont
  {Wang}}, \bibinfo {author} {\bibfnamefont {P.~W.}\ \bibnamefont {Zhao}}, \
  and\ \bibinfo {author} {\bibfnamefont {J.}~\bibnamefont {Meng}},\ }\href
  {\doibase 10.1103/PhysRevC.104.014320} {\bibfield  {journal} {\bibinfo
  {journal} {Phys. Rev. C}\ }\textbf {\bibinfo {volume} {104}},\ \bibinfo
  {pages} {014320} (\bibinfo {year} {2021})}\BibitemShut {NoStop}%
\bibitem [{\citenamefont {Coraggio}\ \emph {et~al.}(2022)\citenamefont
  {Coraggio}, \citenamefont {Itaco}, \citenamefont {De~Gregorio}, \citenamefont
  {Gargano}, \citenamefont {Mancino},\ and\ \citenamefont
  {Nowacki}}]{Coraggio:2022}%
  \BibitemOpen
  \bibfield  {author} {\bibinfo {author} {\bibfnamefont {L.}~\bibnamefont
  {Coraggio}}, \bibinfo {author} {\bibfnamefont {N.}~\bibnamefont {Itaco}},
  \bibinfo {author} {\bibfnamefont {G.}~\bibnamefont {De~Gregorio}}, \bibinfo
  {author} {\bibfnamefont {A.}~\bibnamefont {Gargano}}, \bibinfo {author}
  {\bibfnamefont {R.}~\bibnamefont {Mancino}}, \ and\ \bibinfo {author}
  {\bibfnamefont {F.}~\bibnamefont {Nowacki}},\ }\href@noop {} {\  (\bibinfo
  {year} {2022})},\ \Eprint {http://arxiv.org/abs/2203.01013} {arXiv:2203.01013
  [nucl-th]} \BibitemShut {NoStop}%
\bibitem [{\citenamefont {Men\'endez}\ \emph {et~al.}(2014)\citenamefont
  {Men\'endez}, \citenamefont {Rodr\'{\i}guez}, \citenamefont
  {Mart\'{\i}nez-Pinedo},\ and\ \citenamefont {Poves}}]{Menendez:2014}%
  \BibitemOpen
  \bibfield  {author} {\bibinfo {author} {\bibfnamefont {J.}~\bibnamefont
  {Men\'endez}}, \bibinfo {author} {\bibfnamefont {T.~R.}\ \bibnamefont
  {Rodr\'{\i}guez}}, \bibinfo {author} {\bibfnamefont {G.}~\bibnamefont
  {Mart\'{\i}nez-Pinedo}}, \ and\ \bibinfo {author} {\bibfnamefont
  {A.}~\bibnamefont {Poves}},\ }\href {\doibase 10.1103/PhysRevC.90.024311}
  {\bibfield  {journal} {\bibinfo  {journal} {Phys. Rev. C}\ }\textbf {\bibinfo
  {volume} {90}},\ \bibinfo {pages} {024311} (\bibinfo {year}
  {2014})}\BibitemShut {NoStop}%
\bibitem [{\citenamefont {Men\'endez}\ \emph {et~al.}(2016)\citenamefont
  {Men\'endez}, \citenamefont {Hinohara}, \citenamefont {Engel}, \citenamefont
  {Mart\'{\i}nez-Pinedo},\ and\ \citenamefont
  {Rodr\'{\i}guez}}]{Menendez:2016}%
  \BibitemOpen
  \bibfield  {author} {\bibinfo {author} {\bibfnamefont {J.}~\bibnamefont
  {Men\'endez}}, \bibinfo {author} {\bibfnamefont {N.}~\bibnamefont
  {Hinohara}}, \bibinfo {author} {\bibfnamefont {J.}~\bibnamefont {Engel}},
  \bibinfo {author} {\bibfnamefont {G.}~\bibnamefont {Mart\'{\i}nez-Pinedo}}, \
  and\ \bibinfo {author} {\bibfnamefont {T.~R.}\ \bibnamefont
  {Rodr\'{\i}guez}},\ }\href {\doibase 10.1103/PhysRevC.93.014305} {\bibfield
  {journal} {\bibinfo  {journal} {Phys. Rev. C}\ }\textbf {\bibinfo {volume}
  {93}},\ \bibinfo {pages} {014305} (\bibinfo {year} {2016})}\BibitemShut
  {NoStop}%
\bibitem [{\citenamefont {Geesaman}\ \emph {et~al.}(2015)\citenamefont
  {Geesaman}, \citenamefont {Cirigliano}, \citenamefont {Deshpande},
  \citenamefont {Fahey}, \citenamefont {Hardy}, \citenamefont {Heeger},
  \citenamefont {Hobart}, \citenamefont {Lapi}, \citenamefont {Nagle},
  \citenamefont {Nunes}, \citenamefont {Ormand}, \citenamefont {Piekarewicz},
  \citenamefont {Rossi}, \citenamefont {Schukraft}, \citenamefont {Scholberg},
  \citenamefont {Shepherd}, \citenamefont {Venugopalan}, \citenamefont
  {Wiescher},\ and\ \citenamefont {Wilkerson}}]{LongRangePlan2015}%
  \BibitemOpen
  \bibfield  {author} {\bibinfo {author} {\bibfnamefont {D.}~\bibnamefont
  {Geesaman}}, \bibinfo {author} {\bibfnamefont {V.}~\bibnamefont
  {Cirigliano}}, \bibinfo {author} {\bibfnamefont {A.}~\bibnamefont
  {Deshpande}}, \bibinfo {author} {\bibfnamefont {F.}~\bibnamefont {Fahey}},
  \bibinfo {author} {\bibfnamefont {J.}~\bibnamefont {Hardy}}, \bibinfo
  {author} {\bibfnamefont {K.}~\bibnamefont {Heeger}}, \bibinfo {author}
  {\bibfnamefont {D.}~\bibnamefont {Hobart}}, \bibinfo {author} {\bibfnamefont
  {S.}~\bibnamefont {Lapi}}, \bibinfo {author} {\bibfnamefont {J.}~\bibnamefont
  {Nagle}}, \bibinfo {author} {\bibfnamefont {F.}~\bibnamefont {Nunes}},
  \bibinfo {author} {\bibfnamefont {E.}~\bibnamefont {Ormand}}, \bibinfo
  {author} {\bibfnamefont {J.}~\bibnamefont {Piekarewicz}}, \bibinfo {author}
  {\bibfnamefont {P.}~\bibnamefont {Rossi}}, \bibinfo {author} {\bibfnamefont
  {J.}~\bibnamefont {Schukraft}}, \bibinfo {author} {\bibfnamefont
  {K.}~\bibnamefont {Scholberg}}, \bibinfo {author} {\bibfnamefont
  {M.}~\bibnamefont {Shepherd}}, \bibinfo {author} {\bibfnamefont
  {R.}~\bibnamefont {Venugopalan}}, \bibinfo {author} {\bibfnamefont
  {M.}~\bibnamefont {Wiescher}}, \ and\ \bibinfo {author} {\bibfnamefont
  {J.}~\bibnamefont {Wilkerson}},\ }\href {https://science.osti.gov/np/nsac}
  {\emph {\bibinfo {title} {{Reaching for the horizon: 2015 long range plan for
  nuclear science}}}},\ \bibinfo {type} {Tech. Rep.}\ (\bibinfo  {institution}
  {{U.S. Department of Energy}},\ \bibinfo {year} {2015})\BibitemShut {NoStop}%
\bibitem [{\citenamefont {Engel}\ and\ \citenamefont
  {Men{\'{e}}ndez}(2017)}]{Engel:2017}%
  \BibitemOpen
  \bibfield  {author} {\bibinfo {author} {\bibfnamefont {J.}~\bibnamefont
  {Engel}}\ and\ \bibinfo {author} {\bibfnamefont {J.}~\bibnamefont
  {Men{\'{e}}ndez}},\ }\href {\doibase 10.1088/1361-6633/aa5bc5} {\bibfield
  {journal} {\bibinfo  {journal} {Rep. Prog. Phys.}\ }\textbf {\bibinfo
  {volume} {80}},\ \bibinfo {pages} {046301} (\bibinfo {year}
  {2017})}\BibitemShut {NoStop}%
\bibitem [{\citenamefont {Yao}\ \emph {et~al.}(2022)\citenamefont {Yao},
  \citenamefont {Meng}, \citenamefont {Niu},\ and\ \citenamefont
  {Ring}}]{Yao:2021Review}%
  \BibitemOpen
  \bibfield  {author} {\bibinfo {author} {\bibfnamefont {J.~M.}\ \bibnamefont
  {Yao}}, \bibinfo {author} {\bibfnamefont {J.}~\bibnamefont {Meng}}, \bibinfo
  {author} {\bibfnamefont {Y.~F.}\ \bibnamefont {Niu}}, \ and\ \bibinfo
  {author} {\bibfnamefont {P.}~\bibnamefont {Ring}},\ }\href {\doibase
  https://doi.org/10.1016/j.ppnp.2022.103965} {\bibfield  {journal} {\bibinfo
  {journal} {Prog. Part. Nucl. Phys.}\ }\textbf {\bibinfo {volume} {126}},\
  \bibinfo {pages} {103965} (\bibinfo {year} {2022})}\BibitemShut {NoStop}%
\bibitem [{\citenamefont {Agostini}\ \emph {et~al.}(2022)\citenamefont
  {Agostini}, \citenamefont {Benato}, \citenamefont {Detwiler}, \citenamefont
  {Men\'endez},\ and\ \citenamefont {Vissani}}]{Agostini:2022RMP}%
  \BibitemOpen
  \bibfield  {author} {\bibinfo {author} {\bibfnamefont {M.}~\bibnamefont
  {Agostini}}, \bibinfo {author} {\bibfnamefont {G.}~\bibnamefont {Benato}},
  \bibinfo {author} {\bibfnamefont {J.~A.}\ \bibnamefont {Detwiler}}, \bibinfo
  {author} {\bibfnamefont {J.}~\bibnamefont {Men\'endez}}, \ and\ \bibinfo
  {author} {\bibfnamefont {F.}~\bibnamefont {Vissani}},\ }\href@noop {} {\
  (\bibinfo {year} {2022})},\ \Eprint {http://arxiv.org/abs/2202.01787}
  {arXiv:2202.01787 [hep-ex]} \BibitemShut {NoStop}%
\bibitem [{\citenamefont {Hergert}(2020)}]{Hergert:2020}%
  \BibitemOpen
  \bibfield  {author} {\bibinfo {author} {\bibfnamefont {H.}~\bibnamefont
  {Hergert}},\ }\href {\doibase 10.3389/fphy.2020.00379} {\bibfield  {journal}
  {\bibinfo  {journal} {Front. in Phys.}\ }\textbf {\bibinfo {volume} {8}},\
  \bibinfo {pages} {379} (\bibinfo {year} {2020})},\ \Eprint
  {http://arxiv.org/abs/2008.05061} {arXiv:2008.05061 [nucl-th]} \BibitemShut
  {NoStop}%
\bibitem [{\citenamefont {Pastore}\ \emph {et~al.}(2018)\citenamefont
  {Pastore}, \citenamefont {Carlson}, \citenamefont {Cirigliano}, \citenamefont
  {Dekens}, \citenamefont {Mereghetti},\ and\ \citenamefont
  {Wiringa}}]{Pastore:2018}%
  \BibitemOpen
  \bibfield  {author} {\bibinfo {author} {\bibfnamefont {S.}~\bibnamefont
  {Pastore}}, \bibinfo {author} {\bibfnamefont {J.}~\bibnamefont {Carlson}},
  \bibinfo {author} {\bibfnamefont {V.}~\bibnamefont {Cirigliano}}, \bibinfo
  {author} {\bibfnamefont {W.}~\bibnamefont {Dekens}}, \bibinfo {author}
  {\bibfnamefont {E.}~\bibnamefont {Mereghetti}}, \ and\ \bibinfo {author}
  {\bibfnamefont {R.~B.}\ \bibnamefont {Wiringa}},\ }\href {\doibase
  10.1103/PhysRevC.97.014606} {\bibfield  {journal} {\bibinfo  {journal} {Phys.
  Rev. C}\ }\textbf {\bibinfo {volume} {97}},\ \bibinfo {pages} {014606}
  (\bibinfo {year} {2018})}\BibitemShut {NoStop}%
\bibitem [{\citenamefont {Cirigliano}\ \emph {et~al.}(2019)\citenamefont
  {Cirigliano}, \citenamefont {Dekens}, \citenamefont {de~Vries}, \citenamefont
  {Graesser}, \citenamefont {Mereghetti}, \citenamefont {Pastore},
  \citenamefont {Piarulli}, \citenamefont {van Kolck},\ and\ \citenamefont
  {Wiringa}}]{Cirigliano:2019PRC}%
  \BibitemOpen
  \bibfield  {author} {\bibinfo {author} {\bibfnamefont {V.}~\bibnamefont
  {Cirigliano}}, \bibinfo {author} {\bibfnamefont {W.}~\bibnamefont {Dekens}},
  \bibinfo {author} {\bibfnamefont {J.}~\bibnamefont {de~Vries}}, \bibinfo
  {author} {\bibfnamefont {M.~L.}\ \bibnamefont {Graesser}}, \bibinfo {author}
  {\bibfnamefont {E.}~\bibnamefont {Mereghetti}}, \bibinfo {author}
  {\bibfnamefont {S.}~\bibnamefont {Pastore}}, \bibinfo {author} {\bibfnamefont
  {M.}~\bibnamefont {Piarulli}}, \bibinfo {author} {\bibfnamefont
  {U.}~\bibnamefont {van Kolck}}, \ and\ \bibinfo {author} {\bibfnamefont
  {R.~B.}\ \bibnamefont {Wiringa}},\ }\href {\doibase
  10.1103/PhysRevC.100.055504} {\bibfield  {journal} {\bibinfo  {journal}
  {Phys. Rev. C}\ }\textbf {\bibinfo {volume} {100}},\ \bibinfo {pages}
  {055504} (\bibinfo {year} {2019})}\BibitemShut {NoStop}%
\bibitem [{\citenamefont {Basili}\ \emph {et~al.}(2020)\citenamefont {Basili},
  \citenamefont {Yao}, \citenamefont {Engel}, \citenamefont {Hergert},
  \citenamefont {Lockner}, \citenamefont {Maris},\ and\ \citenamefont
  {Vary}}]{Basili2020}%
  \BibitemOpen
  \bibfield  {author} {\bibinfo {author} {\bibfnamefont {R.~A.~M.}\
  \bibnamefont {Basili}}, \bibinfo {author} {\bibfnamefont {J.~M.}\
  \bibnamefont {Yao}}, \bibinfo {author} {\bibfnamefont {J.}~\bibnamefont
  {Engel}}, \bibinfo {author} {\bibfnamefont {H.}~\bibnamefont {Hergert}},
  \bibinfo {author} {\bibfnamefont {M.}~\bibnamefont {Lockner}}, \bibinfo
  {author} {\bibfnamefont {P.}~\bibnamefont {Maris}}, \ and\ \bibinfo {author}
  {\bibfnamefont {J.~P.}\ \bibnamefont {Vary}},\ }\href {\doibase
  10.1103/PhysRevC.102.014302} {\bibfield  {journal} {\bibinfo  {journal}
  {Phys. Rev. C}\ }\textbf {\bibinfo {volume} {102}},\ \bibinfo {pages}
  {014302} (\bibinfo {year} {2020})}\BibitemShut {NoStop}%
\bibitem [{\citenamefont {Yao}\ \emph {et~al.}(2021)\citenamefont {Yao},
  \citenamefont {Belley}, \citenamefont {Wirth}, \citenamefont {Miyagi},
  \citenamefont {Payne}, \citenamefont {Stroberg}, \citenamefont {Hergert},\
  and\ \citenamefont {Holt}}]{Yao:2021PRC}%
  \BibitemOpen
  \bibfield  {author} {\bibinfo {author} {\bibfnamefont {J.~M.}\ \bibnamefont
  {Yao}}, \bibinfo {author} {\bibfnamefont {A.}~\bibnamefont {Belley}},
  \bibinfo {author} {\bibfnamefont {R.}~\bibnamefont {Wirth}}, \bibinfo
  {author} {\bibfnamefont {T.}~\bibnamefont {Miyagi}}, \bibinfo {author}
  {\bibfnamefont {C.~G.}\ \bibnamefont {Payne}}, \bibinfo {author}
  {\bibfnamefont {S.~R.}\ \bibnamefont {Stroberg}}, \bibinfo {author}
  {\bibfnamefont {H.}~\bibnamefont {Hergert}}, \ and\ \bibinfo {author}
  {\bibfnamefont {J.~D.}\ \bibnamefont {Holt}},\ }\href {\doibase
  10.1103/PhysRevC.103.014315} {\bibfield  {journal} {\bibinfo  {journal}
  {Phys. Rev. C}\ }\textbf {\bibinfo {volume} {103}},\ \bibinfo {pages}
  {014315} (\bibinfo {year} {2021})}\BibitemShut {NoStop}%
\bibitem [{\citenamefont {Yao}\ \emph {et~al.}(2020)\citenamefont {Yao},
  \citenamefont {Bally}, \citenamefont {Engel}, \citenamefont {Wirth},
  \citenamefont {Rodr\'{\i}guez},\ and\ \citenamefont {Hergert}}]{Yao:2020PRL}%
  \BibitemOpen
  \bibfield  {author} {\bibinfo {author} {\bibfnamefont {J.~M.}\ \bibnamefont
  {Yao}}, \bibinfo {author} {\bibfnamefont {B.}~\bibnamefont {Bally}}, \bibinfo
  {author} {\bibfnamefont {J.}~\bibnamefont {Engel}}, \bibinfo {author}
  {\bibfnamefont {R.}~\bibnamefont {Wirth}}, \bibinfo {author} {\bibfnamefont
  {T.~R.}\ \bibnamefont {Rodr\'{\i}guez}}, \ and\ \bibinfo {author}
  {\bibfnamefont {H.}~\bibnamefont {Hergert}},\ }\href {\doibase
  10.1103/PhysRevLett.124.232501} {\bibfield  {journal} {\bibinfo  {journal}
  {Phys. Rev. Lett.}\ }\textbf {\bibinfo {volume} {124}},\ \bibinfo {pages}
  {232501} (\bibinfo {year} {2020})}\BibitemShut {NoStop}%
\bibitem [{\citenamefont {Novario}\ \emph {et~al.}(2021)\citenamefont
  {Novario}, \citenamefont {Gysbers}, \citenamefont {Engel}, \citenamefont
  {Hagen}, \citenamefont {Jansen}, \citenamefont {Morris}, \citenamefont
  {Navr\'atil}, \citenamefont {Papenbrock},\ and\ \citenamefont
  {Quaglioni}}]{Novario:2021PRL}%
  \BibitemOpen
  \bibfield  {author} {\bibinfo {author} {\bibfnamefont {S.}~\bibnamefont
  {Novario}}, \bibinfo {author} {\bibfnamefont {P.}~\bibnamefont {Gysbers}},
  \bibinfo {author} {\bibfnamefont {J.}~\bibnamefont {Engel}}, \bibinfo
  {author} {\bibfnamefont {G.}~\bibnamefont {Hagen}}, \bibinfo {author}
  {\bibfnamefont {G.~R.}\ \bibnamefont {Jansen}}, \bibinfo {author}
  {\bibfnamefont {T.~D.}\ \bibnamefont {Morris}}, \bibinfo {author}
  {\bibfnamefont {P.}~\bibnamefont {Navr\'atil}}, \bibinfo {author}
  {\bibfnamefont {T.}~\bibnamefont {Papenbrock}}, \ and\ \bibinfo {author}
  {\bibfnamefont {S.}~\bibnamefont {Quaglioni}},\ }\href {\doibase
  10.1103/PhysRevLett.126.182502} {\bibfield  {journal} {\bibinfo  {journal}
  {Phys. Rev. Lett.}\ }\textbf {\bibinfo {volume} {126}},\ \bibinfo {pages}
  {182502} (\bibinfo {year} {2021})},\ \Eprint
  {http://arxiv.org/abs/2008.09696} {arXiv:2008.09696 [nucl-th]} \BibitemShut
  {NoStop}%
\bibitem [{\citenamefont {Belley}\ \emph {et~al.}(2021)\citenamefont {Belley},
  \citenamefont {Payne}, \citenamefont {Stroberg}, \citenamefont {Miyagi},\
  and\ \citenamefont {Holt}}]{Belley2021PRL}%
  \BibitemOpen
  \bibfield  {author} {\bibinfo {author} {\bibfnamefont {A.}~\bibnamefont
  {Belley}}, \bibinfo {author} {\bibfnamefont {C.~G.}\ \bibnamefont {Payne}},
  \bibinfo {author} {\bibfnamefont {S.~R.}\ \bibnamefont {Stroberg}}, \bibinfo
  {author} {\bibfnamefont {T.}~\bibnamefont {Miyagi}}, \ and\ \bibinfo {author}
  {\bibfnamefont {J.~D.}\ \bibnamefont {Holt}},\ }\href {\doibase
  10.1103/PhysRevLett.126.042502} {\bibfield  {journal} {\bibinfo  {journal}
  {Phys. Rev. Lett.}\ }\textbf {\bibinfo {volume} {126}},\ \bibinfo {pages}
  {042502} (\bibinfo {year} {2021})},\ \Eprint
  {http://arxiv.org/abs/2008.06588} {arXiv:2008.06588 [nucl-th]} \BibitemShut
  {NoStop}%
\bibitem [{\citenamefont {Shimizu}\ \emph {et~al.}(2018)\citenamefont
  {Shimizu}, \citenamefont {Men\'endez},\ and\ \citenamefont
  {Yako}}]{Shimizu:2018PRL}%
  \BibitemOpen
  \bibfield  {author} {\bibinfo {author} {\bibfnamefont {N.}~\bibnamefont
  {Shimizu}}, \bibinfo {author} {\bibfnamefont {J.}~\bibnamefont {Men\'endez}},
  \ and\ \bibinfo {author} {\bibfnamefont {K.}~\bibnamefont {Yako}},\ }\href
  {\doibase 10.1103/PhysRevLett.120.142502} {\bibfield  {journal} {\bibinfo
  {journal} {Phys. Rev. Lett.}\ }\textbf {\bibinfo {volume} {120}},\ \bibinfo
  {pages} {142502} (\bibinfo {year} {2018})}\BibitemShut {NoStop}%
\bibitem [{\citenamefont {Vogel}\ \emph {et~al.}(1988)\citenamefont {Vogel},
  \citenamefont {Ericson},\ and\ \citenamefont {Vergados}}]{Vogel:1988}%
  \BibitemOpen
  \bibfield  {author} {\bibinfo {author} {\bibfnamefont {P.}~\bibnamefont
  {Vogel}}, \bibinfo {author} {\bibfnamefont {M.}~\bibnamefont {Ericson}}, \
  and\ \bibinfo {author} {\bibfnamefont {J.}~\bibnamefont {Vergados}},\ }\href
  {\doibase https://doi.org/10.1016/0370-2693(88)91313-5} {\bibfield  {journal}
  {\bibinfo  {journal} {Phys. Lett. B}\ }\textbf {\bibinfo {volume} {212}},\
  \bibinfo {pages} {259} (\bibinfo {year} {1988})}\BibitemShut {NoStop}%
\bibitem [{\citenamefont {Auerbach}\ \emph {et~al.}(1989)\citenamefont
  {Auerbach}, \citenamefont {Zamick},\ and\ \citenamefont
  {Zheng}}]{Auerbach:1989}%
  \BibitemOpen
  \bibfield  {author} {\bibinfo {author} {\bibfnamefont {N.}~\bibnamefont
  {Auerbach}}, \bibinfo {author} {\bibfnamefont {L.}~\bibnamefont {Zamick}}, \
  and\ \bibinfo {author} {\bibfnamefont {D.~C.}\ \bibnamefont {Zheng}},\ }\href
  {\doibase https://doi.org/10.1016/0003-4916(89)90117-6} {\bibfield  {journal}
  {\bibinfo  {journal} {Ann. Phys.}\ }\textbf {\bibinfo {volume} {192}},\
  \bibinfo {pages} {77} (\bibinfo {year} {1989})}\BibitemShut {NoStop}%
\bibitem [{\citenamefont {Zheng}\ \emph {et~al.}(1989)\citenamefont {Zheng},
  \citenamefont {Zamick},\ and\ \citenamefont {Auerbach}}]{Zheng:1989}%
  \BibitemOpen
  \bibfield  {author} {\bibinfo {author} {\bibfnamefont {D.~C.}\ \bibnamefont
  {Zheng}}, \bibinfo {author} {\bibfnamefont {L.}~\bibnamefont {Zamick}}, \
  and\ \bibinfo {author} {\bibfnamefont {N.}~\bibnamefont {Auerbach}},\ }\href
  {\doibase 10.1103/PhysRevC.40.936} {\bibfield  {journal} {\bibinfo  {journal}
  {Phys. Rev. C}\ }\textbf {\bibinfo {volume} {40}},\ \bibinfo {pages} {936}
  (\bibinfo {year} {1989})}\BibitemShut {NoStop}%
\bibitem [{\citenamefont {Zheng}\ \emph {et~al.}(1990)\citenamefont {Zheng},
  \citenamefont {Zamick},\ and\ \citenamefont {Auerbach}}]{Zheng:1990}%
  \BibitemOpen
  \bibfield  {author} {\bibinfo {author} {\bibfnamefont {D.-C.}\ \bibnamefont
  {Zheng}}, \bibinfo {author} {\bibfnamefont {L.}~\bibnamefont {Zamick}}, \
  and\ \bibinfo {author} {\bibfnamefont {N.}~\bibnamefont {Auerbach}},\ }\href
  {\doibase https://doi.org/10.1016/0003-4916(90)90215-A} {\bibfield  {journal}
  {\bibinfo  {journal} {Annals of Physics}\ }\textbf {\bibinfo {volume}
  {197}},\ \bibinfo {pages} {343} (\bibinfo {year} {1990})}\BibitemShut
  {NoStop}%
\bibitem [{\citenamefont {Men\'endez}(2018)}]{Menendez:2018JPS}%
  \BibitemOpen
  \bibfield  {author} {\bibinfo {author} {\bibfnamefont {J.}~\bibnamefont
  {Men\'endez}},\ }\href {\doibase 10.7566/JPSCP.23.012036} {\bibfield
  {journal} {\bibinfo  {journal} {JPS Conf. Proc.}\ }\textbf {\bibinfo {volume}
  {23}},\ \bibinfo {pages} {012036} (\bibinfo {year} {2018})},\ \Eprint
  {http://arxiv.org/abs/1804.02102} {arXiv:1804.02102 [nucl-th]} \BibitemShut
  {NoStop}%
\bibitem [{\citenamefont {Men\'endez}\ \emph {et~al.}(2018)\citenamefont
  {Men\'endez}, \citenamefont {Shimizu},\ and\ \citenamefont
  {Yako}}]{Menendez:2017JPCS}%
  \BibitemOpen
  \bibfield  {author} {\bibinfo {author} {\bibfnamefont {J.}~\bibnamefont
  {Men\'endez}}, \bibinfo {author} {\bibfnamefont {N.}~\bibnamefont {Shimizu}},
  \ and\ \bibinfo {author} {\bibfnamefont {K.}~\bibnamefont {Yako}},\ }\href
  {\doibase 10.1088/1742-6596/1056/1/012037} {\bibfield  {journal} {\bibinfo
  {journal} {J. Phys. Conf. Ser.}\ }\textbf {\bibinfo {volume} {1056}},\
  \bibinfo {pages} {012037} (\bibinfo {year} {2018})},\ \Eprint
  {http://arxiv.org/abs/1712.08691} {arXiv:1712.08691 [nucl-th]} \BibitemShut
  {NoStop}%
\bibitem [{\citenamefont {Takaki}\ \emph {et~al.}(2015)\citenamefont {Takaki}
  \emph {et~al.}}]{Takaki:2015}%
  \BibitemOpen
  \bibfield  {author} {\bibinfo {author} {\bibfnamefont {M.}~\bibnamefont
  {Takaki}} \emph {et~al.},\ }\href {\doibase 10.7566/JPSCP.6.020038}
  {\bibfield  {journal} {\bibinfo  {journal} {JPS Conf. Proc.}\ }\textbf
  {\bibinfo {volume} {6}},\ \bibinfo {pages} {020038} (\bibinfo {year}
  {2015})}\BibitemShut {NoStop}%
\bibitem [{\citenamefont {Takahisa}\ \emph {et~al.}(2017)\citenamefont
  {Takahisa}, \citenamefont {Ejiri}, \citenamefont {Akimune}, \citenamefont
  {Fujita}, \citenamefont {Matumiya}, \citenamefont {Ohta}, \citenamefont
  {Shima}, \citenamefont {Tanaka},\ and\ \citenamefont
  {Yosoi}}]{Takahisa:2017}%
  \BibitemOpen
  \bibfield  {author} {\bibinfo {author} {\bibfnamefont {K.}~\bibnamefont
  {Takahisa}}, \bibinfo {author} {\bibfnamefont {H.}~\bibnamefont {Ejiri}},
  \bibinfo {author} {\bibfnamefont {H.}~\bibnamefont {Akimune}}, \bibinfo
  {author} {\bibfnamefont {H.}~\bibnamefont {Fujita}}, \bibinfo {author}
  {\bibfnamefont {R.}~\bibnamefont {Matumiya}}, \bibinfo {author}
  {\bibfnamefont {T.}~\bibnamefont {Ohta}}, \bibinfo {author} {\bibfnamefont
  {T.}~\bibnamefont {Shima}}, \bibinfo {author} {\bibfnamefont
  {M.}~\bibnamefont {Tanaka}}, \ and\ \bibinfo {author} {\bibfnamefont
  {M.}~\bibnamefont {Yosoi}},\ }\href@noop {} {\  (\bibinfo {year} {2017})},\
  \Eprint {http://arxiv.org/abs/1703.08264} {arXiv:1703.08264 [nucl-ex]}
  \BibitemShut {NoStop}%
\bibitem [{\citenamefont {Cappuzzello}\ \emph {et~al.}(2018)\citenamefont
  {Cappuzzello} \emph {et~al.}}]{Cappuzzello:2018}%
  \BibitemOpen
  \bibfield  {author} {\bibinfo {author} {\bibfnamefont {F.}~\bibnamefont
  {Cappuzzello}} \emph {et~al.},\ }\href {\doibase 10.1140/epja/i2018-12509-3}
  {\bibfield  {journal} {\bibinfo  {journal} {Eur. Phys. J. A}\ }\textbf
  {\bibinfo {volume} {54}},\ \bibinfo {pages} {72} (\bibinfo {year} {2018})},\
  \Eprint {http://arxiv.org/abs/1811.08693} {arXiv:1811.08693 [nucl-ex]}
  \BibitemShut {NoStop}%
\bibitem [{\citenamefont {Santopinto}\ \emph {et~al.}(2018)\citenamefont
  {Santopinto}, \citenamefont {Garc\'{\i}a-Tecocoatzi}, \citenamefont {Maga\~na
  Vsevolodovna},\ and\ \citenamefont {Ferretti}}]{Santopinto:2018}%
  \BibitemOpen
  \bibfield  {author} {\bibinfo {author} {\bibfnamefont {E.}~\bibnamefont
  {Santopinto}}, \bibinfo {author} {\bibfnamefont {H.}~\bibnamefont
  {Garc\'{\i}a-Tecocoatzi}}, \bibinfo {author} {\bibfnamefont {R.~I.}\
  \bibnamefont {Maga\~na Vsevolodovna}}, \ and\ \bibinfo {author}
  {\bibfnamefont {J.}~\bibnamefont {Ferretti}} (\bibinfo {collaboration} {NUMEN
  Collaboration}),\ }\href {\doibase 10.1103/PhysRevC.98.061601} {\bibfield
  {journal} {\bibinfo  {journal} {Phys. Rev. C}\ }\textbf {\bibinfo {volume}
  {98}},\ \bibinfo {pages} {061601} (\bibinfo {year} {2018})}\BibitemShut
  {NoStop}%
\bibitem [{\citenamefont {Brase}\ \emph {et~al.}(2021)\citenamefont {Brase},
  \citenamefont {Men\'endez}, \citenamefont {P\'erez},\ and\ \citenamefont
  {Schwenk}}]{Brase:2021}%
  \BibitemOpen
  \bibfield  {author} {\bibinfo {author} {\bibfnamefont {C.}~\bibnamefont
  {Brase}}, \bibinfo {author} {\bibfnamefont {J.}~\bibnamefont {Men\'endez}},
  \bibinfo {author} {\bibfnamefont {E.~A.~C.}\ \bibnamefont {P\'erez}}, \ and\
  \bibinfo {author} {\bibfnamefont {A.}~\bibnamefont {Schwenk}},\ }\href@noop
  {} {\  (\bibinfo {year} {2021})},\ \Eprint {http://arxiv.org/abs/2108.11805}
  {arXiv:2108.11805 [nucl-th]} \BibitemShut {NoStop}%
\bibitem [{\citenamefont {\ifmmode~\check{S}\else \v{S}\fi{}imkovic}\ \emph
  {et~al.}(2018)\citenamefont {\ifmmode~\check{S}\else \v{S}\fi{}imkovic},
  \citenamefont {Smetana},\ and\ \citenamefont {Vogel}}]{Simkovic:2018}%
  \BibitemOpen
  \bibfield  {author} {\bibinfo {author} {\bibfnamefont {F.}~\bibnamefont
  {\ifmmode~\check{S}\else \v{S}\fi{}imkovic}}, \bibinfo {author}
  {\bibfnamefont {A.}~\bibnamefont {Smetana}}, \ and\ \bibinfo {author}
  {\bibfnamefont {P.}~\bibnamefont {Vogel}},\ }\href {\doibase
  10.1103/PhysRevC.98.064325} {\bibfield  {journal} {\bibinfo  {journal} {Phys.
  Rev. C}\ }\textbf {\bibinfo {volume} {98}},\ \bibinfo {pages} {064325}
  (\bibinfo {year} {2018})}\BibitemShut {NoStop}%
\bibitem [{\citenamefont {Faessler}\ and\ \citenamefont
  {Simkovic}(1998)}]{Faessler:1998JPG}%
  \BibitemOpen
  \bibfield  {author} {\bibinfo {author} {\bibfnamefont {A.}~\bibnamefont
  {Faessler}}\ and\ \bibinfo {author} {\bibfnamefont {F.}~\bibnamefont
  {Simkovic}},\ }\href {\doibase 10.1088/0954-3899/24/12/001} {\bibfield
  {journal} {\bibinfo  {journal} {Journal of Physics G: Nuclear and Particle
  Physics}\ }\textbf {\bibinfo {volume} {24}},\ \bibinfo {pages} {2139}
  (\bibinfo {year} {1998})}\BibitemShut {NoStop}%
\bibitem [{\citenamefont {Rodin}\ \emph {et~al.}(2006)\citenamefont {Rodin},
  \citenamefont {Faessler}, \citenamefont {Šimkovic},\ and\ \citenamefont
  {Vogel}}]{Rodin:2006}%
  \BibitemOpen
  \bibfield  {author} {\bibinfo {author} {\bibfnamefont {V.}~\bibnamefont
  {Rodin}}, \bibinfo {author} {\bibfnamefont {A.}~\bibnamefont {Faessler}},
  \bibinfo {author} {\bibfnamefont {F.}~\bibnamefont {Šimkovic}}, \ and\
  \bibinfo {author} {\bibfnamefont {P.}~\bibnamefont {Vogel}},\ }\href
  {\doibase https://doi.org/10.1016/j.nuclphysa.2005.12.004} {\bibfield
  {journal} {\bibinfo  {journal} {Nuclear Physics A}\ }\textbf {\bibinfo
  {volume} {766}},\ \bibinfo {pages} {107} (\bibinfo {year}
  {2006})}\BibitemShut {NoStop}%
\bibitem [{\citenamefont {\v{S}imkovic}\ \emph {et~al.}(2013)\citenamefont
  {\v{S}imkovic}, \citenamefont {Rodin}, \citenamefont {Faessler},\ and\
  \citenamefont {Vogel}}]{Simkovic:2013}%
  \BibitemOpen
  \bibfield  {author} {\bibinfo {author} {\bibfnamefont {F.}~\bibnamefont
  {\v{S}imkovic}}, \bibinfo {author} {\bibfnamefont {V.}~\bibnamefont {Rodin}},
  \bibinfo {author} {\bibfnamefont {A.}~\bibnamefont {Faessler}}, \ and\
  \bibinfo {author} {\bibfnamefont {P.}~\bibnamefont {Vogel}},\ }\href
  {\doibase 10.1103/PhysRevC.87.045501} {\bibfield  {journal} {\bibinfo
  {journal} {Phys. Rev. C}\ }\textbf {\bibinfo {volume} {87}},\ \bibinfo
  {pages} {045501} (\bibinfo {year} {2013})},\ \Eprint
  {http://arxiv.org/abs/1302.1509} {arXiv:1302.1509 [nucl-th]} \BibitemShut
  {NoStop}%
\bibitem [{\citenamefont {Terasaki}(2015)}]{Terasaki:2015PRC}%
  \BibitemOpen
  \bibfield  {author} {\bibinfo {author} {\bibfnamefont {J.}~\bibnamefont
  {Terasaki}},\ }\href {\doibase 10.1103/PhysRevC.91.034318} {\bibfield
  {journal} {\bibinfo  {journal} {Phys. Rev. C}\ }\textbf {\bibinfo {volume}
  {91}},\ \bibinfo {pages} {034318} (\bibinfo {year} {2015})},\ \Eprint
  {http://arxiv.org/abs/1408.1545} {arXiv:1408.1545 [nucl-th]} \BibitemShut
  {NoStop}%
\bibitem [{\citenamefont {Cappuzzello}\ \emph {et~al.}(2020)\citenamefont
  {Cappuzzello} \emph {et~al.}}]{Cappuzzello:2020}%
  \BibitemOpen
  \bibfield  {author} {\bibinfo {author} {\bibfnamefont {F.}~\bibnamefont
  {Cappuzzello}} \emph {et~al.},\ }\href {\doibase
  10.1088/1742-6596/1643/1/012074} {\bibfield  {journal} {\bibinfo  {journal}
  {Journal of Physics: Conference Series}\ }\textbf {\bibinfo {volume}
  {1643}},\ \bibinfo {pages} {012074} (\bibinfo {year} {2020})}\BibitemShut
  {NoStop}%
\bibitem [{\citenamefont {Cirigliano}\ \emph
  {et~al.}(2018{\natexlab{a}})\citenamefont {Cirigliano}, \citenamefont
  {Dekens}, \citenamefont {de~Vries}, \citenamefont {Graesser}, \citenamefont
  {Mereghetti}, \citenamefont {Pastore},\ and\ \citenamefont {van
  Kolck}}]{Cirigliano:2018}%
  \BibitemOpen
  \bibfield  {author} {\bibinfo {author} {\bibfnamefont {V.}~\bibnamefont
  {Cirigliano}}, \bibinfo {author} {\bibfnamefont {W.}~\bibnamefont {Dekens}},
  \bibinfo {author} {\bibfnamefont {J.}~\bibnamefont {de~Vries}}, \bibinfo
  {author} {\bibfnamefont {M.~L.}\ \bibnamefont {Graesser}}, \bibinfo {author}
  {\bibfnamefont {E.}~\bibnamefont {Mereghetti}}, \bibinfo {author}
  {\bibfnamefont {S.}~\bibnamefont {Pastore}}, \ and\ \bibinfo {author}
  {\bibfnamefont {U.}~\bibnamefont {van Kolck}},\ }\href {\doibase
  10.1103/PhysRevLett.120.202001} {\bibfield  {journal} {\bibinfo  {journal}
  {Phys. Rev. Lett.}\ }\textbf {\bibinfo {volume} {120}},\ \bibinfo {pages}
  {202001} (\bibinfo {year} {2018}{\natexlab{a}})}\BibitemShut {NoStop}%
\bibitem [{\citenamefont {Cirigliano}\ \emph {et~al.}(2021)\citenamefont
  {Cirigliano}, \citenamefont {Dekens}, \citenamefont {de~Vries}, \citenamefont
  {Hoferichter},\ and\ \citenamefont {Mereghetti}}]{Cirigliano:2021PRL}%
  \BibitemOpen
  \bibfield  {author} {\bibinfo {author} {\bibfnamefont {V.}~\bibnamefont
  {Cirigliano}}, \bibinfo {author} {\bibfnamefont {W.}~\bibnamefont {Dekens}},
  \bibinfo {author} {\bibfnamefont {J.}~\bibnamefont {de~Vries}}, \bibinfo
  {author} {\bibfnamefont {M.}~\bibnamefont {Hoferichter}}, \ and\ \bibinfo
  {author} {\bibfnamefont {E.}~\bibnamefont {Mereghetti}},\ }\href {\doibase
  10.1103/PhysRevLett.126.172002} {\bibfield  {journal} {\bibinfo  {journal}
  {Phys. Rev. Lett.}\ }\textbf {\bibinfo {volume} {126}},\ \bibinfo {pages}
  {172002} (\bibinfo {year} {2021})},\ \Eprint
  {http://arxiv.org/abs/2012.11602} {arXiv:2012.11602 [nucl-th]} \BibitemShut
  {NoStop}%
\bibitem [{\citenamefont {Wirth}\ \emph {et~al.}(2021)\citenamefont {Wirth},
  \citenamefont {Yao},\ and\ \citenamefont {Hergert}}]{Wirth:2021}%
  \BibitemOpen
  \bibfield  {author} {\bibinfo {author} {\bibfnamefont {R.}~\bibnamefont
  {Wirth}}, \bibinfo {author} {\bibfnamefont {J.~M.}\ \bibnamefont {Yao}}, \
  and\ \bibinfo {author} {\bibfnamefont {H.}~\bibnamefont {Hergert}},\ }\href
  {\doibase 10.1103/PhysRevLett.127.242502} {\bibfield  {journal} {\bibinfo
  {journal} {Phys. Rev. Lett.}\ }\textbf {\bibinfo {volume} {127}},\ \bibinfo
  {pages} {242502} (\bibinfo {year} {2021})},\ \Eprint
  {http://arxiv.org/abs/2105.05415} {arXiv:2105.05415 [nucl-th]} \BibitemShut
  {NoStop}%
\bibitem [{\citenamefont {Haxton}\ and\ \citenamefont
  {Stephenson}(1984)}]{Haxton1984PPNP}%
  \BibitemOpen
  \bibfield  {author} {\bibinfo {author} {\bibfnamefont {W.}~\bibnamefont
  {Haxton}}\ and\ \bibinfo {author} {\bibfnamefont {G.}~\bibnamefont
  {Stephenson}},\ }\href {\doibase 10.1016/0146-6410(84)90006-1} {\bibfield
  {journal} {\bibinfo  {journal} {Prog. Part. Nucl. Phys.}\ }\textbf {\bibinfo
  {volume} {12}},\ \bibinfo {pages} {409 } (\bibinfo {year}
  {1984})}\BibitemShut {NoStop}%
\bibitem [{\citenamefont {Cirigliano}\ \emph
  {et~al.}(2018{\natexlab{b}})\citenamefont {Cirigliano}, \citenamefont
  {Dekens}, \citenamefont {Mereghetti},\ and\ \citenamefont
  {Walker-Loud}}]{Cirigliano:2018PRC}%
  \BibitemOpen
  \bibfield  {author} {\bibinfo {author} {\bibfnamefont {V.}~\bibnamefont
  {Cirigliano}}, \bibinfo {author} {\bibfnamefont {W.}~\bibnamefont {Dekens}},
  \bibinfo {author} {\bibfnamefont {E.}~\bibnamefont {Mereghetti}}, \ and\
  \bibinfo {author} {\bibfnamefont {A.}~\bibnamefont {Walker-Loud}},\ }\href
  {\doibase 10.1103/PhysRevC.97.065501} {\bibfield  {journal} {\bibinfo
  {journal} {Phys. Rev. C}\ }\textbf {\bibinfo {volume} {97}},\ \bibinfo
  {pages} {065501} (\bibinfo {year} {2018}{\natexlab{b}})}\BibitemShut
  {NoStop}%
\bibitem [{\citenamefont {\ifmmode~\check{S}\else \v{S}\fi{}imkovic}\ \emph
  {et~al.}(1999)\citenamefont {\ifmmode~\check{S}\else \v{S}\fi{}imkovic},
  \citenamefont {Pantis}, \citenamefont {Vergados},\ and\ \citenamefont
  {Faessler}}]{Simkovic:1999}%
  \BibitemOpen
  \bibfield  {author} {\bibinfo {author} {\bibfnamefont {F.}~\bibnamefont
  {\ifmmode~\check{S}\else \v{S}\fi{}imkovic}}, \bibinfo {author}
  {\bibfnamefont {G.}~\bibnamefont {Pantis}}, \bibinfo {author} {\bibfnamefont
  {J.~D.}\ \bibnamefont {Vergados}}, \ and\ \bibinfo {author} {\bibfnamefont
  {A.}~\bibnamefont {Faessler}},\ }\href {\doibase 10.1103/PhysRevC.60.055502}
  {\bibfield  {journal} {\bibinfo  {journal} {Phys. Rev. C}\ }\textbf {\bibinfo
  {volume} {60}},\ \bibinfo {pages} {055502} (\bibinfo {year}
  {1999})}\BibitemShut {NoStop}%
\bibitem [{\citenamefont {\ifmmode~\check{S}\else \v{S}\fi{}imkovic}\ \emph
  {et~al.}(2009)\citenamefont {\ifmmode~\check{S}\else \v{S}\fi{}imkovic},
  \citenamefont {Faessler}, \citenamefont {M\"uther}, \citenamefont {Rodin},\
  and\ \citenamefont {Stauf}}]{Simkovic:2009PRC}%
  \BibitemOpen
  \bibfield  {author} {\bibinfo {author} {\bibfnamefont {F.}~\bibnamefont
  {\ifmmode~\check{S}\else \v{S}\fi{}imkovic}}, \bibinfo {author}
  {\bibfnamefont {A.}~\bibnamefont {Faessler}}, \bibinfo {author}
  {\bibfnamefont {H.}~\bibnamefont {M\"uther}}, \bibinfo {author}
  {\bibfnamefont {V.}~\bibnamefont {Rodin}}, \ and\ \bibinfo {author}
  {\bibfnamefont {M.}~\bibnamefont {Stauf}},\ }\href {\doibase
  10.1103/PhysRevC.79.055501} {\bibfield  {journal} {\bibinfo  {journal} {Phys.
  Rev. C}\ }\textbf {\bibinfo {volume} {79}},\ \bibinfo {pages} {055501}
  (\bibinfo {year} {2009})}\BibitemShut {NoStop}%
\bibitem [{\citenamefont {Greuling}\ and\ \citenamefont
  {Whitten}(1960)}]{Greuling:1960}%
  \BibitemOpen
  \bibfield  {author} {\bibinfo {author} {\bibfnamefont {E.}~\bibnamefont
  {Greuling}}\ and\ \bibinfo {author} {\bibfnamefont {R.}~\bibnamefont
  {Whitten}},\ }\href {\doibase https://doi.org/10.1016/0003-4916(60)90010-5}
  {\bibfield  {journal} {\bibinfo  {journal} {Ann. Phys.}\ }\textbf {\bibinfo
  {volume} {11}},\ \bibinfo {pages} {510} (\bibinfo {year} {1960})}\BibitemShut
  {NoStop}%
\bibitem [{\citenamefont {\ifmmode~\check{S}\else \v{S}\fi{}imkovic}\ \emph
  {et~al.}(2008)\citenamefont {\ifmmode~\check{S}\else \v{S}\fi{}imkovic},
  \citenamefont {Faessler}, \citenamefont {Rodin}, \citenamefont {Vogel},\ and\
  \citenamefont {Engel}}]{Simkovic:2008}%
  \BibitemOpen
  \bibfield  {author} {\bibinfo {author} {\bibfnamefont {F.}~\bibnamefont
  {\ifmmode~\check{S}\else \v{S}\fi{}imkovic}}, \bibinfo {author}
  {\bibfnamefont {A.}~\bibnamefont {Faessler}}, \bibinfo {author}
  {\bibfnamefont {V.}~\bibnamefont {Rodin}}, \bibinfo {author} {\bibfnamefont
  {P.}~\bibnamefont {Vogel}}, \ and\ \bibinfo {author} {\bibfnamefont
  {J.}~\bibnamefont {Engel}},\ }\href {\doibase 10.1103/PhysRevC.77.045503}
  {\bibfield  {journal} {\bibinfo  {journal} {Phys. Rev. C}\ }\textbf {\bibinfo
  {volume} {77}},\ \bibinfo {pages} {045503} (\bibinfo {year}
  {2008})}\BibitemShut {NoStop}%
\bibitem [{\citenamefont {Roth}(2009)}]{Roth:2009}%
  \BibitemOpen
  \bibfield  {author} {\bibinfo {author} {\bibfnamefont {R.}~\bibnamefont
  {Roth}},\ }\href {\doibase 10.1103/PhysRevC.79.064324} {\bibfield  {journal}
  {\bibinfo  {journal} {Phys. Rev. C}\ }\textbf {\bibinfo {volume} {79}},\
  \bibinfo {pages} {064324} (\bibinfo {year} {2009})}\BibitemShut {NoStop}%
\bibitem [{\citenamefont {Stroberg}\ \emph {et~al.}(2019)\citenamefont
  {Stroberg}, \citenamefont {Hergert}, \citenamefont {Bogner},\ and\
  \citenamefont {Holt}}]{Stroberg:2019}%
  \BibitemOpen
  \bibfield  {author} {\bibinfo {author} {\bibfnamefont {S.~R.}\ \bibnamefont
  {Stroberg}}, \bibinfo {author} {\bibfnamefont {H.}~\bibnamefont {Hergert}},
  \bibinfo {author} {\bibfnamefont {S.~K.}\ \bibnamefont {Bogner}}, \ and\
  \bibinfo {author} {\bibfnamefont {J.~D.}\ \bibnamefont {Holt}},\ }\href
  {\doibase 10.1146/annurev-nucl-101917-021120} {\bibfield  {journal} {\bibinfo
   {journal} {Annu. Rev. Nucl. Part. Sci.}\ }\textbf {\bibinfo {volume} {69}},\
  \bibinfo {pages} {307} (\bibinfo {year} {2019})}\BibitemShut {NoStop}%
\bibitem [{\citenamefont {Yao}\ \emph {et~al.}(2018)\citenamefont {Yao},
  \citenamefont {Engel}, \citenamefont {Wang}, \citenamefont {Jiao},\ and\
  \citenamefont {Hergert}}]{Yao:2018wq}%
  \BibitemOpen
  \bibfield  {author} {\bibinfo {author} {\bibfnamefont {J.~M.}\ \bibnamefont
  {Yao}}, \bibinfo {author} {\bibfnamefont {J.}~\bibnamefont {Engel}}, \bibinfo
  {author} {\bibfnamefont {L.~J.}\ \bibnamefont {Wang}}, \bibinfo {author}
  {\bibfnamefont {C.~F.}\ \bibnamefont {Jiao}}, \ and\ \bibinfo {author}
  {\bibfnamefont {H.}~\bibnamefont {Hergert}},\ }\href {\doibase
  10.1103/PhysRevC.98.054311} {\bibfield  {journal} {\bibinfo  {journal} {Phys.
  Rev. C}\ }\textbf {\bibinfo {volume} {98}},\ \bibinfo {pages} {054311}
  (\bibinfo {year} {2018})}\BibitemShut {NoStop}%
\bibitem [{\citenamefont {Hergert}\ \emph {et~al.}(2016)\citenamefont
  {Hergert}, \citenamefont {Bogner}, \citenamefont {Morris}, \citenamefont
  {Schwenk},\ and\ \citenamefont {Tsukiyama}}]{Hergert:2016jk}%
  \BibitemOpen
  \bibfield  {author} {\bibinfo {author} {\bibfnamefont {H.}~\bibnamefont
  {Hergert}}, \bibinfo {author} {\bibfnamefont {S.~K.}\ \bibnamefont {Bogner}},
  \bibinfo {author} {\bibfnamefont {T.~D.}\ \bibnamefont {Morris}}, \bibinfo
  {author} {\bibfnamefont {A.}~\bibnamefont {Schwenk}}, \ and\ \bibinfo
  {author} {\bibfnamefont {K.}~\bibnamefont {Tsukiyama}},\ }\bibfield
  {booktitle} {\emph {\bibinfo {booktitle} {Memorial Volume in Honor of Gerald
  E. Brown}},\ }\href {\doibase
  http://dx.doi.org/10.1016/j.physrep.2015.12.007} {\bibfield  {journal}
  {\bibinfo  {journal} {Physics Reports}\ }\textbf {\bibinfo {volume} {621}},\
  \bibinfo {pages} {165} (\bibinfo {year} {2016})}\BibitemShut {NoStop}%
\bibitem [{\citenamefont {Entem}\ and\ \citenamefont
  {Machleidt}(2003)}]{Entem:2003}%
  \BibitemOpen
  \bibfield  {author} {\bibinfo {author} {\bibfnamefont {D.~R.}\ \bibnamefont
  {Entem}}\ and\ \bibinfo {author} {\bibfnamefont {R.}~\bibnamefont
  {Machleidt}},\ }\href {\doibase 10.1103/PhysRevC.68.041001} {\bibfield
  {journal} {\bibinfo  {journal} {Phys. Rev. C}\ }\textbf {\bibinfo {volume}
  {68}},\ \bibinfo {pages} {041001} (\bibinfo {year} {2003})}\BibitemShut
  {NoStop}%
\bibitem [{\citenamefont {Bogner}\ \emph {et~al.}(2010)\citenamefont {Bogner},
  \citenamefont {Furnstahl},\ and\ \citenamefont {Schwenk}}]{Bogner:2010}%
  \BibitemOpen
  \bibfield  {author} {\bibinfo {author} {\bibfnamefont {S.}~\bibnamefont
  {Bogner}}, \bibinfo {author} {\bibfnamefont {R.}~\bibnamefont {Furnstahl}}, \
  and\ \bibinfo {author} {\bibfnamefont {A.}~\bibnamefont {Schwenk}},\ }\href
  {\doibase 10.1016/j.ppnp.2010.03.001} {\bibfield  {journal} {\bibinfo
  {journal} {Prog. Part. Nucl. Phys.}\ }\textbf {\bibinfo {volume} {65}},\
  \bibinfo {pages} {94 } (\bibinfo {year} {2010})}\BibitemShut {NoStop}%
\bibitem [{\citenamefont {Hebeler}\ \emph {et~al.}(2011)\citenamefont
  {Hebeler}, \citenamefont {Bogner}, \citenamefont {Furnstahl}, \citenamefont
  {Nogga},\ and\ \citenamefont {Schwenk}}]{Hebeler:2011}%
  \BibitemOpen
  \bibfield  {author} {\bibinfo {author} {\bibfnamefont {K.}~\bibnamefont
  {Hebeler}}, \bibinfo {author} {\bibfnamefont {S.~K.}\ \bibnamefont {Bogner}},
  \bibinfo {author} {\bibfnamefont {R.~J.}\ \bibnamefont {Furnstahl}}, \bibinfo
  {author} {\bibfnamefont {A.}~\bibnamefont {Nogga}}, \ and\ \bibinfo {author}
  {\bibfnamefont {A.}~\bibnamefont {Schwenk}},\ }\href {\doibase
  10.1103/PhysRevC.83.031301} {\bibfield  {journal} {\bibinfo  {journal} {Phys.
  Rev. C}\ }\textbf {\bibinfo {volume} {83}},\ \bibinfo {pages} {031301}
  (\bibinfo {year} {2011})}\BibitemShut {NoStop}%
\bibitem [{\citenamefont {Nogga}\ \emph {et~al.}(2004)\citenamefont {Nogga},
  \citenamefont {Bogner},\ and\ \citenamefont {Schwenk}}]{Nogga:2004il}%
  \BibitemOpen
  \bibfield  {author} {\bibinfo {author} {\bibfnamefont {A.}~\bibnamefont
  {Nogga}}, \bibinfo {author} {\bibfnamefont {S.~K.}\ \bibnamefont {Bogner}}, \
  and\ \bibinfo {author} {\bibfnamefont {A.}~\bibnamefont {Schwenk}},\ }\href
  {\doibase 10.1103/PhysRevC.70.061002} {\bibfield  {journal} {\bibinfo
  {journal} {Phys. Rev. C}\ }\textbf {\bibinfo {volume} {70}},\ \bibinfo
  {pages} {061002} (\bibinfo {year} {2004})}\BibitemShut {NoStop}%
\bibitem [{\citenamefont {Jiang}\ \emph {et~al.}(2020)\citenamefont {Jiang},
  \citenamefont {Ekstr{\"{o}}m}, \citenamefont {Forss{\'{e}}n}, \citenamefont
  {Hagen}, \citenamefont {Jansen},\ and\ \citenamefont
  {Papenbrock}}]{Jiang2020}%
  \BibitemOpen
  \bibfield  {author} {\bibinfo {author} {\bibfnamefont {W.~G.}\ \bibnamefont
  {Jiang}}, \bibinfo {author} {\bibfnamefont {A.}~\bibnamefont
  {Ekstr{\"{o}}m}}, \bibinfo {author} {\bibfnamefont {C.}~\bibnamefont
  {Forss{\'{e}}n}}, \bibinfo {author} {\bibfnamefont {G.}~\bibnamefont
  {Hagen}}, \bibinfo {author} {\bibfnamefont {G.~R.}\ \bibnamefont {Jansen}}, \
  and\ \bibinfo {author} {\bibfnamefont {T.}~\bibnamefont {Papenbrock}},\
  }\href {\doibase 10.1103/PhysRevC.102.054301} {\bibfield  {journal} {\bibinfo
   {journal} {Phys. Rev. C}\ }\textbf {\bibinfo {volume} {102}},\ \bibinfo
  {pages} {054301} (\bibinfo {year} {2020})},\ \Eprint
  {http://arxiv.org/abs/2006.16774} {arXiv:2006.16774} \BibitemShut {NoStop}%
\bibitem [{\citenamefont {Bowley}(1928)}]{Bowley1928}%
  \BibitemOpen
  \bibfield  {author} {\bibinfo {author} {\bibfnamefont {A.~L.}\ \bibnamefont
  {Bowley}},\ }\href {\doibase 10.1080/01621459.1928.10502991} {\bibfield
  {journal} {\bibinfo  {journal} {Journal of the American Statistical
  Association}\ }\textbf {\bibinfo {volume} {23}},\ \bibinfo {pages} {31}
  (\bibinfo {year} {1928})},\ \Eprint
  {http://arxiv.org/abs/https://www.tandfonline.com/doi/pdf/10.1080/01621459.1928.10502991}
  {https://www.tandfonline.com/doi/pdf/10.1080/01621459.1928.10502991}
  \BibitemShut {NoStop}%
\bibitem [{\citenamefont {Rodriguez}\ and\ \citenamefont
  {Martinez-Pinedo}(2013)}]{Rodriguez:2013PLB}%
  \BibitemOpen
  \bibfield  {author} {\bibinfo {author} {\bibfnamefont {T.~R.}\ \bibnamefont
  {Rodriguez}}\ and\ \bibinfo {author} {\bibfnamefont {G.}~\bibnamefont
  {Martinez-Pinedo}},\ }\href {\doibase 10.1016/j.physletb.2012.12.063}
  {\bibfield  {journal} {\bibinfo  {journal} {Phys. Lett. B}\ }\textbf
  {\bibinfo {volume} {719}},\ \bibinfo {pages} {174} (\bibinfo {year}
  {2013})},\ \Eprint {http://arxiv.org/abs/1210.3225} {arXiv:1210.3225
  [nucl-th]} \BibitemShut {NoStop}%
\bibitem [{\citenamefont {Anderson}\ \emph {et~al.}(2010)\citenamefont
  {Anderson}, \citenamefont {Bogner}, \citenamefont {Furnstahl},\ and\
  \citenamefont {Perry}}]{Anderson:2010}%
  \BibitemOpen
  \bibfield  {author} {\bibinfo {author} {\bibfnamefont {E.~R.}\ \bibnamefont
  {Anderson}}, \bibinfo {author} {\bibfnamefont {S.~K.}\ \bibnamefont
  {Bogner}}, \bibinfo {author} {\bibfnamefont {R.~J.}\ \bibnamefont
  {Furnstahl}}, \ and\ \bibinfo {author} {\bibfnamefont {R.~J.}\ \bibnamefont
  {Perry}},\ }\href {\doibase 10.1103/PhysRevC.82.054001} {\bibfield  {journal}
  {\bibinfo  {journal} {Phys. Rev. C}\ }\textbf {\bibinfo {volume} {82}},\
  \bibinfo {pages} {054001} (\bibinfo {year} {2010})}\BibitemShut {NoStop}%
\bibitem [{\citenamefont {Bogner}\ and\ \citenamefont
  {Roscher}(2012)}]{Bogner:2012}%
  \BibitemOpen
  \bibfield  {author} {\bibinfo {author} {\bibfnamefont {S.~K.}\ \bibnamefont
  {Bogner}}\ and\ \bibinfo {author} {\bibfnamefont {D.}~\bibnamefont
  {Roscher}},\ }\href {\doibase 10.1103/PhysRevC.86.064304} {\bibfield
  {journal} {\bibinfo  {journal} {Phys. Rev. C}\ }\textbf {\bibinfo {volume}
  {86}},\ \bibinfo {pages} {064304} (\bibinfo {year} {2012})}\BibitemShut
  {NoStop}%
\bibitem [{\citenamefont {Cruz-Torres}\ \emph {et~al.}(2021)\citenamefont
  {Cruz-Torres} \emph {et~al.}}]{Cruz-Torres:2021}%
  \BibitemOpen
  \bibfield  {author} {\bibinfo {author} {\bibfnamefont {R.}~\bibnamefont
  {Cruz-Torres}} \emph {et~al.},\ }\href {\doibase 10.1038/s41567-020-01053-7}
  {\bibfield  {journal} {\bibinfo  {journal} {Nature Phys.}\ }\textbf {\bibinfo
  {volume} {17}},\ \bibinfo {pages} {306} (\bibinfo {year} {2021})},\ \Eprint
  {http://arxiv.org/abs/1907.03658} {arXiv:1907.03658 [nucl-th]} \BibitemShut
  {NoStop}%
\bibitem [{\citenamefont {Weiss}\ \emph {et~al.}(2021)\citenamefont {Weiss},
  \citenamefont {Soriano}, \citenamefont {Lovato}, \citenamefont {Menendez},\
  and\ \citenamefont {Wiringa}}]{Weiss:2021}%
  \BibitemOpen
  \bibfield  {author} {\bibinfo {author} {\bibfnamefont {R.}~\bibnamefont
  {Weiss}}, \bibinfo {author} {\bibfnamefont {P.}~\bibnamefont {Soriano}},
  \bibinfo {author} {\bibfnamefont {A.}~\bibnamefont {Lovato}}, \bibinfo
  {author} {\bibfnamefont {J.}~\bibnamefont {Menendez}}, \ and\ \bibinfo
  {author} {\bibfnamefont {R.~B.}\ \bibnamefont {Wiringa}},\ }\href@noop {} {\
  (\bibinfo {year} {2021})},\ \Eprint {http://arxiv.org/abs/2112.08146}
  {arXiv:2112.08146 [nucl-th]} \BibitemShut {NoStop}%
\bibitem [{\citenamefont {Weiss}\ \emph {et~al.}(2019)\citenamefont {Weiss},
  \citenamefont {Schmidt}, \citenamefont {Miller},\ and\ \citenamefont
  {Barnea}}]{Weiss:2018}%
  \BibitemOpen
  \bibfield  {author} {\bibinfo {author} {\bibfnamefont {R.}~\bibnamefont
  {Weiss}}, \bibinfo {author} {\bibfnamefont {A.}~\bibnamefont {Schmidt}},
  \bibinfo {author} {\bibfnamefont {G.~A.}\ \bibnamefont {Miller}}, \ and\
  \bibinfo {author} {\bibfnamefont {N.}~\bibnamefont {Barnea}},\ }\href
  {\doibase https://doi.org/10.1016/j.physletb.2019.01.053} {\bibfield
  {journal} {\bibinfo  {journal} {Phys. Lett. B}\ }\textbf {\bibinfo {volume}
  {790}},\ \bibinfo {pages} {484} (\bibinfo {year} {2019})}\BibitemShut
  {NoStop}%
\bibitem [{\citenamefont {Tropiano}\ \emph {et~al.}(2021)\citenamefont
  {Tropiano}, \citenamefont {Bogner},\ and\ \citenamefont
  {Furnstahl}}]{Tropiano:2021}%
  \BibitemOpen
  \bibfield  {author} {\bibinfo {author} {\bibfnamefont {A.~J.}\ \bibnamefont
  {Tropiano}}, \bibinfo {author} {\bibfnamefont {S.~K.}\ \bibnamefont
  {Bogner}}, \ and\ \bibinfo {author} {\bibfnamefont {R.~J.}\ \bibnamefont
  {Furnstahl}},\ }\href {\doibase 10.1103/PhysRevC.104.034311} {\bibfield
  {journal} {\bibinfo  {journal} {Phys. Rev. C}\ }\textbf {\bibinfo {volume}
  {104}},\ \bibinfo {pages} {034311} (\bibinfo {year} {2021})}\BibitemShut
  {NoStop}%
\bibitem [{\citenamefont {Cruz-Torres}\ \emph {et~al.}(2018)\citenamefont
  {Cruz-Torres}, \citenamefont {Schmidt}, \citenamefont {Miller}, \citenamefont
  {Weinstein}, \citenamefont {Barnea}, \citenamefont {Weiss}, \citenamefont
  {Piasetzky},\ and\ \citenamefont {Hen}}]{Cruz:2018PLB}%
  \BibitemOpen
  \bibfield  {author} {\bibinfo {author} {\bibfnamefont {R.}~\bibnamefont
  {Cruz-Torres}}, \bibinfo {author} {\bibfnamefont {A.}~\bibnamefont
  {Schmidt}}, \bibinfo {author} {\bibfnamefont {G.}~\bibnamefont {Miller}},
  \bibinfo {author} {\bibfnamefont {L.}~\bibnamefont {Weinstein}}, \bibinfo
  {author} {\bibfnamefont {N.}~\bibnamefont {Barnea}}, \bibinfo {author}
  {\bibfnamefont {R.}~\bibnamefont {Weiss}}, \bibinfo {author} {\bibfnamefont
  {E.}~\bibnamefont {Piasetzky}}, \ and\ \bibinfo {author} {\bibfnamefont
  {O.}~\bibnamefont {Hen}},\ }\href {\doibase
  https://doi.org/10.1016/j.physletb.2018.07.069} {\bibfield  {journal}
  {\bibinfo  {journal} {Phys. Lett. B}\ }\textbf {\bibinfo {volume} {785}},\
  \bibinfo {pages} {304} (\bibinfo {year} {2018})}\BibitemShut {NoStop}%
\bibitem [{\citenamefont {Brown}\ and\ \citenamefont
  {Richter}(2006)}]{Brown:2006PRC}%
  \BibitemOpen
  \bibfield  {author} {\bibinfo {author} {\bibfnamefont {B.~A.}\ \bibnamefont
  {Brown}}\ and\ \bibinfo {author} {\bibfnamefont {W.~A.}\ \bibnamefont
  {Richter}},\ }\href {\doibase 10.1103/PhysRevC.74.034315} {\bibfield
  {journal} {\bibinfo  {journal} {Phys. Rev. C}\ }\textbf {\bibinfo {volume}
  {74}},\ \bibinfo {pages} {034315} (\bibinfo {year} {2006})}\BibitemShut
  {NoStop}%
\bibitem [{\citenamefont {Honma}\ \emph {et~al.}(2005)\citenamefont {Honma},
  \citenamefont {Otsuka}, \citenamefont {Brown},\ and\ \citenamefont
  {Mizusaki}}]{Honma2005}%
  \BibitemOpen
  \bibfield  {author} {\bibinfo {author} {\bibfnamefont {M.}~\bibnamefont
  {Honma}}, \bibinfo {author} {\bibfnamefont {T.}~\bibnamefont {Otsuka}},
  \bibinfo {author} {\bibfnamefont {B.~A.}\ \bibnamefont {Brown}}, \ and\
  \bibinfo {author} {\bibfnamefont {T.}~\bibnamefont {Mizusaki}},\ }\href
  {\doibase 10.1140/epjad/i2005-06-032-2} {\bibfield  {journal} {\bibinfo
  {journal} {The European Physical Journal A - Hadrons and Nuclei}\ }\textbf
  {\bibinfo {volume} {25}},\ \bibinfo {pages} {499} (\bibinfo {year}
  {2005})}\BibitemShut {NoStop}%
\bibitem [{\citenamefont {Hu}\ \emph {et~al.}(2021)\citenamefont {Hu} \emph
  {et~al.}}]{Hu:2021}%
  \BibitemOpen
  \bibfield  {author} {\bibinfo {author} {\bibfnamefont {B.}~\bibnamefont {Hu}}
  \emph {et~al.},\ }\href@noop {} {\  (\bibinfo {year} {2021})},\ \Eprint
  {http://arxiv.org/abs/2112.01125} {arXiv:2112.01125 [nucl-th]} \BibitemShut
  {NoStop}%
\bibitem [{\citenamefont {Romeo}\ \emph {et~al.}(2022)\citenamefont {Romeo},
  \citenamefont {Men\'endez},\ and\ \citenamefont {Pe\~na Garay}}]{Romeo:2021}%
  \BibitemOpen
  \bibfield  {author} {\bibinfo {author} {\bibfnamefont {B.}~\bibnamefont
  {Romeo}}, \bibinfo {author} {\bibfnamefont {J.}~\bibnamefont {Men\'endez}}, \
  and\ \bibinfo {author} {\bibfnamefont {C.}~\bibnamefont {Pe\~na Garay}},\
  }\href {\doibase 10.1016/j.physletb.2022.136965} {\bibfield  {journal}
  {\bibinfo  {journal} {Phys. Lett. B}\ }\textbf {\bibinfo {volume} {827}},\
  \bibinfo {pages} {136965} (\bibinfo {year} {2022})},\ \Eprint
  {http://arxiv.org/abs/2102.11101} {arXiv:2102.11101 [nucl-th]} \BibitemShut
  {NoStop}%
\end{thebibliography}
 
%

\end{document}